\documentclass[%
 reprint,
superscriptaddress,
nofootinbib,
 amsmath,amssymb,
 aps,
floatfix,
]{revtex4-1}

\usepackage{graphicx}
\usepackage{dcolumn}
\usepackage{bm}
\usepackage{hyperref}
\usepackage{float}

\begin{document}

\preprint{APS/123-QED}

\title{Discrimination of electronic recoils from nuclear recoils\\in two-phase xenon time projection chambers}

\author{D.S.~Akerib} \affiliation{SLAC National Accelerator Laboratory, 2575 Sand Hill Road, Menlo Park, CA 94205, USA} \affiliation{Kavli Institute for Particle Astrophysics and Cosmology, Stanford University, 452 Lomita Mall, Stanford, CA 94309, USA} 
\author{S.~Alsum} \affiliation{University of Wisconsin-Madison, Department of Physics, 1150 University Ave., Madison, WI 53706, USA}  
\author{H.M.~Ara\'{u}jo} \affiliation{Imperial College London, High Energy Physics, Blackett Laboratory, London SW7 2BZ, United Kingdom}  
\author{X.~Bai} \affiliation{South Dakota School of Mines and Technology, 501 East St Joseph St., Rapid City, SD 57701, USA}  
\author{J.~Balajthy} \affiliation{University of California Davis, Department of Physics, One Shields Ave., Davis, CA 95616, USA}  
\author{A.~Baxter} \affiliation{University of Liverpool, Department of Physics, Liverpool L69 7ZE, UK}  
\author{E.P.~Bernard} \affiliation{University of California Berkeley, Department of Physics, Berkeley, CA 94720, USA}  
\author{A.~Bernstein} \affiliation{Lawrence Livermore National Laboratory, 7000 East Ave., Livermore, CA 94551, USA}  
\author{T.P.~Biesiadzinski} \affiliation{SLAC National Accelerator Laboratory, 2575 Sand Hill Road, Menlo Park, CA 94205, USA} \affiliation{Kavli Institute for Particle Astrophysics and Cosmology, Stanford University, 452 Lomita Mall, Stanford, CA 94309, USA} 
\author{E.M.~Boulton} \affiliation{University of California Berkeley, Department of Physics, Berkeley, CA 94720, USA} \affiliation{Lawrence Berkeley National Laboratory, 1 Cyclotron Rd., Berkeley, CA 94720, USA} \affiliation{Yale University, Department of Physics, 217 Prospect St., New Haven, CT 06511, USA}
\author{B.~Boxer} \affiliation{University of Liverpool, Department of Physics, Liverpool L69 7ZE, UK}  
\author{P.~Br\'as} \affiliation{LIP-Coimbra, Department of Physics, University of Coimbra, Rua Larga, 3004-516 Coimbra, Portugal}  
\author{S.~Burdin} \affiliation{University of Liverpool, Department of Physics, Liverpool L69 7ZE, UK}  
\author{D.~Byram} \affiliation{University of South Dakota, Department of Physics, 414E Clark St., Vermillion, SD 57069, USA} \affiliation{South Dakota Science and Technology Authority, Sanford Underground Research Facility, Lead, SD 57754, USA} 
\author{M.C.~Carmona-Benitez} \affiliation{Pennsylvania State University, Department of Physics, 104 Davey Lab, University Park, PA  16802-6300, USA}  
\author{C.~Chan} \affiliation{Brown University, Department of Physics, 182 Hope St., Providence, RI 02912, USA}  
\author{J.E.~Cutter} \affiliation{University of California Davis, Department of Physics, One Shields Ave., Davis, CA 95616, USA}  
\author{L.~de\,Viveiros}  \affiliation{Pennsylvania State University, Department of Physics, 104 Davey Lab, University Park, PA  16802-6300, USA}  
\author{E.~Druszkiewicz} \affiliation{University of Rochester, Department of Physics and Astronomy, Rochester, NY 14627, USA}  
\author{A.~Fan} \affiliation{SLAC National Accelerator Laboratory, 2575 Sand Hill Road, Menlo Park, CA 94205, USA} \affiliation{Kavli Institute for Particle Astrophysics and Cosmology, Stanford University, 452 Lomita Mall, Stanford, CA 94309, USA} 
\author{S.~Fiorucci} \affiliation{Lawrence Berkeley National Laboratory, 1 Cyclotron Rd., Berkeley, CA 94720, USA} \affiliation{Brown University, Department of Physics, 182 Hope St., Providence, RI 02912, USA} 
\author{R.J.~Gaitskell} \affiliation{Brown University, Department of Physics, 182 Hope St., Providence, RI 02912, USA}  
\author{C.~Ghag} \affiliation{Department of Physics and Astronomy, University College London, Gower Street, London WC1E 6BT, United Kingdom}  
\author{M.G.D.~Gilchriese} \affiliation{Lawrence Berkeley National Laboratory, 1 Cyclotron Rd., Berkeley, CA 94720, USA}  
\author{C.~Gwilliam} \affiliation{University of Liverpool, Department of Physics, Liverpool L69 7ZE, UK}  
\author{C.R.~Hall} \affiliation{University of Maryland, Department of Physics, College Park, MD 20742, USA}  
\author{S.J.~Haselschwardt} \affiliation{University of California Santa Barbara, Department of Physics, Santa Barbara, CA 93106, USA}  
\author{S.A.~Hertel} \affiliation{University of Massachusetts, Amherst Center for Fundamental Interactions and Department of Physics, Amherst, MA 01003-9337 USA} \affiliation{Lawrence Berkeley National Laboratory, 1 Cyclotron Rd., Berkeley, CA 94720, USA} 
\author{D.P.~Hogan} \affiliation{University of California Berkeley, Department of Physics, Berkeley, CA 94720, USA}  
\author{M.~Horn} \affiliation{South Dakota Science and Technology Authority, Sanford Underground Research Facility, Lead, SD 57754, USA} \affiliation{University of California Berkeley, Department of Physics, Berkeley, CA 94720, USA} 
\author{D.Q.~Huang} \affiliation{Brown University, Department of Physics, 182 Hope St., Providence, RI 02912, USA}  
\author{C.M.~Ignarra} \affiliation{SLAC National Accelerator Laboratory, 2575 Sand Hill Road, Menlo Park, CA 94205, USA} \affiliation{Kavli Institute for Particle Astrophysics and Cosmology, Stanford University, 452 Lomita Mall, Stanford, CA 94309, USA} 
\author{R.G.~Jacobsen} \affiliation{University of California Berkeley, Department of Physics, Berkeley, CA 94720, USA}  
\author{O.~Jahangir} \affiliation{Department of Physics and Astronomy, University College London, Gower Street, London WC1E 6BT, United Kingdom}  
\author{W.~Ji} \affiliation{SLAC National Accelerator Laboratory, 2575 Sand Hill Road, Menlo Park, CA 94205, USA} \affiliation{Kavli Institute for Particle Astrophysics and Cosmology, Stanford University, 452 Lomita Mall, Stanford, CA 94309, USA} 
\author{K.~Kamdin} \affiliation{University of California Berkeley, Department of Physics, Berkeley, CA 94720, USA} \affiliation{Lawrence Berkeley National Laboratory, 1 Cyclotron Rd., Berkeley, CA 94720, USA} 
\author{K.~Kazkaz} \affiliation{Lawrence Livermore National Laboratory, 7000 East Ave., Livermore, CA 94551, USA}  
\author{D.~Khaitan} \affiliation{University of Rochester, Department of Physics and Astronomy, Rochester, NY 14627, USA}  
\author{E.V.~Korolkova} \affiliation{University of Sheffield, Department of Physics and Astronomy, Sheffield, S3 7RH, United Kingdom}  
\author{S.~Kravitz} \affiliation{Lawrence Berkeley National Laboratory, 1 Cyclotron Rd., Berkeley, CA 94720, USA}  
\author{V.A.~Kudryavtsev} \affiliation{University of Sheffield, Department of Physics and Astronomy, Sheffield, S3 7RH, United Kingdom}  
\author{E.~Leason} \affiliation{SUPA, School of Physics and Astronomy, University of Edinburgh, Edinburgh EH9 3FD, United Kingdom}  
\author{B.G.~Lenardo} \affiliation{University of California Davis, Department of Physics, One Shields Ave., Davis, CA 95616, USA} \affiliation{Lawrence Livermore National Laboratory, 7000 East Ave., Livermore, CA 94551, USA} 
\author{K.T.~Lesko} \affiliation{Lawrence Berkeley National Laboratory, 1 Cyclotron Rd., Berkeley, CA 94720, USA}  
\author{J.~Liao} \affiliation{Brown University, Department of Physics, 182 Hope St., Providence, RI 02912, USA}  
\author{J.~Lin} \affiliation{University of California Berkeley, Department of Physics, Berkeley, CA 94720, USA}  
\author{A.~Lindote} \affiliation{LIP-Coimbra, Department of Physics, University of Coimbra, Rua Larga, 3004-516 Coimbra, Portugal}  
\author{M.I.~Lopes} \affiliation{LIP-Coimbra, Department of Physics, University of Coimbra, Rua Larga, 3004-516 Coimbra, Portugal}  
\author{A.~Manalaysay} \affiliation{Lawrence Berkeley National Laboratory, 1 Cyclotron Rd., Berkeley, CA 94720, USA} \affiliation{University of California Davis, Department of Physics, One Shields Ave., Davis, CA 95616, USA} 
\author{R.L.~Mannino} \affiliation{Texas A \& M University, Department of Physics, College Station, TX 77843, USA} \affiliation{University of Wisconsin-Madison, Department of Physics, 1150 University Ave., Madison, WI 53706, USA} 
\author{N.~Marangou} \affiliation{Imperial College London, High Energy Physics, Blackett Laboratory, London SW7 2BZ, United Kingdom}  
\author{D.N.~McKinsey} \affiliation{University of California Berkeley, Department of Physics, Berkeley, CA 94720, USA} \affiliation{Lawrence Berkeley National Laboratory, 1 Cyclotron Rd., Berkeley, CA 94720, USA} 
\author{D.-M.~Mei} \affiliation{University of South Dakota, Department of Physics, 414E Clark St., Vermillion, SD 57069, USA}  
\author{M.~Moongweluwan} \affiliation{University of Rochester, Department of Physics and Astronomy, Rochester, NY 14627, USA}  
\author{J.A.~Morad} \affiliation{University of California Davis, Department of Physics, One Shields Ave., Davis, CA 95616, USA}  
\author{A.St.J.~Murphy} \affiliation{SUPA, School of Physics and Astronomy, University of Edinburgh, Edinburgh EH9 3FD, United Kingdom}  
\author{A.~Naylor} \affiliation{University of Sheffield, Department of Physics and Astronomy, Sheffield, S3 7RH, United Kingdom}  
\author{C.~Nehrkorn} \affiliation{University of California Santa Barbara, Department of Physics, Santa Barbara, CA 93106, USA}  
\author{H.N.~Nelson} \affiliation{University of California Santa Barbara, Department of Physics, Santa Barbara, CA 93106, USA}  
\author{F.~Neves} \affiliation{LIP-Coimbra, Department of Physics, University of Coimbra, Rua Larga, 3004-516 Coimbra, Portugal}  
\author{A.~Nilima} \affiliation{SUPA, School of Physics and Astronomy, University of Edinburgh, Edinburgh EH9 3FD, United Kingdom}  
\author{K.C.~Oliver-Mallory} \affiliation{University of California Berkeley, Department of Physics, Berkeley, CA 94720, USA} \affiliation{Lawrence Berkeley National Laboratory, 1 Cyclotron Rd., Berkeley, CA 94720, USA} 
\author{K.J.~Palladino} \affiliation{University of Wisconsin-Madison, Department of Physics, 1150 University Ave., Madison, WI 53706, USA}  
\author{E.K.~Pease} \affiliation{University of California Berkeley, Department of Physics, Berkeley, CA 94720, USA} \affiliation{Lawrence Berkeley National Laboratory, 1 Cyclotron Rd., Berkeley, CA 94720, USA} 
\author{Q.~Riffard} \affiliation{University of California Berkeley, Department of Physics, Berkeley, CA 94720, USA} \affiliation{Lawrence Berkeley National Laboratory, 1 Cyclotron Rd., Berkeley, CA 94720, USA} 
\author{G.R.C.~Rischbieter} \affiliation{University at Albany, State University of New York, Department of Physics, 1400 Washington Ave., Albany, NY 12222, USA}  
\author{C.~Rhyne} \affiliation{Brown University, Department of Physics, 182 Hope St., Providence, RI 02912, USA}  
\author{P.~Rossiter} \affiliation{University of Sheffield, Department of Physics and Astronomy, Sheffield, S3 7RH, United Kingdom}  
\author{S.~Shaw} \affiliation{University of California Santa Barbara, Department of Physics, Santa Barbara, CA 93106, USA} \affiliation{Department of Physics and Astronomy, University College London, Gower Street, London WC1E 6BT, United Kingdom} 
\author{T.A.~Shutt} \affiliation{SLAC National Accelerator Laboratory, 2575 Sand Hill Road, Menlo Park, CA 94205, USA} \affiliation{Kavli Institute for Particle Astrophysics and Cosmology, Stanford University, 452 Lomita Mall, Stanford, CA 94309, USA} 
\author{C.~Silva} \affiliation{LIP-Coimbra, Department of Physics, University of Coimbra, Rua Larga, 3004-516 Coimbra, Portugal}  
\author{M.~Solmaz} \affiliation{University of California Santa Barbara, Department of Physics, Santa Barbara, CA 93106, USA}  
\author{V.N.~Solovov} \affiliation{LIP-Coimbra, Department of Physics, University of Coimbra, Rua Larga, 3004-516 Coimbra, Portugal}  
\author{P.~Sorensen} \affiliation{Lawrence Berkeley National Laboratory, 1 Cyclotron Rd., Berkeley, CA 94720, USA}  
\author{T.J.~Sumner} \affiliation{Imperial College London, High Energy Physics, Blackett Laboratory, London SW7 2BZ, United Kingdom}  
\author{M.~Szydagis} \affiliation{University at Albany, State University of New York, Department of Physics, 1400 Washington Ave., Albany, NY 12222, USA}  
\author{D.J.~Taylor} \affiliation{South Dakota Science and Technology Authority, Sanford Underground Research Facility, Lead, SD 57754, USA}  
\author{R.~Taylor} \affiliation{Imperial College London, High Energy Physics, Blackett Laboratory, London SW7 2BZ, United Kingdom}  
\author{W.C.~Taylor} \affiliation{Brown University, Department of Physics, 182 Hope St., Providence, RI 02912, USA}  
\author{B.P.~Tennyson} \affiliation{Yale University, Department of Physics, 217 Prospect St., New Haven, CT 06511, USA}  
\author{P.A.~Terman} \affiliation{Texas A \& M University, Department of Physics, College Station, TX 77843, USA}  
\author{D.R.~Tiedt} \affiliation{University of Maryland, Department of Physics, College Park, MD 20742, USA}  
\author{W.H.~To} \affiliation{California State University Stanislaus, Department of Physics, 1 University Circle, Turlock, CA 95382, USA}  
\author{L.~Tvrznikova} \affiliation{University of California Berkeley, Department of Physics, Berkeley, CA 94720, USA} \affiliation{Lawrence Berkeley National Laboratory, 1 Cyclotron Rd., Berkeley, CA 94720, USA} \affiliation{Yale University, Department of Physics, 217 Prospect St., New Haven, CT 06511, USA}
\author{U.~Utku} \affiliation{Department of Physics and Astronomy, University College London, Gower Street, London WC1E 6BT, United Kingdom}  
\author{S.~Uvarov} \affiliation{University of California Davis, Department of Physics, One Shields Ave., Davis, CA 95616, USA}  
\author{A.~Vacheret} \affiliation{Imperial College London, High Energy Physics, Blackett Laboratory, London SW7 2BZ, United Kingdom}  
\author{V.~Velan} \email[Corresponding author, ] {vvelan@berkeley.edu} \affiliation{University of California Berkeley, Department of Physics, Berkeley, CA 94720, USA}  
\author{R.C.~Webb} \affiliation{Texas A \& M University, Department of Physics, College Station, TX 77843, USA}  
\author{J.T.~White} \affiliation{Texas A \& M University, Department of Physics, College Station, TX 77843, USA}  
\author{T.J.~Whitis} \affiliation{SLAC National Accelerator Laboratory, 2575 Sand Hill Road, Menlo Park, CA 94205, USA} \affiliation{Kavli Institute for Particle Astrophysics and Cosmology, Stanford University, 452 Lomita Mall, Stanford, CA 94309, USA} 
\author{M.S.~Witherell} \affiliation{Lawrence Berkeley National Laboratory, 1 Cyclotron Rd., Berkeley, CA 94720, USA}  
\author{F.L.H.~Wolfs} \affiliation{University of Rochester, Department of Physics and Astronomy, Rochester, NY 14627, USA}  
\author{D.~Woodward} \affiliation{Pennsylvania State University, Department of Physics, 104 Davey Lab, University Park, PA  16802-6300, USA}  
\author{J.~Xu} \affiliation{Lawrence Livermore National Laboratory, 7000 East Ave., Livermore, CA 94551, USA}  
\author{C.~Zhang} \affiliation{University of South Dakota, Department of Physics, 414E Clark St., Vermillion, SD 57069, USA}  

\collaboration{The LUX Collaboration}


\date{\today}

\begin{abstract}

We present a comprehensive analysis of electronic recoil vs.~nuclear recoil discrimination in liquid/gas xenon time projection chambers, using calibration data from the 2013 and 2014--16 runs of the Large Underground Xenon (LUX) experiment. We observe strong charge-to-light discrimination enhancement with increased event energy. For events with S1~=~120 detected photons, i.e.~equivalent to a nuclear recoil energy of $\sim$100~keV, we observe an electronic recoil background acceptance of \textless{$10^{-5}$} at a nuclear recoil signal acceptance of 50\%. We also observe modest electric field dependence of the discrimination power, which peaks at a field of around 300~V/cm over the range of fields explored in this study (50--500~V/cm). In the WIMP search region of S1~=~1--80~phd, the minimum electronic recoil leakage we observe is ${(7.3\pm0.6)\times10^{-4}}$, which is obtained for a drift field of 240--290~V/cm. Pulse shape discrimination is utilized to improve our results, and we find that, at low energies and low fields, there is an additional reduction in background leakage by a factor of up to 3. We develop an empirical model for recombination fluctuations which, when used alongside the Noble Element Scintillation Technique (NEST) simulation package, correctly reproduces the skewness of the electronic recoil data. We use this updated simulation to study the width of the electronic recoil band, finding that its dominant contribution comes from electron-ion recombination fluctuations, followed in magnitude of contribution by fluctuations in the S1 signal, fluctuations in the S2 signal, and fluctuations in the total number of quanta produced for a given energy deposition.

\end{abstract}

\pacs{Valid PACS appear here}

\maketitle

\renewcommand{\thefootnote}{\arabic{footnote}}



\section*{Introduction}
\label{sec:Introduction}

Over the past fifteen years, two-phase (liquid/gas) noble element time projection chambers (TPCs) have emerged as a critical tool for rare event searches, most notably the direct detection of dark matter. In particular, xenon detectors, including the Large Underground Xenon (LUX) experiment, XENON1T, and PandaX-II, have set world-leading constraints on spin-independent dark matter-nucleon elastic scattering for particle masses above a few GeV/c$^2$ \cite{Akerib2017_LUX_Combined_DM_Limit, Aprile2018_XENON1T_DM_Limit, Cui2017_PANDAX_II_DM_Limit}, and have set competitive limits on sub-GeV/c$^2$ dark matter \cite{Aprile2019_XENON1T_Migdal_Brem, Akerib2019_LUX_Migdal_Brem, Aprile2019_XENON1T_S2_Only} and spin-dependent elastic scattering \cite{Aprile2019_XENON1T_Spin_Dependent, Xia2019_PANDAX_II_Spin_Dependent, Akerib2017_LUX_Spin_Dependent}. Future two-phase xenon experiments will be able to test an even greater extent of dark matter parameter space \cite{Akerib2020_LZ_Sensitivity, Zhang2019_PANDAX_4T_Sensitivity, Aprile2020_XENONnT_Sensitivity}.

The xenon TPC is an attractive instrument for dark matter searches for a variety of reasons, including the high density of the liquid xenon target, self-shielding, scalability, and three-dimensional (3D) position reconstruction \cite{Chepel2013_Liquid_Noble_Detectors}. In addition to these, a critical trait of this technology is its ability to \textit{discriminate}, or distinguish, between two types of energy depositions: those creating electronic recoils (ERs), in which energy is transferred to an atomic electron, and those generating nuclear recoils (NRs), in which energy is initially transferred to a xenon nucleus. Discrimination is necessary for a xenon-based dark matter experiment because the canonical signal is a weakly interacting massive particle (WIMP)-induced nuclear recoil, while the dominant background rate is from electronic recoils. These backgrounds include $\gamma$-rays and $\beta^-$ particles from the detector materials, namely, from early-chain decays of $^{238}$U and $^{232}$Th daughters; radioactive contaminants such as $^{222}$Rn, $^{220}$Rn, $^{85}$Kr, and $^{136}$Xe in the liquid xenon volume; and solar neutrinos \cite{Akerib2015_LUX_Radiogenic_Backgrounds, Akerib2020_LZ_Sensitivity}. A xenon TPC is able to discriminate based on two principles. First, the ratio of charge to light leaving the recoil site is different for nuclear recoils and electronic recoils \cite{Aprile2006_NR_Yields_Columbia_and_Case, Dahl2009_Thesis}. Second, the ratio of singlet to triplet excimers is different for nuclear recoils and electronic recoils; since these have different decay times, discrimination is possible based on primary scintillation pulse shape \cite{Akerib2018_LUX_PSD, Abe2018_XMASS_NR_Scintillation_And_PSD, Hogenbirk2018_Scintillation_Pulse_Shape}.

Backgrounds from detector construction materials and surface contaminants will be a relatively small issue in upcoming and future experiments, due to a combination of tonne-scale self-shielding and aggressive campaigns to ensure the cleanliness of the detector. Instead, the dominant backgrounds will be from internal liquid xenon contamination and irreducible neutrino backgrounds. For example, the LUX-ZEPLIN (LZ) sensitivity projection \cite{Akerib2020_LZ_Sensitivity} predicts that 95\% of the electronic recoil background over the energy range 1.5--6.5~keVee\footnote{As defined in Eq.~\ref{eq:keVee}} (equivalent to 88\% of the total background in that energy range) is from Xe contaminants ($^{220}$Rn, $^{222}$Rn, $^{85}$Kr, and $^{39}$Ar), electron scattering by $pp$ solar neutrinos, and $^{136}$Xe two-neutrino double beta decay. The internal backgrounds are difficult to eliminate without enormous further efforts in xenon purification and detector cleanliness. These backgrounds arise from detector material impurities (dominantly $^{238}$U and $^{232}$Th), but unlike the early-chain and surface backgrounds, these contaminants can leak into the xenon volume, rendering self-shielding ineffective. Meanwhile, the neutrino background is impossible to remove. Discrimination is effectively the only strategy to suppress these backgrounds, allowing an experiment to probe a greater region of dark matter parameter space.

In this paper, we examine electronic recoil vs.~nuclear recoil discrimination in close detail. Using data from the two primary runs of LUX, we are able to characterize how charge-to-light discrimination is affected by the drift electric field and the detector's light collection efficiency, and we observe how pulse shape discrimination can enhance this effect. We also develop an understanding of the microphysics of discrimination, based on a marriage of LUX data with the Noble Element Scintillation Technique (NEST) \cite{Szydagis2019_NEST_v2.0.1} simulation code.

\section{The Large Underground Xenon (LUX) Experiment}
\label{sec:LUX}

\subsection{About the detector}
\label{sec:LUXAboutDetector}

The LUX experiment was a two-phase liquid/gas xenon time projection chamber that operated at the 4850' level of the Davis Cavern at the Sanford Underground Research Facility in Lead, South Dakota. It had two primary science runs, from April to August 2013 (referred to here as WS2013), and another from September~2014 to August~2016 (\mbox{WS2014--16}). The active mass was 250~kg of liquid xenon, while the fiducial mass for the dark matter search was about 100~kg. There was an additional 1~cm of gaseous xenon above the liquid that converted the ionization response into an optical signal via electroluminescence. The detector was instrumented with 122 5.6-cm diameter Hamamatsu R8778 photomultiplier tubes (PMTs), with 61 PMTs at the top of the detector (in the gas phase) and 61 at the bottom (immersed in the liquid phase). Furthermore, the detector was instrumented with three wire grids to control the electric field in the liquid and the gas---a cathode at the bottom of the detector, a gate slightly below the liquid level, and an anode in the xenon gas above the liquid level---and two grids in front of the PMT arrays to prevent stray fields from affecting the PMT photocathodes. Full technical details of the experiment's configuration can be found in \cite{Akerib2018_LUX_Run3_Comprehensive}. Here we focus on how signals are produced and detected.

Any energy deposited in the liquid will be transferred to xenon atoms in three modes: heat, atomic excitation, and ionization. The heat is unobservable in a xenon TPC, and for electronic recoils, the fraction of recoil energy going into the heat channel is constant with recoil energy. The atomic excitation leads to the formation of \textit{excimers}, diatomic xenon molecules that deexcite to repulsive ground states with emission of 175~nm photons. These photons are detected by the PMTs, resulting in a signal called ``S1''; the average number of photons detected for each photon leaving the recoil site is called $g_1$. Since the S1 pulse is relatively small in this analysis, up to 120 photons detected, we can measure S1 in two ways: by integrating the full pulse area or by counting the number of photoelectron ``spikes'' recorded in each PMT. The ionization electrons are drifted through the electric field in liquid (i.e.~the \textit{drift field}), extracted into the gas phase by a stronger field, and produce secondary scintillation light which is detected by the PMTs. This signal is called ``S2,'' and the number of photons detected from a single ionization electron is called $g_2$. The units of both S1 and S2 are \textit{photons detected}, which we abbreviate to \textit{phd}. The drift time, i.e.~the time between S1 and S2, gives the $z$-position (depth) of the recoil. The pattern of S2 light in the top PMT array is used to reconstruct $x$ and $y$. Most of the light is detected in PMTs located near the site where the ionization electron cloud is extracted into the gas phase, so the distribution of pulse areas can be used to determine the ($x$,~$y$)-position of the recoil site \cite{Akerib2018_LUX_Run3_Comprehensive, Solovov2011_Mercury}.

Furthermore, the S1 and S2 variables are adjusted based on the position of the event. The S1 adjustment is primarily based on the variation of light collection efficiency in the detector; most of the S1 light is detected by the bottom PMTs, so S1 light collection is higher for lower regions of the detector than for higher regions. The adjustment is calculated such that the corrected S1 corresponds to the scintillation light for an equivalent event at the center of the liquid volume. The S2 adjustment is primarily based on the fact that if the electrons drift for a longer time in the liquid signal, they are more likely to attach onto an electronegative impurity. This adjustment is calculated such that the corrected S2 corresponds to the charge signal for an equivalent event at the liquid/gas surface. In this paper, we use the following conventions, unless otherwise noted. S1c and S2c refer to the position-corrected variables, and S1 and S2 refer to the position-uncorrected variables. The position corrections are dependent on $z$ only in \mbox{WS2014--16} data and on the full $xyz$ position in WS2013 data. However, the WS2013 corrections are dominantly $z$-dependent, and when we compare results between the two science runs, we use $z$-dependent position corrections for WS2013 data. S1 or S1c refers to spike count if the pulse area is less than 80 detected photons, and it refers to pulse area otherwise. This ``hybrid'' variable is used because spike counting leads to better discrimination at low energies, but it cannot be reliably determined for large photon statistics at higher energies.

If the energy deposition comes from an electronic recoil, the combined energy from scintillation and ionization is given by $E_{\text{ee}}$ in Eq.~\ref{eq:keVee}, where $W$ is the average energy required to generate a quantum of response leaving the recoil site (either a photon or electron). As a result, we refer to $E_{\text{ee}}$ as the \textit{electronic equivalent energy}. From data \cite{Dahl2009_Thesis}, we know that ${W=13.7\pm0.2}$~eV.\footnote{The EXO-200 collaboration recently measured ${W=11.5\pm0.5}$~eV in electronic recoils using 1.2--2.6~MeV~$\gamma$ calibrations \cite{Anton2019_EXO200_MeV_Response}. The discrepancy is not yet understood. As EXO-200 is a single-phase TPC and uses avalanche photodiodes to detect photons instead of PMTs, we use ${W=13.7\pm0.2}$~eV to be consistent with other dual-phase xenon TPCs.}
\begin{equation}
E_{\text{ee}} = W \left(\frac{S1c}{g_1} + \frac{S2c}{g_2} \right) \,.
\label{eq:keVee}
\end{equation}

Meanwhile, if we assume that the energy deposition is a nuclear recoil, we need to consider the additional energy lost to heat and its energy dependence. We find the total energy of a nuclear recoil $E_{\text{nr}}$ can be related to its electronic equivalent energy by Eq.~\ref{eq:keVnr}.
\begin{equation}
\begin{gathered}
E_{\text{ee}} = A \, E^{\, \gamma}_{\text{nr}} \: , \\
A = 0.173 \text{ and } \gamma = 1.05 \,.
\end{gathered}
\label{eq:keVnr}
\end{equation}
\noindent We have confirmed that, by using this relationship, we are able to match LUX D-D nuclear recoil calibration data to its theoretical energy spectrum. The reader should note that since $\gamma \approx 1$, Eq.~\ref{eq:keVnr} is comparable to a linear scaling. This model is similar but not identical to the Lindhard model \cite{Lindhard1963} often used to describe nuclear recoils in liquid xenon. The discrepancy is reasonable because the Lindhard model does not perfectly reproduce the nuclear recoil energy scale across all energies; see e.g.~Fig.~15 of \cite{Akerib2016_LUX_DD_Calibration_Run3}.

\subsection{Calibrations}
\label{sec:LUXCalibrations}

LUX underwent several calibration campaigns throughout WS2013 and \mbox{WS2014--16} to understand the detector's response to different types of energy depositions. Both runs featured three specific calibrations that we focus on here. First, we injected a tritiated methane source into the xenon \cite{Akerib2016_LUX_Tritium_Calibration_Run3, Akerib2019_LUX_Beta_Calibrations_Combined}; this is a molecule that is chemically similar to methane, CH$_4$, but with one of the hydrogen atoms replaced by tritium. Tritium is a $\beta^-$ emitter with a half-life of 12.3~years and an end point of 18.6~keV, making it useful for calibrating low-energy electronic recoils. It also filled the entire detector volume, allowing us to examine effects in different locations. Second, we ran nuclear recoil calibration campaigns by generating 2.5~MeV neutrons from deuterium-deuterium fusion (referred to as a D-D calibration), which deposit up to 74~keV on a xenon nucleus \cite{Akerib2016_LUX_DD_Calibration_Run3, Verbus2017_DD_Proposal}. These were produced by a neutron generator placed outside the xenon volume, and the height of this generator varied during \mbox{WS2014--16}. Third, we regularly (approximately weekly) calibrated the detector with $^{83\text{m}}$Kr, a 41.6~keVee source that filled the detector volume uniformly and decayed with a 1.83-hour half-life \cite{Akerib2017_LUX_Kr83m_Calibration}. In addition to these, LUX ran a $^{14}$C calibration campaign after the final WIMP search, in August 2016; we injected a $^{14}$CH$_4$ methane molecule, which allowed us to calibrate the detector up to 156.5~keVee \cite{Akerib2019_LUX_Beta_Calibrations_Combined, Balajthy2018_Thesis}.

In this paper, we use data from all of these calibration campaigns, focusing only on single scatter events (events with one S1, followed by one S2 within an appropriate time window). We do apply some additional quality cuts to the data, most of which are described in past literature \cite{Akerib2016_LUX_Tritium_Calibration_Run3, Akerib2016_LUX_DD_Calibration_Run3, Akerib2018_LUX_Run3_Comprehensive, Akerib2017_LUX_Combined_DM_Limit, Akerib2016_LUX_Run3_Reanalysis}. To summarize, these include cuts on event position to select recoils in the central region of the liquid volume, or in the path of the beam for D-D nuclear recoils; cuts on S1 and S2 area to select events in the appropriate energy range; cuts on the S1 and S2 pulse shapes; and a cut to remove multiple scatters that are misclassified as single scatters.

\subsection{Electric field variation}
\label{sec:LUXFieldVariation}

\begin{figure}
{\includegraphics[width=3.25in]{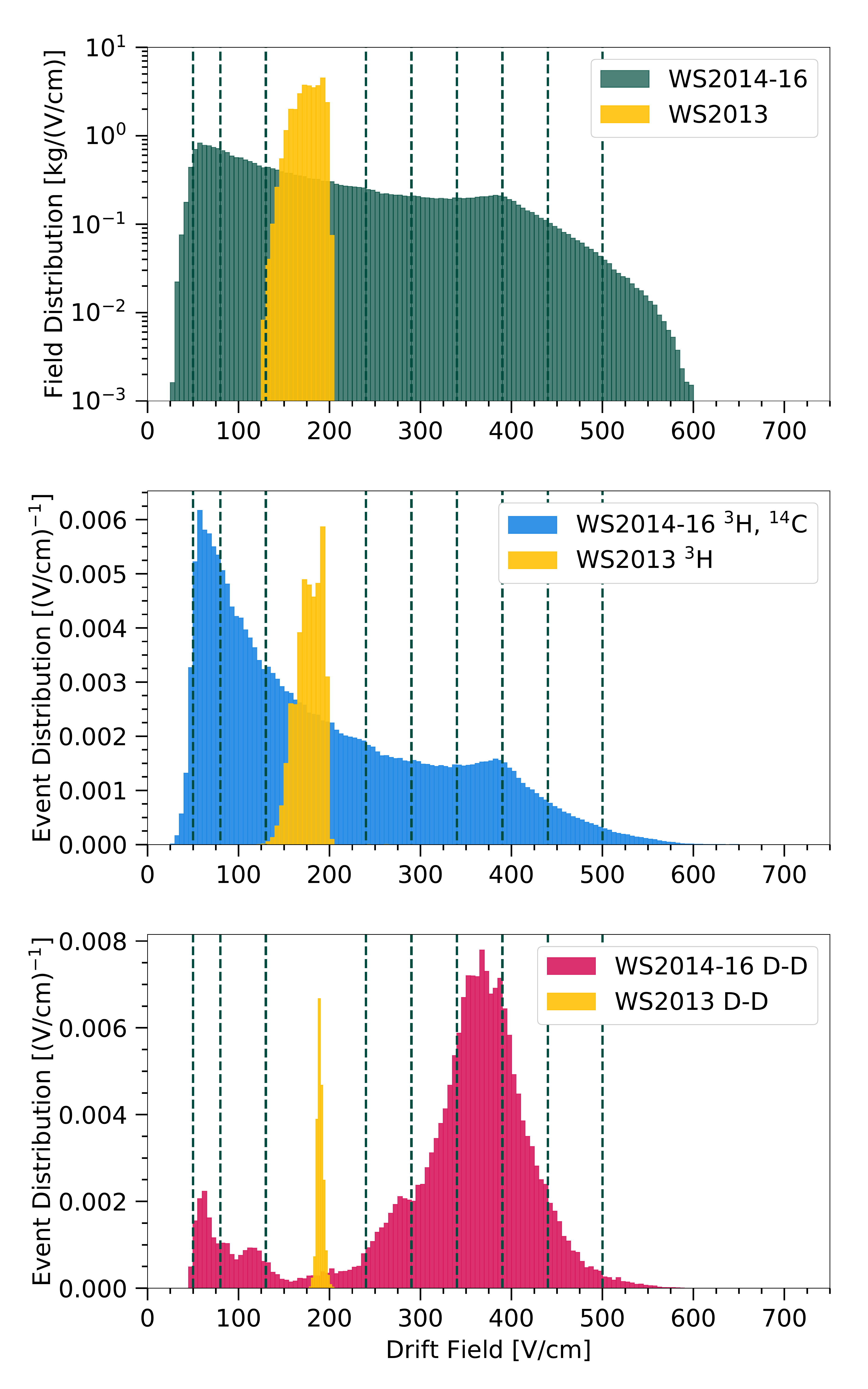}}
\caption{The distribution of drift fields in the LUX datasets. \textit{(Top)} The mass distribution of field within the LUX fiducial volume. In this analysis, we define the WS2013 fiducial volume as $r$~\textless~20~cm and 38~\textless~$\left(t_{drift}/\mu s\right)$~\textless~305, and the \mbox{WS2014--16} fiducial volume as $r$~\textless~20~cm and 40~\textless~$\left(t_{drift}/\mu s\right)$~\textless~300. \textit{(Middle)} For each electronic recoil, the field at the recoil site is calculated using the results of \cite{Akerib2017_LUX_Fields}, and we plot a normalized histogram of the results. The $^3$H and $^{14}$C datasets are combined for \mbox{WS2014--16} because they both fill the entire detector volume, and thus have identical distributions. Black dashed lines are used to indicate the field bins used in Section~\ref{sec:Bands}. The WS2013 and \mbox{WS2014--16} histograms are normalized separately in order to visualize the data effectively, so the relative heights of the blue and yellow histograms should not be considered an expression of the number of events in each dataset. \textit{(Bottom)} The same as the middle panel, but for nuclear recoils.}
\label{fig:Field_Distributions}
\end{figure}

In WS2013, the drift field was fairly uniform across the liquid xenon target region at $177~\pm~14$~V/cm. However, in \mbox{WS2014--16}, the drift field varied significantly throughout the detector from 30~V/cm at the bottom of the fiducial region to 600~V/cm at the top. In \cite{Akerib2017_LUX_Fields}, the LUX Collaboration hypothesized that the drift field variation was created by net charge buildup within the polytetrafluoroethylene (PTFE) detector walls and that this buildup of charge was induced by the strong VUV fluxes experienced during grid conditioning. A method for converting an event's 3D position to the electric field at the recoil site was described in that publication. This was a complication for the WIMP search analysis, but it provides us with an opportunity to examine how discrimination is affected by electric field. Figure~\ref{fig:Field_Distributions} shows the distribution of field in the LUX fiducial volume, as well as the field distribution of events in the calibrations mentioned in Section~\ref{sec:LUXCalibrations}; the reader may observe the dramatic difference between the two runs. The uncertainty on the electric field magnitude is estimated to be $\sim$10\%, based on comparisons between light and charge yields in simulation and data \cite{Tvrznikova2019_Thesis}.

\section{Electronic and Nuclear Recoil Bands}
\label{sec:Bands}

\subsection{Electronic Recoils}
\label{sec:ER_Band}

For each electronic recoil in the dataset, the LUX detector observes a single S1 signal, followed by a single S2 signal. As has been widely observed by liquid xenon experiments \cite{Aprile2006_NR_Yields_Columbia_and_Case, Dahl2009_Thesis, Aprile2018_XENON100_Discrimination, Akerib2017_LUX_Signal_Yields, Akerib2017_LUX_Combined_DM_Limit}, one can plot these recoils on axes of log\textsubscript{10}(S2c/S1c) vs.~S1c to obtain a ``band'' of events. We will refer to this as the \textit{ER band}, as is common in the literature.

\begin{figure}
{\includegraphics[width=3.25in]{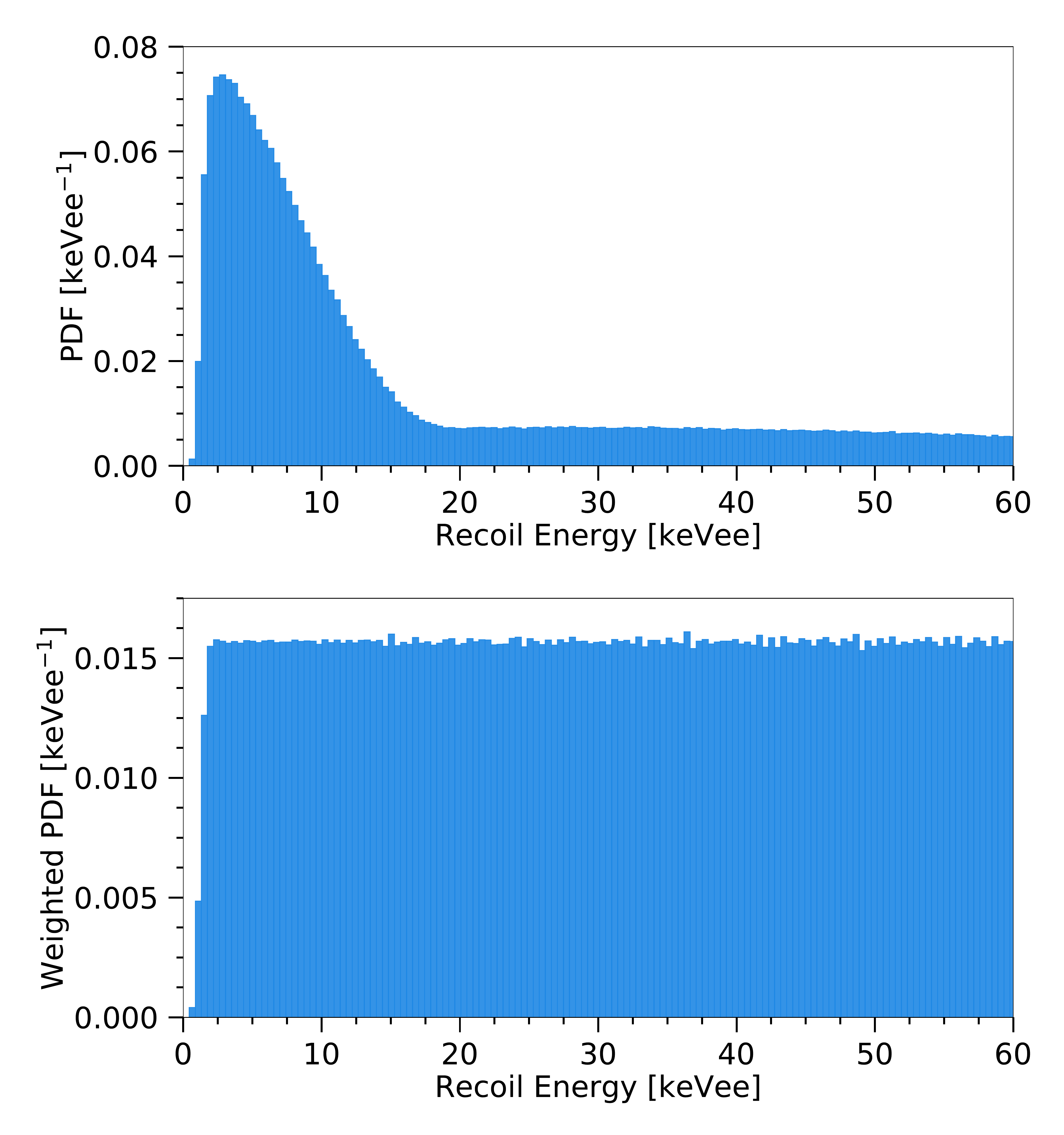}}
\caption{\textit{(Top)} The recoil energy spectrum of the \mbox{WS2014--16} electronic recoil dataset, including $^3$H and $^{14}$C decays. \textit{(Bottom)} The same energy spectrum, but with weights applied such that the spectrum is flat with a threshold at low energy.}
\label{fig:Energy_Weighting}
\end{figure}

We calculate relevant quantities characterizing the ER band in the following way. First, we account for the irregular energy spectrum of the dataset, which includes both $^3$H and $^{14}$C $\beta^-$ decays. For each event, a weight is calculated such that the weighted energy distribution is proportional to $f(E)$ in Eq.~\ref{eq:Smooth_Step_Function}, in which $E$ is the recoil energy determined with Eq.~\ref{eq:keVee}.
\begin{equation}
f(E) = \frac{1}{2} \left[ 1 + \text{erf} \left( \frac{E - E_\mu}{E_\sigma \sqrt{2}} \right) \right] \,.
\label{eq:Smooth_Step_Function}
\end{equation}
The parameters $E_\mu$ and $E_\sigma$ are determined by fitting the $^3$H and $^{14}$C energy distributions to their beta decay spectra multiplied by $f(E)$. They are fit to about 1~keVee and 0.3~keVee, respectively. Effectively, $E_\mu$ is the energy threshold for measuring electronic recoils, and $E_\sigma$ is the ``width'' of this threshold. In this way, the energy spectrum of the dataset is transformed into a flat distribution, apart from the threshold behavior at low energy. See Fig.~\ref{fig:Energy_Weighting} for a depiction of this weighting.

This procedure allows us to calculate an ER band that is universal for electronic recoils. Furthermore, it yields a result that is relevant for future xenon dark matter experiments. These experiments (as explained in Section~\ref{sec:LUXAboutDetector}) are prone to backgrounds from $pp$ neutrinos and daughters of $^{220}$Rn and $^{222}$Rn, which are relatively constant in energy over the range of energies relevant for dark matter direct detection.

We then split the electronic recoil data into small bins of S1c. Within S1c bins, the distribution of log\textsubscript{10}(S2c/S1c) is often \cite{Akerib2016_LUX_Tritium_Calibration_Run3, Akerib2018_LUX_Run3_Comprehensive, Alner2007_ZEPLIN_II_WS} assumed to be Gaussian, but we observe that a skew-Gaussian distribution is a better fit for the electronic recoil data, as also observed in \cite{Lebedenko2009_ZEPLIN_III_First_Science_Run}. A skew-Gaussian distribution follows the probability density function (PDF) in Eq.~\ref{eq:skewgaus}. This distribution is similar to a Gaussian distribution, if we identify $\xi$ and $\omega$ with the mean and standard deviation. However, the skew-Gaussian distribution is modified by a parameter $\alpha$, biasing the PDF toward higher values than a Gaussian PDF if $\alpha$~\textgreater~0 and lower values if $\alpha$~\textless~0. As a result, the mean $\mu$ and variance $\sigma^2$ of the skew-Gaussian distribution are given by Equations \ref{eq:skewgaus_mean} and \ref{eq:skewgaus_variance}, respectively \cite{Azzalini1985_Skes_Normal}.
\begin{equation}
f(x) = \frac{1}{\omega \sqrt{2 \pi}} \, e^{-\frac{(x - \xi)^2} {2 \omega^2}}
\left[1 + \text{erf} \left( \frac{\alpha \, (x - \xi)}{\omega \sqrt{2}} \right) \right] \,.
\label{eq:skewgaus}
\end{equation}
\begin{equation}
\mu = \xi + \sqrt{\frac{2}{\pi}} \, \frac{\alpha \omega}{\sqrt{1 + \alpha^2}} \,.
\label{eq:skewgaus_mean}
\end{equation}
\begin{equation}
\sigma^2 = \omega^2 \left(1 - \frac{2}{\pi} \frac{\alpha^2}{1 + \alpha^2} \right) \,.
\label{eq:skewgaus_variance}
\end{equation}

\noindent We will refer to $\alpha$ as the skewness parameter, but it is important to note that $\alpha$ does not correspond to the algebraic skewness of the distribution (i.e. the third standardized moment). Furthermore, when referencing skew-Gaussian fits to distributions of log\textsubscript{10}(S2c/S1c), we denote this parameter as $\alpha_B$. The subscript ``$B$'' identifies this quantity as a trait of the ER (or NR) band.

\begin{figure}
{\includegraphics[width=3.25in]{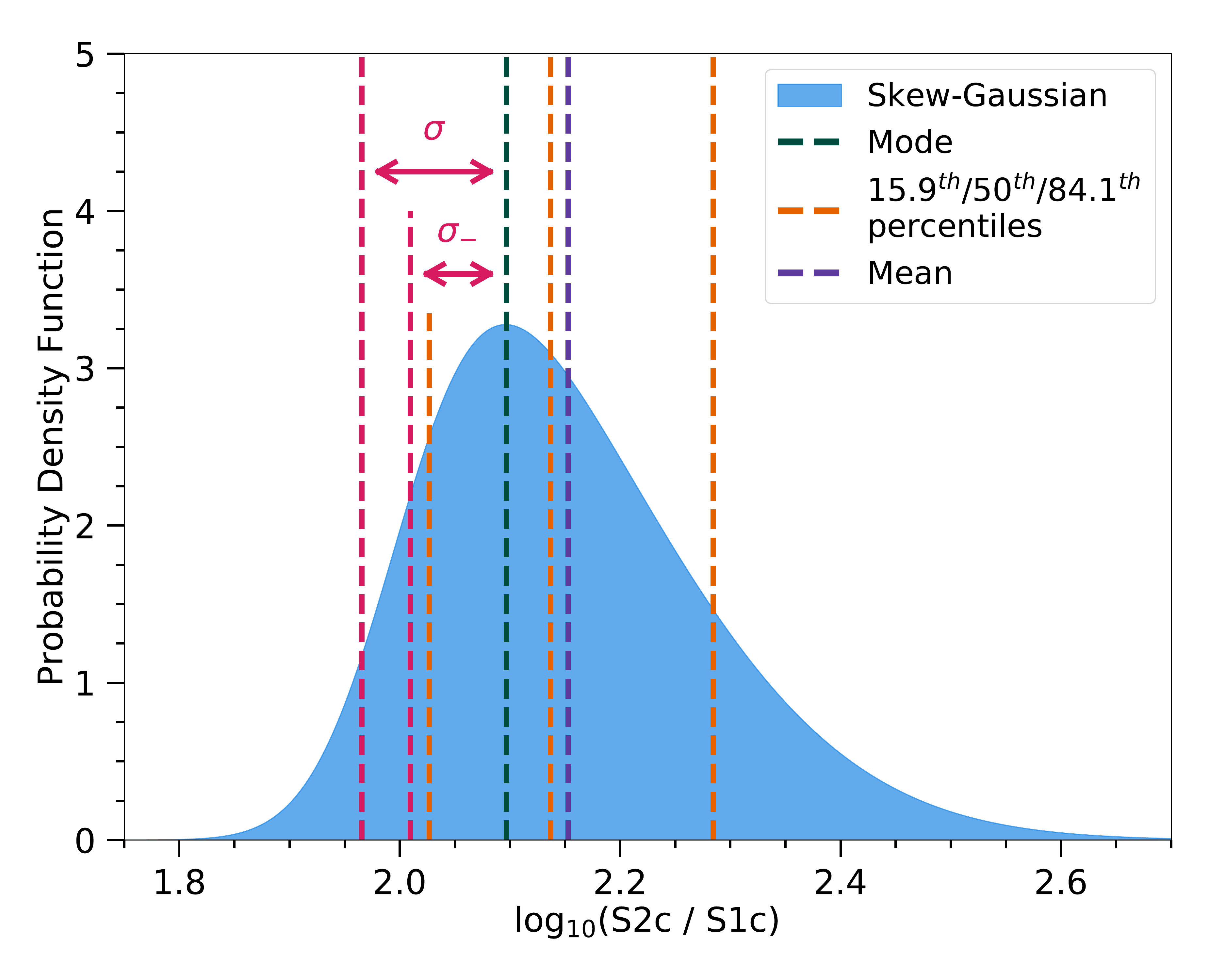}}
\caption{A skew-Gaussian distribution with $\xi = 2$, $\omega = 0.2$, and $\alpha_B = 3$. The mode, median, and mean of the distribution are shown, as well as the 15.9th and 84.1th percentiles. We also graphically show the standard deviation $\sigma$ from Eq.~\ref{eq:skewgaus_variance} and $\sigma_{\text{-}}$ from Eq.~\ref{eq:skewgaus_sigmaminus}, relative to the mode of the distribution. In real log\textsubscript{10}(S2c/S1c) data, $\alpha_B$ is typically smaller than 3, but a high skewness parameter is shown for ease of viewing.}
\label{fig:Skew_Gaussian_Example}
\end{figure}

In our energy range, electronic recoil data nearly always display positive skewness; $\alpha_B$~\textgreater~0. Figure~\ref{fig:Skew_Gaussian_Example} shows the effects of positive skewness; the mean is greater than the median, and both are greater than the mode. We emphasize that positive skewness is not a statistical artifact, such as from Poisson statistics in the S1 signal; it seems to be the result of liquid xenon recombination physics, as we will explore in Section~\ref{sec:Skewness}.

\begin{figure}
{\includegraphics[width=3.25in]{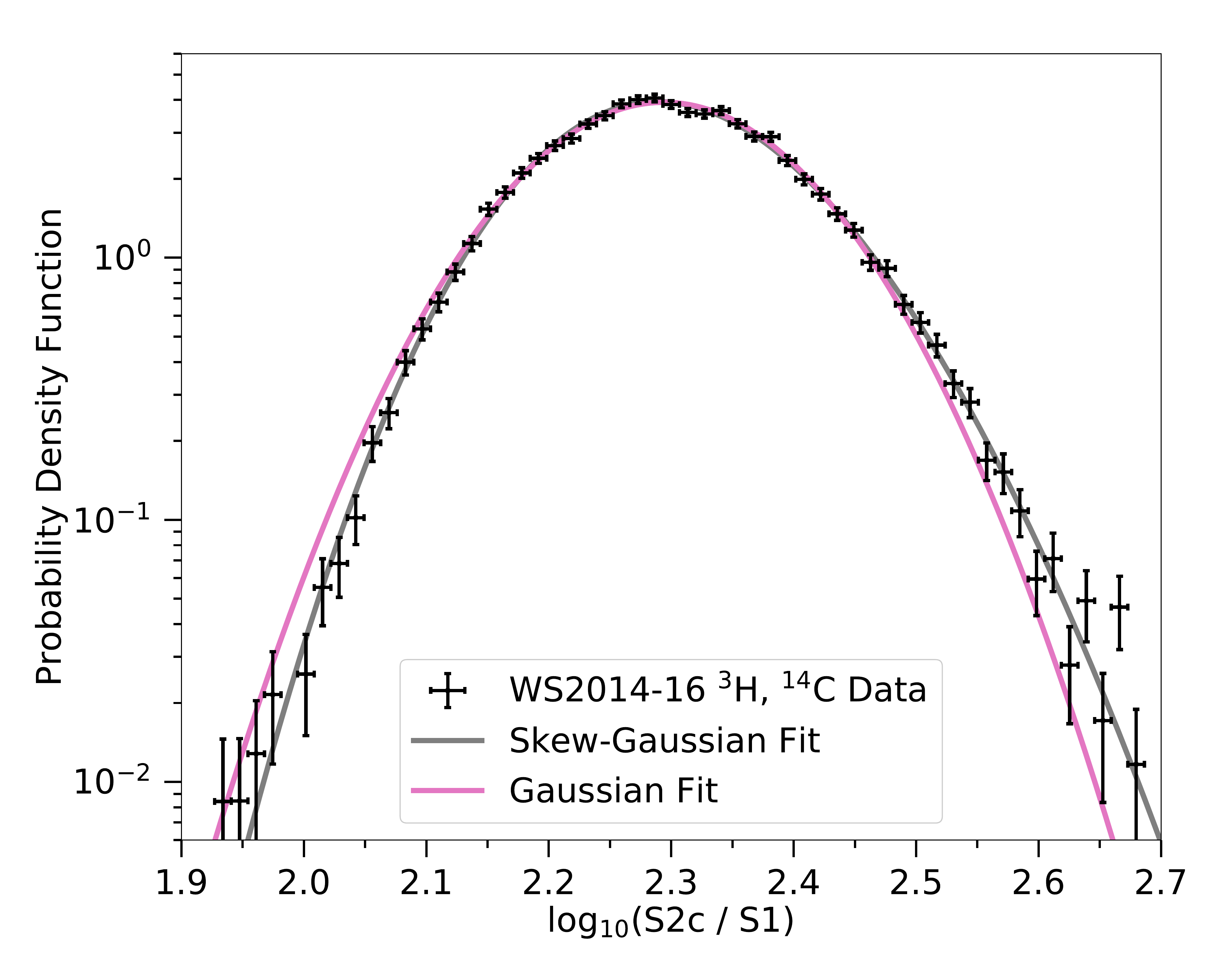}}
\caption{An example histogram of log\textsubscript{10}(S2c/S1) for electronic recoil data, and associated fits to a skew-Gaussian and Gaussian distribution. The events have an S1 signal between 13 and 16 phd and a drift field between 80 and 130~V/cm; they are weighted based on their energy as described in Section \ref{sec:ER_Band}. The best-fit skew-Gaussian parameters are ${\xi=2.222\pm0.003}$, ${\omega=0.128\pm0.002}$, and ${\alpha_B=1.14\pm0.06}$. The best-fit Gaussian parameters are ${\mu=2.2941\pm0.0010}$ and ${\sigma=0.1018\pm0.0010}$.}
\label{fig:Skew_Gaussian_Data}
\end{figure}

In each S1c bin, we fit the weighted histogram of log\textsubscript{10}(S2c/S1c) to a skew-Gaussian distribution, using $\chi^2$~minimization.\footnote{A maximum likelihood fit with a Poisson estimator returns consistent results, but because of our weighting procedure, the uncertainties on the fit parameters are not representative. Therefore, we report the results from $\chi^2$~minimization.} Figure~\ref{fig:Skew_Gaussian_Data} shows an example of this fit; note that the skew-Gaussian fit more closely matches the data than the fit to a Gaussian. The median of the distribution is easily extracted. The width is defined in two ways. First, the width of the total distribution $\sigma$ is obtained by using Eq.~\ref{eq:skewgaus_variance}. Second, we use Eq.~\ref{eq:skewgaus_sigmaminus} to define a quantity that we call $\sigma_{\text{-}}$, which is relevant for discrimination. In log\textsubscript{10}(S2c/S1c) vs.~S1c space, electronic recoils lie above nuclear recoils, so the leakage of electronic recoils into the nuclear recoil region is based only on the lower part of the log\textsubscript{10}(S2c/S1c) distribution. Thus, $\sigma_{\text{-}}$ serves as a measure of the portion of the width due only to downward fluctuations, and it is determined by the condition
\begin{equation}
\begin{gathered}
\int_{m - \sigma_{\text{-}}}^{m} f(x) \, dx = 0.68 \int_{-\infty}^{m} f(x) \, dx \:\:, \\
\text{where} \: m \: \text{is the mode of} \: f(x).
\end{gathered}
\label{eq:skewgaus_sigmaminus}
\end{equation}

\begin{figure}
{\includegraphics[width=3.25in]{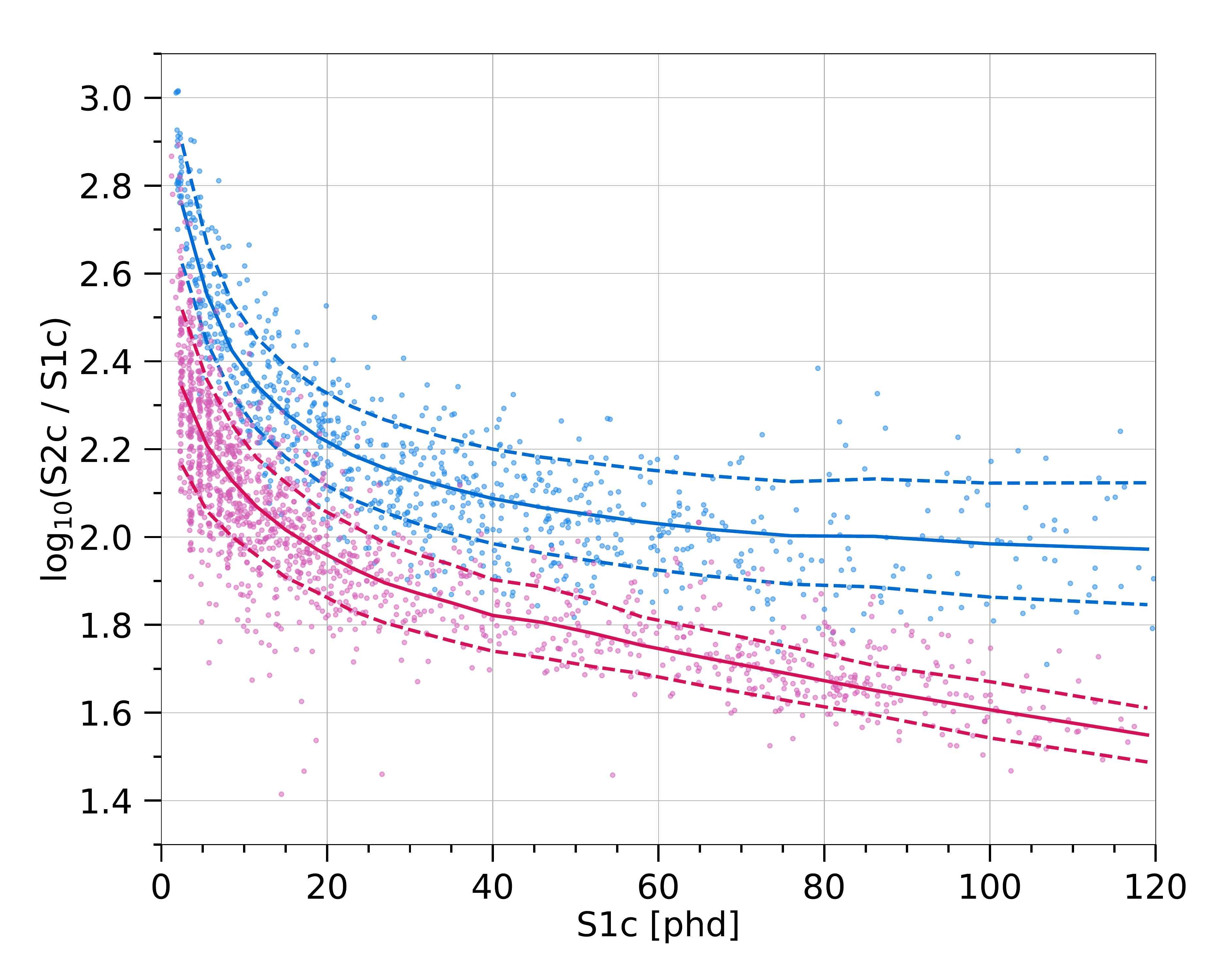}}
\caption{A sample of electronic and nuclear recoils, along with the associated bands. A randomly selected 1500 electronic (nuclear) recoils from \mbox{WS2014--16} are shown in blue (red) dots, the median of the ER (NR) band is shown as a solid blue (red) line, and the 15.9th~percentile and 84.1th~percentile of the ER (NR) band are shown as dashed blue (red) lines. Details on the calculation of the ER and NR band can be found in Sec.~\ref{sec:ER_Band} and Sec.~\ref{sec:NR_Band}, respectively.}
\label{fig:Bands_Example}
\end{figure}

The uncertainties of the skew-Gaussian fit parameters, which are extracted from the $\chi^2$~minimization, are used to estimate the uncertainties of the ER band median and width: ${\delta(\text{median})=\delta \xi}$ and ${\delta(\sigma_{\text{-}})=\delta(\sqrt{\sigma^2})}$. Figure~\ref{fig:Bands_Example} shows a sample of electronic recoils from \mbox{WS2014--16}, as well as the ER band calculated from the entire \mbox{WS2014--16} dataset.

\subsection{Nuclear Recoils}
\label{sec:NR_Band}

Nuclear recoils can be analyzed similarly to electronic recoils, allowing us to define an analogous \textit{NR band}. One modification we make to the procedure outlined in Sec.\ref{sec:ER_Band} is that we eliminate the energy-based event weights. Instead, we use the unweighted D-D calibration data, which has a recoil energy spectrum similar to that of a 50~GeV/$c^2$ WIMP. The other adjustment for nuclear recoils is that in bins of S1c, we assume the distribution of log\textsubscript{10}(S2c/S1c) is Gaussian. As will be described in Sec.~\ref{sec:Skewness}, a skew-Gaussian distribution actually fits the NR data better, but we model the NR band as Gaussian for two reasons. First, due to the low statistics of the NR data, the skew-Gaussian fit often fails to converge or gives large errors on the fit parameters. Second, the Gaussian fit reproduces the same median and width as the skew-Gaussian fit, and these parameters have a greater impact on discrimination and sensitivity than the skewness itself. The uncertainties on the NR band median and width are simply the uncertainties on the Gaussian fit. Figure~\ref{fig:Bands_Example} shows a sample of nuclear recoils from \mbox{WS2014--16}, as well as the NR band calculated from the entire \mbox{WS2014--16} data set. 

We also note a small source of bias in the NR band calculation. To improve data quality, we have removed events with S2~\textless~270 phd (164 phd) in the \mbox{WS2014--16} (WS2013) D-D data. In the lowest S1c bin, this removes up to 10\% of events. When the Gaussian fit is performed, the best-fit mean and width are higher and lower, respectively, than they would be if the dataset contained events with a smaller S2 signal. The shift in these best-fit parameters is expected to be \textless~2\%, as estimated from simulation. The shift is small but could impact electronic recoil discrimination, as will be described in Sec.~\ref{sec:ChargeLightDiscrimination}. This effect is not relevant for higher S1c bins in the nuclear recoil data and any S1c bins in the electronic recoil data, because all events have S2 signals significantly larger than the analysis threshold.

\subsection{\texorpdfstring{Variation with $g_1$}{Variation with g1}}
\label{sec:Bands_g1_Variation}

A key detector parameter in two-phase xenon dark matter experiments is the prompt light collection gain $g_1$, which is primarily dictated by the detector geometry, the reflectivity of the inner surfaces, and the quantum efficiency of the PMTs. In WS2013, the average value of $g_1$ was 0.117 \cite{Akerib2016_LUX_Tritium_Calibration_Run3}, while in \mbox{WS2014--16}, it varied from 0.0974 to 0.0994. The time dependence of $g_1$ could be caused by varying impurity concentration in the Xe bulk or changes in wire grid reflectivity. We expect $g_1$ to have a strong impact on discrimination; as more light is collected, the S1 signal will grow in magnitude, and the relative size of S1 fluctuations will decrease. Thus, $g_1$ should be positively correlated with discrimination power.

This is an effect we can observe in LUX through a novel procedure. For each event, the S1c signal is a sum of the signals in each of LUX's 122~PMTs (adjusted for position-dependent and PMT-dependent effects). By adding together the pulses in only a fraction of the PMTs, we are able to artificially reduce $g_1$. We use $^{83\text{m}}$Kr WS2013 calibration data \cite{Akerib2017_LUX_Kr83m_Calibration} to determine the effective $g_1$ for a given subset of PMTs. The $^{83\text{m}}$Kr decay is a two-step process, emitting 32.1 and 9.4~keV conversion electrons. The time between the two decay steps is exponentially distributed with a half-life of 154 ns and is observed to affect the light yield of the second energy deposit \cite{Aprile2012_Compton_and_Kr83m_Calibrations, Baudis2013_Compton_and_Kr83m_Calibrations, Akerib2017_LUX_Kr83m_Calibration, Singh2020_PIXeY_Kr83m}. However, our analysis only uses events in which the two light signals are merged. In this analysis, that is generally true for events in which the time between the two decay steps is \textless~1200~ns. We can thus treat the $^{83\text{m}}$Kr decay as monoenergetic with a single (field-dependent) S1c and S2c peak. The mean value of the $^{83\text{m}}$Kr S1c peak (in photons detected) is reduced when we add the signals in a subset of the 122~PMTs, relative to its value when using the full LUX detector. The reduction in the value of the S1c peak is proportional to the reduction in $g_1$. For example, one PMT configuration has 105 PMTs, and when S1c is recalculated for all $^{83\text{m}}$Kr events using only the signals detected by these 105 PMTs, the average S1c is reduced by 11\% relative to adding the signals in all 122 PMTs. Thus, we infer that the effective $g_1$ obtained by using these 105 PMTs is ${g_1=0.89\times0.117=0.104}$.

We isolate the effect of $g_1$ on the ER and NR bands by considering only WS2013 data, which have a uniform drift field. First, ten PMT configurations are chosen, and the corresponding $g_1$ values are calculated. We intentionally choose PMT configurations so the resulting $g_1$ values are evenly distributed between 50\% and 100\% of $g_1$ for the full detector. For each configuration of PMTs, we calculate new S1c values for each event in the WS2013 $^3$H and D-D data. The S1 signal is obtained by adding together the signals from only the PMTs in that subset, and this is translated to S1c with the same position-dependent correction factor used in the analysis of all 122~PMTs. Then, we recalculate the ER and NR band.

\begin{figure*}[p!]
{\includegraphics[width=6.5in]{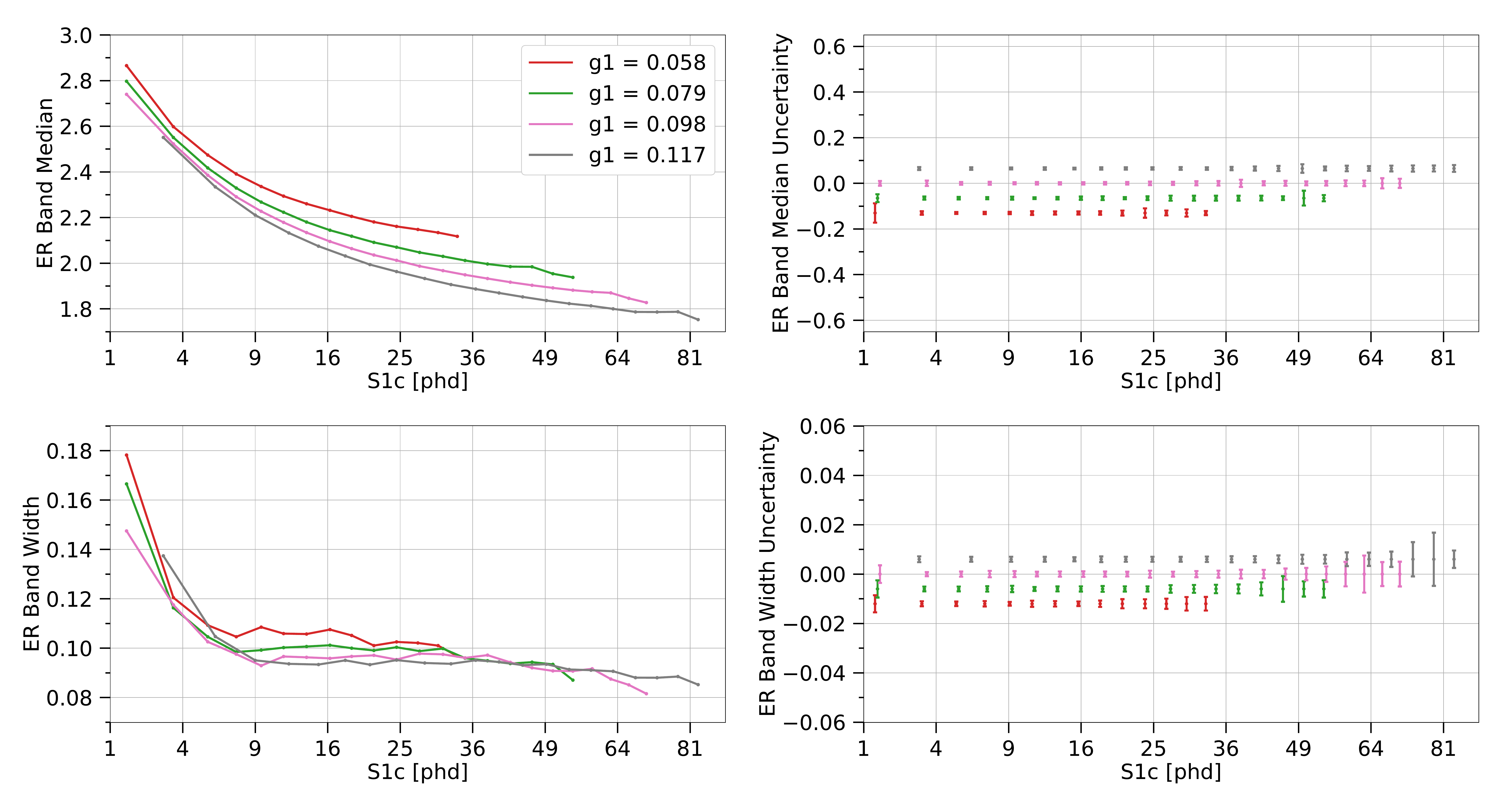}}
\caption{The median and width $\sigma_{\text{-}}$ of the ER band for several values of $g_1$, using WS2013 data. The left plots show measurements, and the right plots display error bars corresponding to these measurements. The S1c axis is proportional to $\sqrt{\text{S1c}}$. In each row, the $y$-axes have the same range; the size of the error bars on the right plot can be directly translated to the points on the left plot. For ease of visualization on the right plots, the S1c values are slightly shifted relative to their true value, and the error bars are centered at a different $y$-value for each $g_1$. Note the S1c range varies for each $g_1$ because as $g_1$ decreases, the $^3$H end point in S1c space decreases. See Fig.~\ref{fig:ER_Medians_Widths_g1_Full} for the ER band median and width for all the $g_1$ values we considered.}
\label{fig:ER_Medians_Widths_g1}
\end{figure*}

The results for the ER band are shown in Fig.~\ref{fig:ER_Medians_Widths_g1}, where we display only four $g_1$ values for ease of visualization. See Fig.~\ref{fig:ER_Medians_Widths_g1_Full} in Appendix~\ref{app:Additional_Figures} for the full set of results. As $g_1$ increases, the median of the ER band shifts down; this is a fairly straightforward result, because a larger $g_1$ implies a larger S1c and thus a lower log\textsubscript{10}(S2c/S1c). Also, as $g_1$ increases, the absolute ER band width decreases, particularly for S1 values less than 30~phd. This also matches our expectations, because as the light collection increases, the relative size of the fluctuations in the number of photons detected decreases. Note that the leftmost point for $g_1 = 0.117$ in the bottom panel of Fig.~\ref{fig:ER_Medians_Widths_g1} appears to be an outlier, showing a different behavior than the other measurements. However, it is not an outlier. Instead, this appearance is due to the changing conversion of energy to S1c as $g_1$ varies. Above 30~phd, the shrinking of the ER band width with $g_1$ plateaus, and we can account for this with three explanations. First, since $^3$H has an end point in our region of interest, the changing $g_1$ changes the maximum S1c, which excludes certain curves at high energy. Second, the number of events in each S1c bin decreases as we near the end point, making the error bars larger and reducing our sensitivity to any small differences. Third, as the number of photons detected increases, the relative fluctuations in the S1 signal become smaller, and the total ER band width is dominated by other $g_1$-independent fluctuations such as recombination.

\begin{figure*}[p!]
{\includegraphics[width=6.5in]{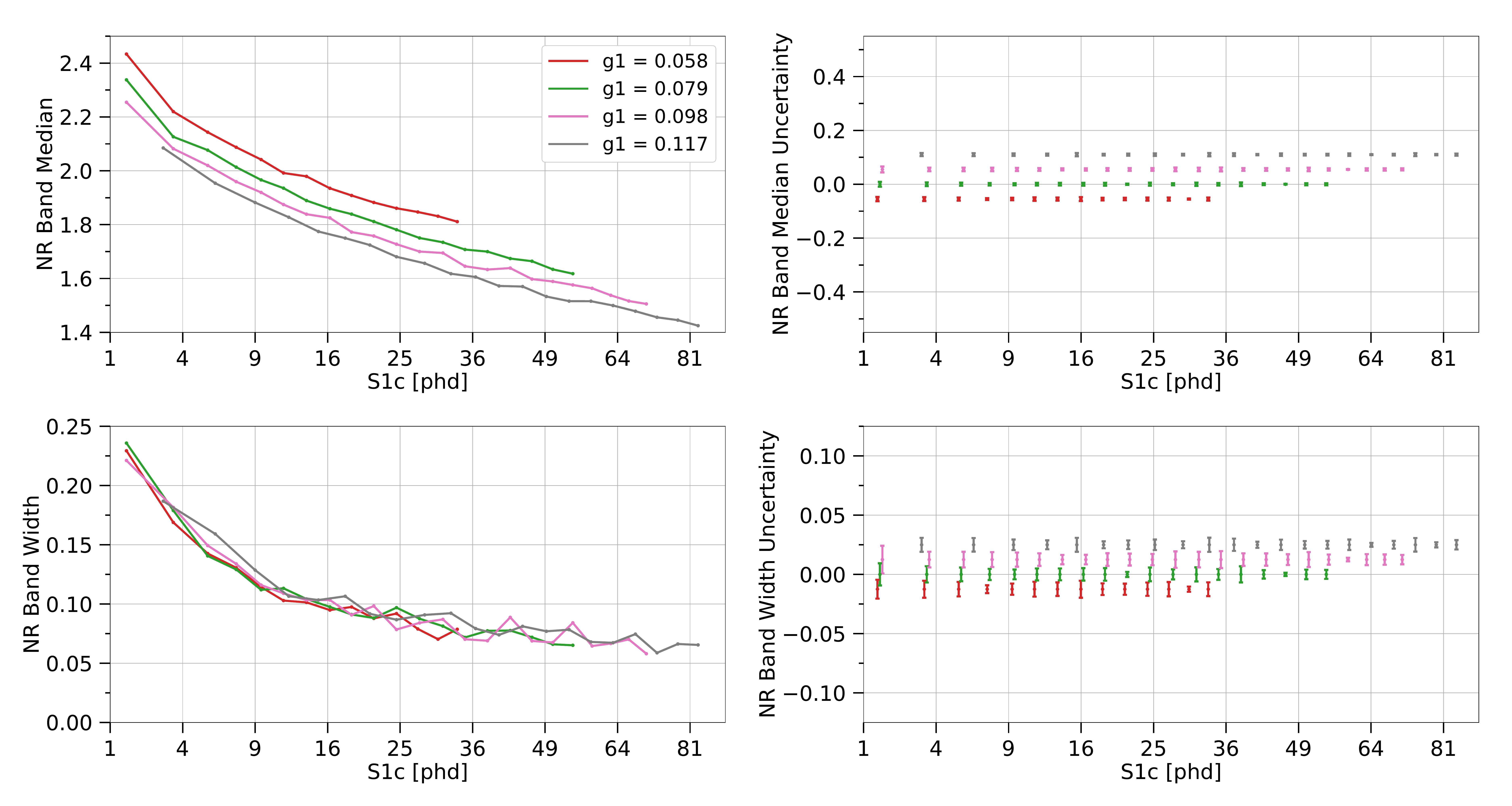}}
\caption{The median and width of the NR band for several values of $g_1$, using WS2013 data. The left plots show measurements, and the right plots display error bars corresponding to these measurements. The S1c axis is proportional to $\sqrt{\text{S1c}}$. In each row, the $y$-axes have the same range; the size of the error bars on the right plot can be directly translated to the points on the left plot. For ease of visualization on the right plots, the S1c values are slightly shifted relative to their true value, and the error bars are centered at a different $y$-value for each $g_1$. Note the S1c range varies for each $g_1$ because as $g_1$ decreases, the D-D end point in S1c space decreases. See Fig.~\ref{fig:NR_Medians_Widths_g1_Full} for the NR band median and width for all the $g_1$ values we considered.}
\label{fig:NR_Medians_Widths_g1}
\end{figure*}

The variation of the NR band with $g_1$, shown in Fig.~\ref{fig:NR_Medians_Widths_g1}, is similar to that of the ER band. It shifts down with $g_1$ for straightforward reasons; as light collection increases, log\textsubscript{10}(S2c/S1c) must decrease. The impact of $g_1$ on the NR band width is more muted, however.

\subsection{Variation with drift field}
\label{sec:Bands_Field_Variation}

Another crucial detector parameter is the drift field. As described in Section~\ref{sec:LUXFieldVariation}, \mbox{WS2014--16} saw significant field variation in the liquid xenon volume; we can use this to study the effect of electric field on the ER and NR bands.

First, we separate the electronic recoil and nuclear recoil data into bins based on the field at the recoil site. For \mbox{WS2014--16} data, the bin boundaries are [50, 80, 130, 240, 290, 340, 390, 440, 500]~V/cm. The bins were chosen to be wide enough such that the number of events in each bin is sufficient for the analysis, but narrow enough to yield precise measurements of field effects; they are overlaid over histograms of the data in Fig.~\ref{fig:Field_Distributions}. For WS2013 data, the data are all collected into a single field bin, leading to nine total field bins. In the LUX detector, electric field variation is degenerate with variation in light collection through $z$-position. Higher (lower drift time) regions of the LUX detector have higher drift field, but also lower light collection due to total internal reflection at the liquid-gas interface. This causes photons produced near the top of the detector to, on average, pass through more liquid xenon and encounter the PTFE surface more times than photons produced near the bottom of the detector. Thus, we then adjust the light collection efficiency in each field bin through the PMT removal procedure described in Section~\ref{sec:Bands_g1_Variation}. The adjustment in light collection, relative to the top of the LUX detector, ranges from 0.787 to 1.000 in \mbox{WS2014--16} and is equal to 0.744 for WS2013. This adjustment effectively accounts for the $z$-dependent position corrections, and so, in this portion of the analysis, we remove position corrections from the S1 variable.

Within each field bin, we calculate the median and width of the ER and NR bands. For the WS2013 results only, we adjust the band medians so that they are consistent with $g_2$ in \mbox{WS2014--16}: $g_2$ = 12.1 for WS2013 \cite{Akerib2016_LUX_Tritium_Calibration_Run3}, and the average $g_2$ = 19.085 for \mbox{WS2014--16}. Thus, the WS2013 band medians are shifted up by $\text{log}_{10}(19.085/12.1) = 0.198$.

\begin{figure*}[p!]
{\includegraphics[width=6.5in]{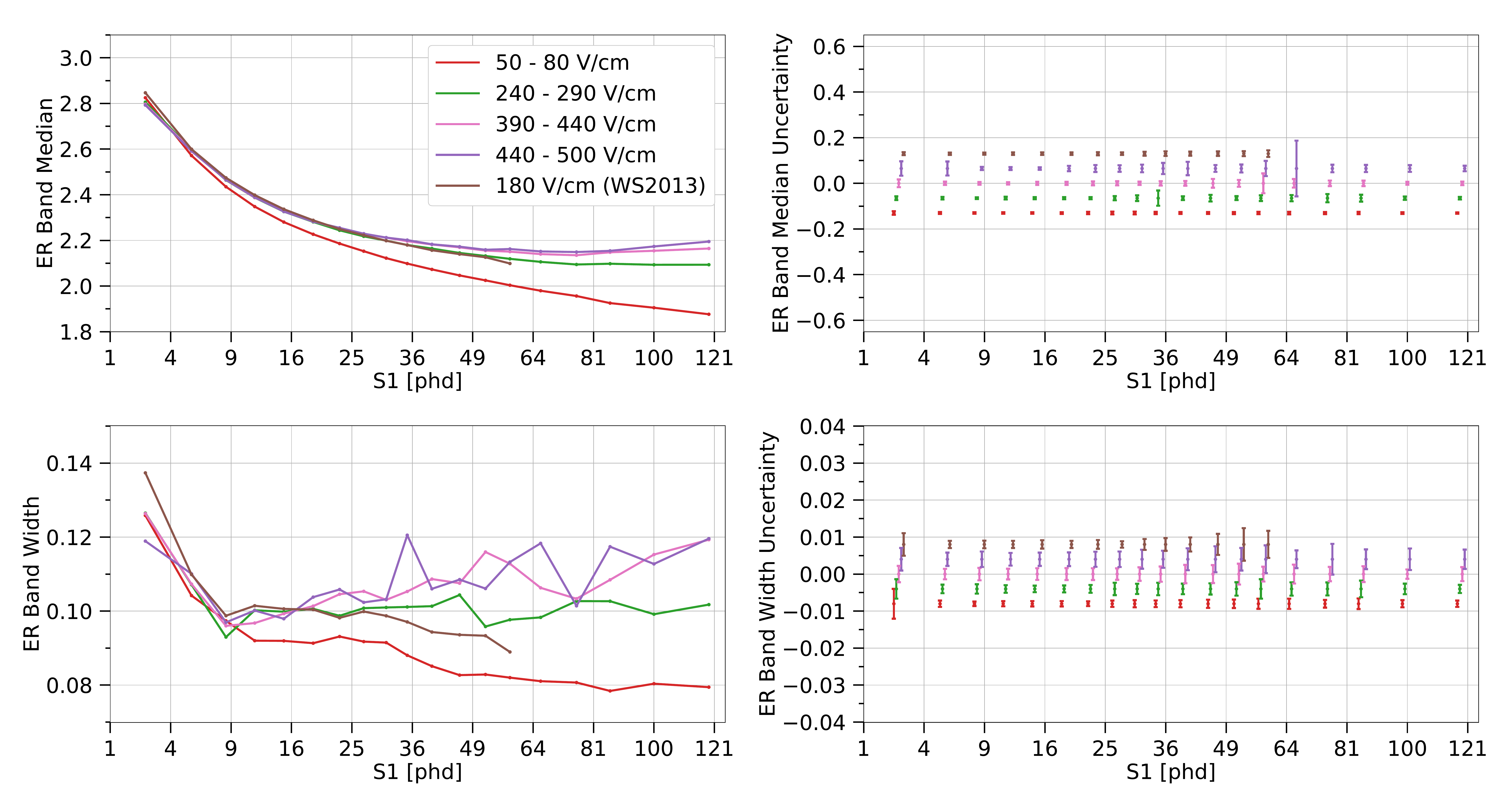}}
\caption{The median and width $\sigma_{\text{-}}$ of the ER band for several drift fields. The left plots show measurements, and the right plots display error bars corresponding to these measurements. The S1 axis is proportional to $\sqrt{\text{S1}}$. The ER band for WS2013 is adjusted so $g_2$ is consistent for the WS2013 and \mbox{WS2014--16} results. In each row, the $y$-axes have the same range; the size of the error bars on the right plot can be directly translated to the points on the left plot. For ease of visualization on the right plots, the S1 values are slightly shifted relative to their true value, and the error bars are centered at a different $y$-value for each field bin. See Fig.~\ref{fig:ER_Medians_Widths_Field_Full} for the ER band median and width for all the field bins we considered.}
\label{fig:ER_Medians_Widths_Field}
\end{figure*}

The results for the ER band in five field bins are shown in Fig.~\ref{fig:ER_Medians_Widths_Field}, where we exclude the other bins for visualization purposes. The results for all nine field bins can be found in Fig.~\ref{fig:ER_Medians_Widths_Field_Full} in Appendix~\ref{app:Additional_Figures}. As the drift field increases, the ER band median and width both increase convincingly. The former effect is expected; a plethora of data \cite{Dahl2009_Thesis, Aprile2006_NR_Yields_Columbia_and_Case, Akerib2019_LUX_Beta_Calibrations_Combined} shows that increasing electric field is correlated with a higher charge signal and smaller light signal, due to lower recombination. The increasing width is a consequence of this---with a lower light signal, the relative size of S1 fluctuations will increase. Crucially, as we will explore later, the width of the ER band is a major factor in discrimination. We note that the outlier width point at 35~phd for the 440--500~V/cm bin is the result of our skew-Gaussian fit converging to a negative skewness, whereas most fits converge to a positive skewness. It is not symptomatic of any trend; in fact, if we consider $\sigma$ rather than $\sigma_{\text{-}}$, this point is no longer an outlier.

\begin{figure*}[p!]
{\includegraphics[width=6.5in]{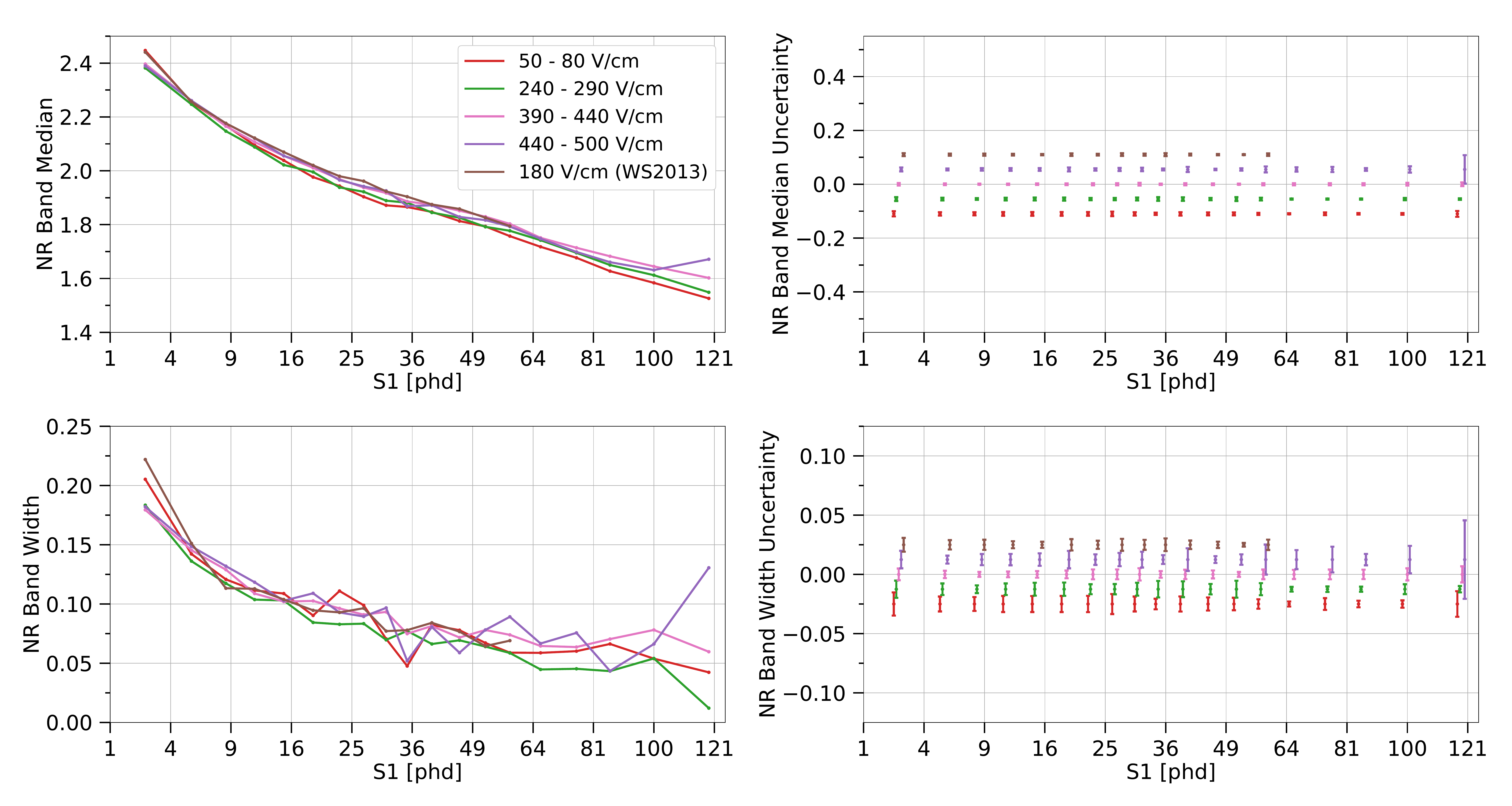}}
\caption{The median and width $\sigma_{\text{-}}$ of the NR band for several drift fields. The left plots show measurements, and the right plots display error bars corresponding to these measurements. The S1 axis is proportional to $\sqrt{\text{S1}}$. The NR band for WS2013 is adjusted so $g_2$ is consistent for the WS2013 and \mbox{WS2014--16} results. In each row, the $y$-axes have the same range; the size of the error bars on the right plot can be directly translated to the points on the left plot. For ease of visualization on the right plots, the S1 values are slightly shifted relative to their true value, and the error bars are centered at a different $y$-value for each field bin. See Fig.~\ref{fig:NR_Medians_Widths_Field_Full} for the NR band median and width for all the field bins we considered.}
\label{fig:NR_Medians_Widths_Field}
\end{figure*}

The variation of the NR band with electric field is shown in Fig.~\ref{fig:NR_Medians_Widths_Field}. The behavior of the NR band as we vary electric field is quite different to that of the ER band, indicating fundamental physical differences in these interactions. Primarily, the NR band is substantially less sensitive to electric field than the ER band, a finding that has been seen by others \cite{Dahl2009_Thesis}. The median moves up with increased electric field, in a statistically significant but small effect. The width has nearly no discernible variation from the electric field, except that the two highest field bins (390--440~V/cm and 440--500~V/cm) appear to have the largest widths across the entire energy range.

\section{Leakage and Discrimination}
\label{sec:Discrimination}

\subsection{Charge-to-light discrimination}
\label{sec:ChargeLightDiscrimination}

Studying the electronic and nuclear recoil bands separately is informative, but the discrimination power is the critical figure-of-merit for studying how detector parameters affect sensitivity. Figure~\ref{fig:Bands_Example} shows charge-to-light discrimination graphically; the electronic recoils lie above nuclear recoils in these axes. This is understood to be for two reasons. First, the initial exciton-to-ion ratio varies: it is approximately 1 for nuclear recoils \cite{Dahl2009_Thesis, Lenardo2015_NEST_Update, Sorensen2011_Nuclear_Recoils} and 0.2 for electronic recoils \cite{Doke2002_Scintillation_Yields, Aprile2007_ER_Anticorrelation, Lin2015_Xenon_Yields}. Second, recombination varies. Electronic recoils follow the Doke-Birks model \cite{Doke1988_Doke_Recombination} at high energies ($\gtrsim$ 10~keVee) \cite{Szydagis2011_NEST_Original, Akerib2017_LUX_Signal_Yields}, in which recombination is based on ionization density; they follow the Thomas-Imel model \cite{Thomas1987_Thomas_Imel_Recombination} at lower energies, in which thermal and diffusive effects smear out the track, and recombination can be considered to take place entirely in a small box of size O($\mu$m). Nuclear recoils are governed solely by the Thomas-Imel model at our energies of interest \cite{Lenardo2015_NEST_Update}. Thus, at these lowest energies, electronic recoils are disparate from nuclear recoils in their initial exciton-to-ion ratio and the fraction of energy lost to heat.

Within each S1c bin, we can calculate the charge-to-light leakage fraction (or alternatively, its inverse: the discrimination power) at 50\% nuclear recoil acceptance in two ways. First, we can count the number of weighted electronic recoils falling below the NR band median. We take the uncertainty on the leakage fraction to be the Poisson error. Second, we can integrate the skew-Gaussian ER distribution below the NR band median. The uncertainty here is found by propagating the errors in the ER band skew-Gaussian fit and the NR band Gaussian fit. The two methods have been confirmed to be consistent with each other, except in the lowest S1c bin where, due to PMT and threshold effects, the distribution of log\textsubscript{10}(S2c/S1c) does not match a skew-Gaussian. The latter method allows us to calculate the leakage fraction even if the number of events in the bin is too low to count the leaked events, so we use it except where specifically mentioned.

Before presenting our results, we discuss sources of potential systematic uncertainty on the leakage fraction. First, $g_1$ and $g_2$ are uncertain at the 1--3\% level; thus, the positions of the ER and NR bands are uncertain at a similar scale. However, this uncertainty will not lead to a systematic error on the leakage fraction, because if the $g_1$ or $g_2$ measurement is offset from its true value, the ER and NR bands will move together by the same amount. An error in $g_1$ could affect the ER band width and thus the electronic recoil leakage fraction, but this effect is insignificant at the level of the uncertainty on $g_1$. Second, when we decrease $g_1$ by using a subset of LUX PMTs, this procedure introduces an extra systematic uncertainty on $g_1$. This uncertainty has been calculated and is \textless0.1\%, so it is negligible. Third, the binning of log\textsubscript{10}(S2c/S1c) will introduce a bias on the ER skew-Gaussian and NR Gaussian fits. We have experimented with different levels of binning and observed that the leakage fraction is not significantly affected by our choice of binning. The only effect of this choice is whether the ER skew-Gaussian fit converges. Fourth, in the lowest S1c bin only, the NR band median is biased slightly upward due to the finite S2 analysis threshold (see Sec.~\ref{sec:NR_Band} for details). This means that the estimated leakage fraction is higher than it would be in a zero-threshold analysis. Using simulations, we have determined that this effect is smaller than the uncertainties on the leakage fraction from statistics and Gaussian-fitting the nuclear recoil data. However, an experiment with a higher S2 threshold could be significantly affected by the shift in the NR band, so caution should be taken if extrapolating our lowest-energy results to such an experiment.

\subsubsection{\texorpdfstring{Variation with $g_1$}{Variation with g1}}
\label{sec:Leakage_g1_Variation}

\begin{figure}
{\includegraphics[width=3.25in]{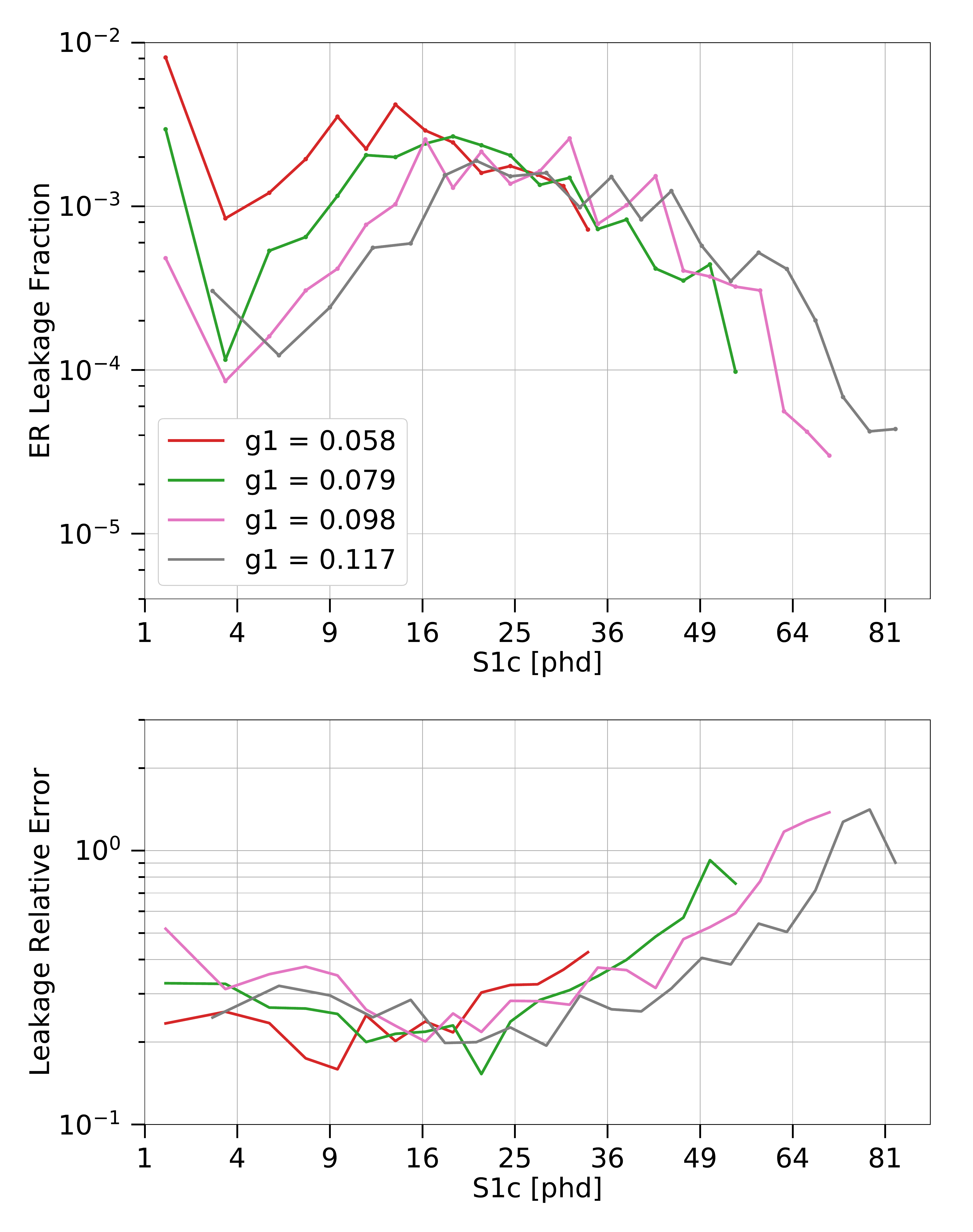}}
\caption{\textit{(Top)} The electronic recoil leakage fraction for a flat energy spectrum in S1c bins, for various values of $g_1$, calculated from a skew-Gaussian extrapolation of the ER band below the NR band median. The S1c axis is proportional to $\sqrt{\text{S1c}}$. The leakage fraction calculated in this way is consistent with the real counted leakage, except in the lowest S1c bin; see Fig.~\ref{fig:Leakage_Lowest_g1} for a comparison between the two leakage calculations in this S1c bin. \textit{(Bottom)} The relative error on these leakage fraction values, defined as: $\text{leakage\_fraction\_error}~/~\text{leakage\_fraction}$. Note that the leakage relative error can be greater than 1, indicating that the leakage fraction is consistent with 0. See Fig.~\ref{fig:Leakage_g1_Full} for the leakage across all the $g_1$ bins in the dataset.}
\label{fig:Leakage_g1}
\end{figure}

Calculating the leakage function in S1c bins with $g_1$ variation gives the results in Fig.~\ref{fig:Leakage_g1}. The most striking effect is that as $g_1$ increases, the leakage decreases. Furthermore, it shares some features with the bottom of Fig.~\ref{fig:ER_Medians_Widths_g1}, namely that the effect is strongest below 25~phd. This suggests that the improvement in discrimination is due to the shrinking of the ER band width. Above 25~phd, the improvement in discrimination with $g_1$ is absent or suppressed, but we do not necessarily conclude that $g_1$ has no effect on discrimination at high energies. Low $^3$H statistics at energies near the 18.6-keV end point give rise to large uncertainties on the leakage fractions. As mentioned, the real (counted) leakage does not match the skew-Gaussian leakage in the lowest S1c bin only; the ratio between the two is plotted in Fig.~\ref{fig:Leakage_Lowest_g1} in the Appendix.

Another way to look at xenon discrimination power is the total leakage in a wide energy range. Using the full set of PMTs and the WS2013 data, we find that the leakage fraction from 0--50~phd, i.e.~the WIMP search region used in the 2013 limit \cite{Akerib2016_LUX_Run3_Reanalysis}, is about 0.1\%.\footnote{Our measurement of 0.1\% is different than the 0.2\% reported in \cite{Akerib2018_LUX_Run3_Comprehensive}. The difference is due to our use of a skew-Gaussian distribution, as well as our energy weighting.}

\begin{figure}
{\includegraphics[width=3.25in]{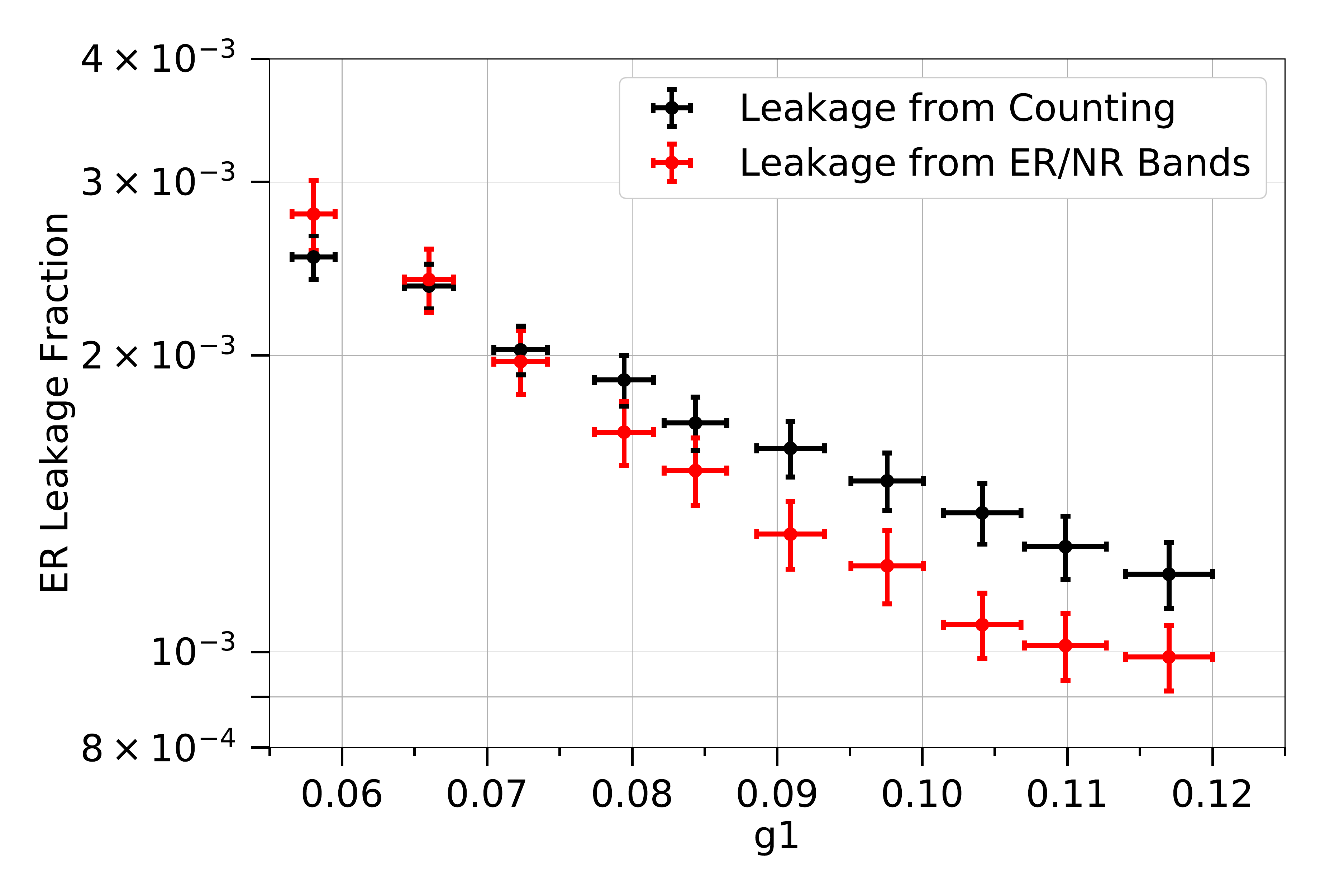}}
\caption{The integrated electronic recoil leakage for a flat recoil energy spectrum from 0--9.7~keVee, while varying $g_1$ in WS2013 data. The max S1c is proportional to 50 photons detected at ${g_1=0.117}$. The leakage is calculated by either counting the number of electronic recoils falling below the NR band (black), or by integrating the electronic recoil skew-Gaussian fits below the NR band (red). The discrepancy between the two methods is explained by a poor fit of the data to a skew-Gaussian distribution in the lowest S1c bin. Statistical errors from Poisson fluctuations are shown.}
\label{fig:Integrated_Leakage_g1}
\end{figure}

If we artificially remove PMTs as described in Section~\ref{sec:Bands_g1_Variation}, we can still calculate the total leakage, but there is an extra step required due to the $^3$H end point. Since the end point is around 85~phd, any setup in which the relative light collection is less than ${50/85=0.59}$ of the full detector will show bizarre behaviors in which the ER band cannot be calculated properly. Thus, we shift the maximum S1c to be proportional to $g_1$; e.g.~S1c${_{\text{max}}=50}$~phd for ${g_1=0.117}$, S1c${_{\text{max}}=25}$~phd for ${g_1=0.0585}$, etc. This effectively keeps the maximum energy constant at 9.7~keVee. The results are shown in Fig.~\ref{fig:Integrated_Leakage_g1}, and they show convincingly that as light collection increases, discrimination improves. The total leakage fraction varies slightly based on the method we use. If we count the weighted number of electronic recoils falling below the NR band median, we generally get a higher leakage than if we use the skew-Gaussian fits; the reverse is true for the lowest $g_1$ values. This discrepancy is almost entirely due to the discrepancy in the lowest S1c bin.

\subsubsection{Variation with drift field}
\label{sec:Leakage_Field_Variation}

Meanwhile, we can also examine the effect of drift field on charge-to-light discrimination, as done in Fig.~\ref{fig:Leakage_Field} (and Fig.~\ref{fig:Leakage_Lowest_Field} in the Appendix for the lowest S1 bin). The effect is mostly muted. Drift field does not provide significant variation in the leakage fraction when we look at individual S1 bins. However, we can note some patterns. Across the entire energy range, the lowest field bin of 50--80~V/cm is among the highest leakages for a given S1 bin. Meanwhile, the highest and second-highest fields (390--440~V/cm and 440--500~V/cm, respectively) also often give the highest leakage. Indeed, there seems to be an effect of the leakage reaching a minimum at 240--290~V/cm in several S1 bins.

The WS2013 results are in line with the \mbox{WS2014--16} results, even though the ER and NR bands separately showed some outlier behavior. A potential explanation for this latter effect is uncertainties in $g_1$, $g_2$, and the drift field at the recoil site. The LUX collaboration has previously shown that in order for simulations to correctly mimic data, these quantities need to be slightly adjusted from their measured values \cite{Akerib2020_LUX_ER_Modeling}.

\begin{figure}
{\includegraphics[width=3.25in]{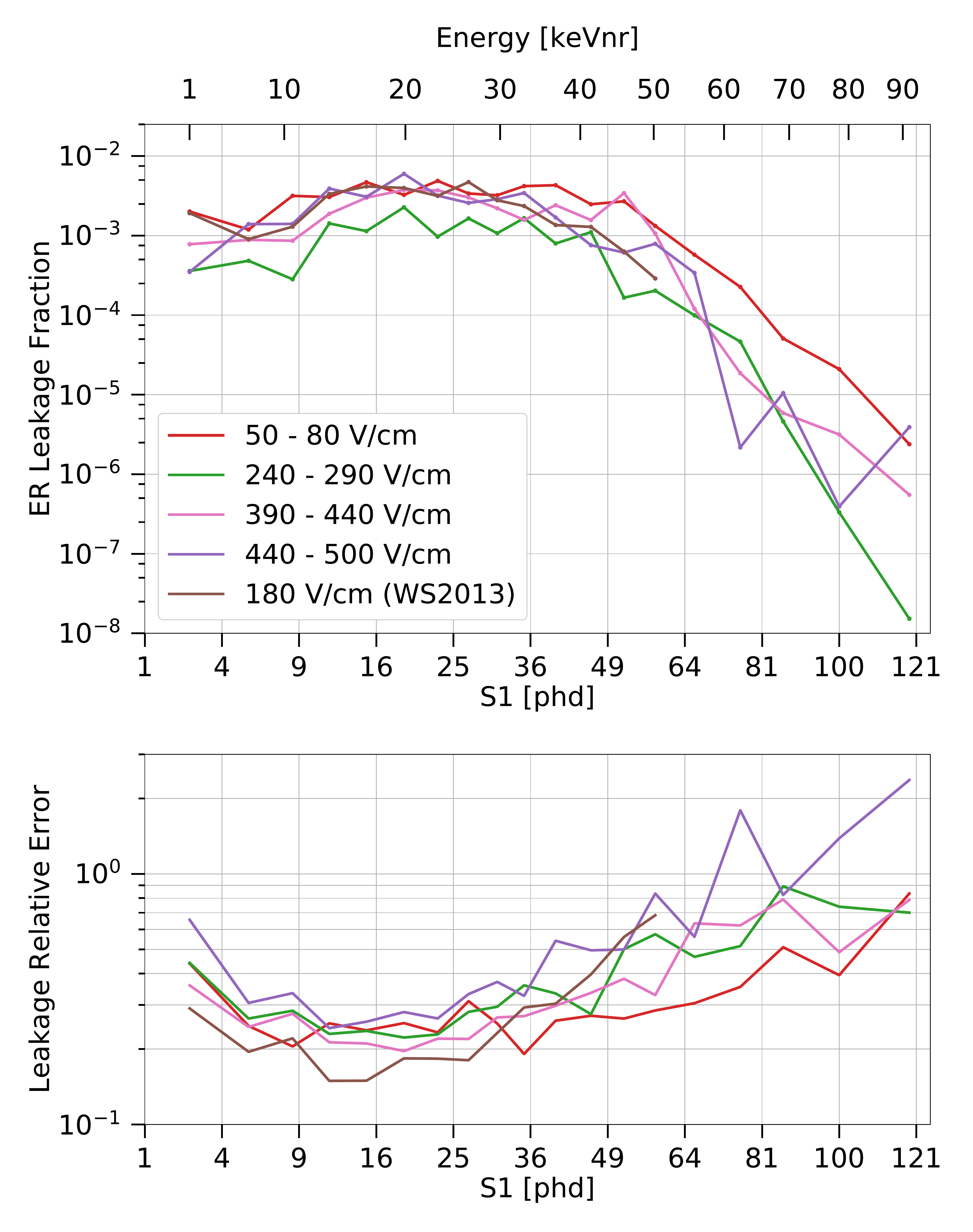}}
\caption{\textit{(Top)} The electronic recoil leakage fraction for a flat energy spectrum in S1 bins, for various values of drift field, calculated from a skew-Gaussian extrapolation of the ER band below the NR band median. The S1 axis is proportional to $\sqrt{\text{S1}}$. The equivalent nuclear recoil energy for an S1 is calculated by using the S1 and S2c at the median of the NR band; this varies by field, but not significantly, so we report the energy averaged over the eight field bins. The leakage fraction calculated in this way is consistent with the real counted leakage, except in the lowest S1 bin; see Fig.~\ref{fig:Leakage_Lowest_Field} for a comparison between the two leakage calculations in this S1 bin. \textit{(Bottom)} The relative error on these leakage fraction values, defined as: $\text{leakage\_fraction\_error}~/~\text{leakage\_fraction}$. Note that the leakage relative error can be greater than 1, indicating that the leakage fraction is consistent with 0. See Fig.~\ref{fig:Leakage_Field_Full} for the leakage across all the field bins in the dataset.}
\label{fig:Leakage_Field}
\end{figure}

\begin{figure}
{\includegraphics[width=3.25in]{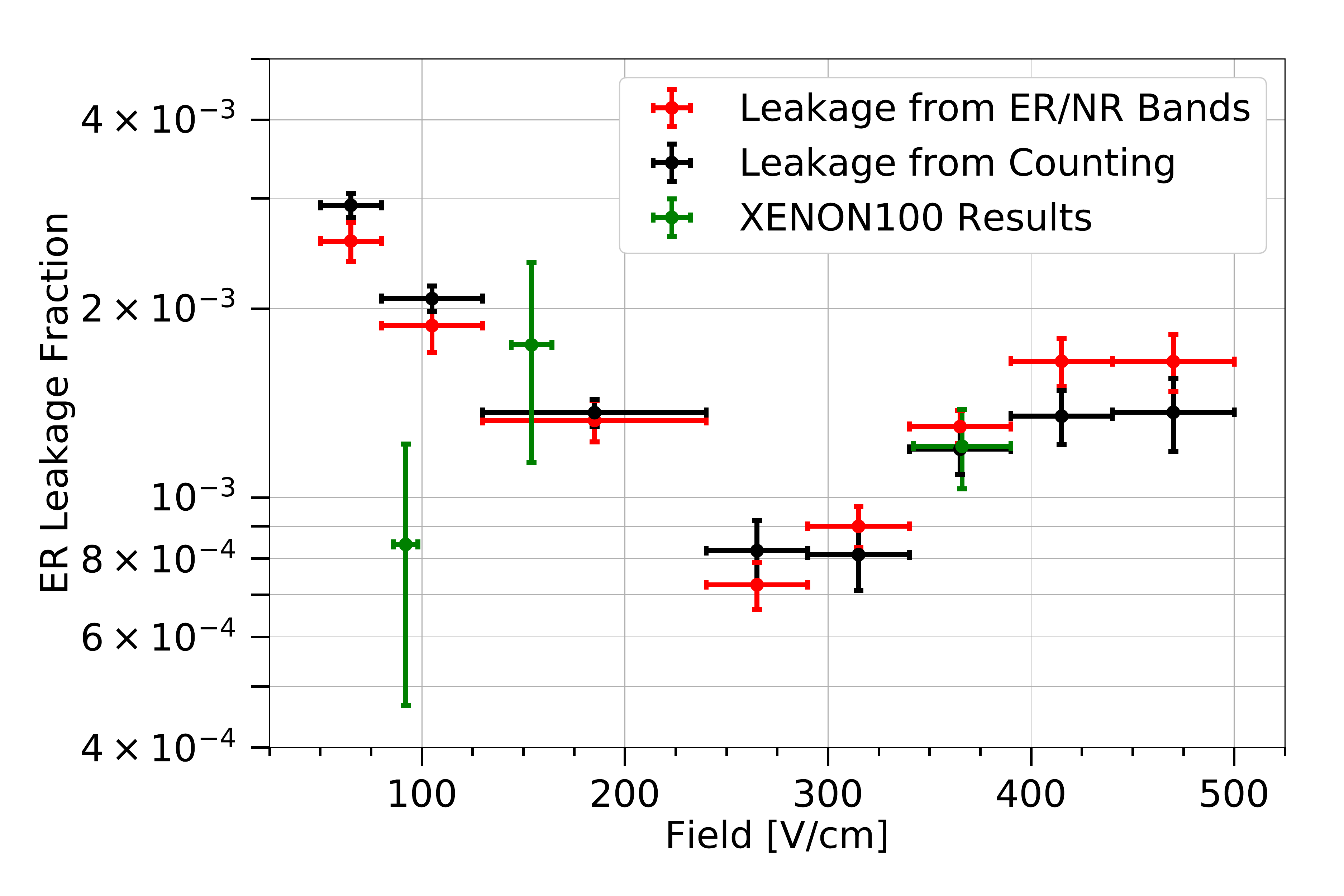}}
\caption{The integrated electronic recoil leakage for a flat recoil energy spectrum from 1--80 S1 photons detected (equivalent to 2--65~keVnr), while varying drift field in \mbox{WS2014--16} data. The leakage is calculated by either counting the number of electronic recoils falling below the NR band (black), or by integrating the electronic recoil skew-Gaussian fits below the NR band (red). Our optimal field over the range examined is $\sim$300~V/cm, which is within the expected drift field range of the forthcoming LZ experiment and matches LZ's design specification of 310~V/cm. However, an exact quantitative prediction of the LZ leakage is impossible because of the higher expected $g_1$ and $g_2$ in LZ \cite{Akerib2020_LZ_Sensitivity}. Results from XENON100 \cite{Aprile2018_XENON100_Discrimination} are shown in green, where we use their leakages at ${g_1=0.081}$ (our results are at ${g_1=0.087}$). The XENON100 leakages correspond to 8--32 photons detected, i.e.~11--34~keVnr.}
\label{fig:Integrated_Leakage_Field}
\end{figure}

We can also calculate the total leakage up to 80~phd, the maximum pulse area considered in the LZ projected sensitivity \cite{Akerib2020_LZ_Sensitivity}. This is done in Fig.~\ref{fig:Integrated_Leakage_Field} and shows strong evidence of discrimination being maximized around 300~V/cm. The existence of an optimal drift field in the range accessible to LUX motivated a reduction in the nominal operating field of LZ. The early designs considered a drift field of 600~V/cm \cite{Akerib2015_LZ_CDR}, while the final design adopts a field of 310~V/cm \cite{Akerib2020_LZ_Sensitivity, Mount2017_LZ_TDR}. We compare these results to those from XENON100 \cite{Aprile2018_XENON100_Discrimination} at similar $g_1$, and we find agreement at the higher fields but a discrepancy at their lowest field of 92~V/cm. However, we emphasize that a direct comparison is impossible, because the two experiments used different S1 thresholds---1~photon detected in LUX and 8~photons detected in XENON100, corresponding to 2~keVnr and 11~keVnr, respectively.

\subsection{Pulse Shape Discrimination}
\label{sec:PulseShapeDiscrimination}

The charge-to-light ratio is undoubtedly the best discriminant in liquid xenon, but under some conditions, its performance can be enhanced with pulse shape information. Xenon excimers are formed in either a singlet or triplet state, and these deexcite on different time scales. The mean lifetime of a singlet excimer is ${\tau=3.27\pm0.66}$~ns, while that of a triplet excimer is ${\tau=23.97\pm0.17}$~ns, as measured by the LUX Collaboration \cite{Akerib2018_LUX_PSD}. The fraction of excimers produced in each state is found to vary based on the incident particle, with nuclear recoils producing a greater fraction of fast-decaying singlets than electronic recoils. In this paper, we build on the LUX collaboration's previous analysis of pulse shape discrimination \cite{Akerib2018_LUX_PSD}. We explore how our ability to discriminate is dependent on drift field and particle energy.

\begin{figure}
{\includegraphics[width=3.25in]{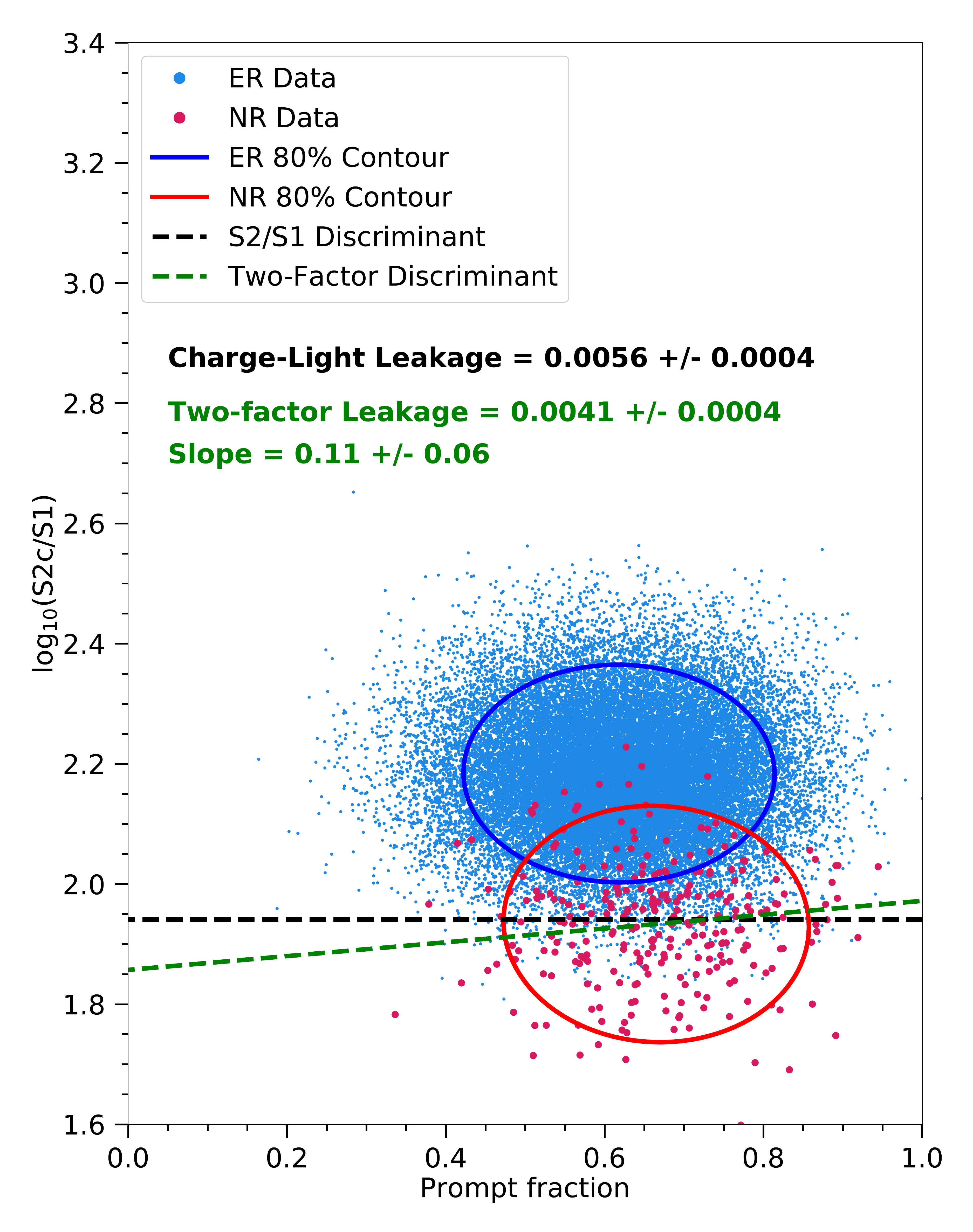}}
\caption{An example of how the two-factor leakage is calculated, using data for 80--130~V/cm and 20--30~phd. The electronic recoil and nuclear recoil data are plotted on axes of charge-to-light vs.~prompt fraction. Ellipses containing 80\% of the data are shown. The black dashed line shows the nuclear recoil median in log\textsubscript{10}(S2c/S1) only, and the black text shows the corresponding electronic recoil leakage fraction. The green dashed line shows the optimized discriminating line between the two distributions; the green text shows the resulting electronic recoil leakage, as well as the slope of this line. We note about 27\% improvement in the leakage fraction. Further details on this calculation can be found in the text.}
\label{fig:PSD_Example}
\end{figure}

Figure~\ref{fig:PSD_Example} shows an example of how this analysis was conducted. Each event is assigned a \textit{prompt fraction} value, based on the shape of its S1 pulse. The exact calculation is detailed in \cite{Akerib2018_LUX_PSD}, but in summary: each S1 pulse is decomposed into its detected photon constituents, these detected photons are adjusted based on PMT-specific effects and the location of the recoil, and the fraction of photons within a particular time window is computed. We make one key adjustment to the calculation, which is effectively the same $g_1$ adjustment described in Section~\ref{sec:Bands_g1_Variation}. Within each electric field bin, we only consider photons that have hit the PMTs used to calculate the ER and NR bands in that bin in order to calculate the prompt fraction. This allows us to adjust for light collection, which we assume accounts for the depth dependence observed in \cite{Akerib2018_LUX_PSD}. This fraction is usually between 0.4 and 0.9, but the distribution of prompt fraction for electronic recoils is somewhat lower than the distribution for nuclear recoils. As a result, pulse shape serves as a moderately effective discriminant on its own, as also seen by the XMASS experiment \cite{Abe2018_XMASS_NR_Scintillation_And_PSD, Ueshima2011_XMASS_PSD}, the ZEPLIN-I experiment \cite{Alner2005_ZEPLIN_I_WS}, and others \cite{Kwong2010_Case_PSD}.

Here, we construct a two-factor discriminant by combining pulse shape with the charge-to-light ratio; this reflects the same strategy as the previous LUX publication and other past analyses \cite{Akimov2010_ZEPLIN_III_Inelastic, Kwong2010_Case_PSD}. Within each bin of drift field and S1, we consider the prompt fraction and log\textsubscript{10}(S2c/S1) in two dimensions. We use maximum likelihood estimation on the ER and NR populations separately to fit the data to a 2D Gaussian distribution. The data are observed to match a 2D Gaussian distribution well except the outermost edges of the electronic recoil data (\textless10\% of the ER distribution). Then, we choose a line in prompt fraction vs.~log\textsubscript{10}(S2c/S1) space to discriminate between the two populations. The line is forced to go through the center of the NR 2D Gaussian fit, but the slope is a free parameter; it is determined by minimizing the ER leakage into the NR region. Note one key difference already from \cite{Akerib2018_LUX_PSD}: the previous analysis forced this line to pass through the NR median prompt fraction and log\textsubscript{10}(S2c/S1), but we find that using the center of the 2D Gaussian gives lower leakage while maintaining 50\% NR acceptance. However, for the lowest S1 bin (0--10~phd), the 2D Gaussian fit is poor, because there is an abundance of events with prompt fraction of exactly 0 or 1.\footnote{If an S1 pulse has only a few photons, there is a significant probability that its prompt fraction is 0 or 1.} This fit is so poor that the resulting two-factor leakage ends up being greater than the charge-to-light leakage. As a result, for this bin only, we continue to use the median in both dimensions.

The second addition we make is to use the bootstrap method to determine the slope of the discriminating line and its uncertainty. First, a random selection of $N$ electronic recoil events is chosen with replacement, where $N$ is the total number of electronic recoil events in this field/S1 bin. This means that it is almost certain that some events will be in the bootstrap sample twice or more often. Then, we calculate the optimal slope on this sample, using the procedure described in the previous paragraph. We do this 100 times to get a distribution of slopes (the number of iterations has been chosen to be high enough such that the resulting distribution of slopes is negligibly affected by the pseudo-random number generation). The slope that we use for the final discriminating line of this field/S1 bin is the mean of this distribution, while the error on that slope is given by the standard deviation of this distribution. Finally, we calculate the two-factor leakage by counting the number of (weighted) electronic recoil events falling below the discriminating line. This procedure allows us to obtain an uncertainty on the slope of the discriminating line, and it serves as a safeguard, preventing the calculation from being too dependent on a single leaked electronic recoil.

The statistical error on the two-factor leakage has two components: the Poisson error on the number of leaked events and the error on the slope of the discriminating line. The total statistical error is not found by adding these in quadrature because they are not independent; the Poisson error is a function of the leakage value, so it is dependent on the discriminating line error. We perform this analysis as follows. Given an S1 and field bin, we calculate the distribution of slopes as described in the previous paragraph. We then draw 100 random slopes, assuming that this distribution is Gaussian with the appropriate mean and standard deviation.\footnote{The Gaussian assumption is accurate for the majority of S1/field bins, although there are a few bins where the distribution has a sharp preference for a slope separate from the main peak. In these, a handful of events bias the minimization toward this value, and the use of a Gaussian distribution smooths out this effect.} For each slope, we calculate the two-factor leakage and its Poisson error. Then, we randomly choose a leakage from a Gaussian distribution with the two-factor leakage as its mean and the Poisson error as its width. Finally, we take the mean and standard deviation of this 100-sample dataset as the average leakage and its error.

\begin{figure}
{\includegraphics[width=3.25in]{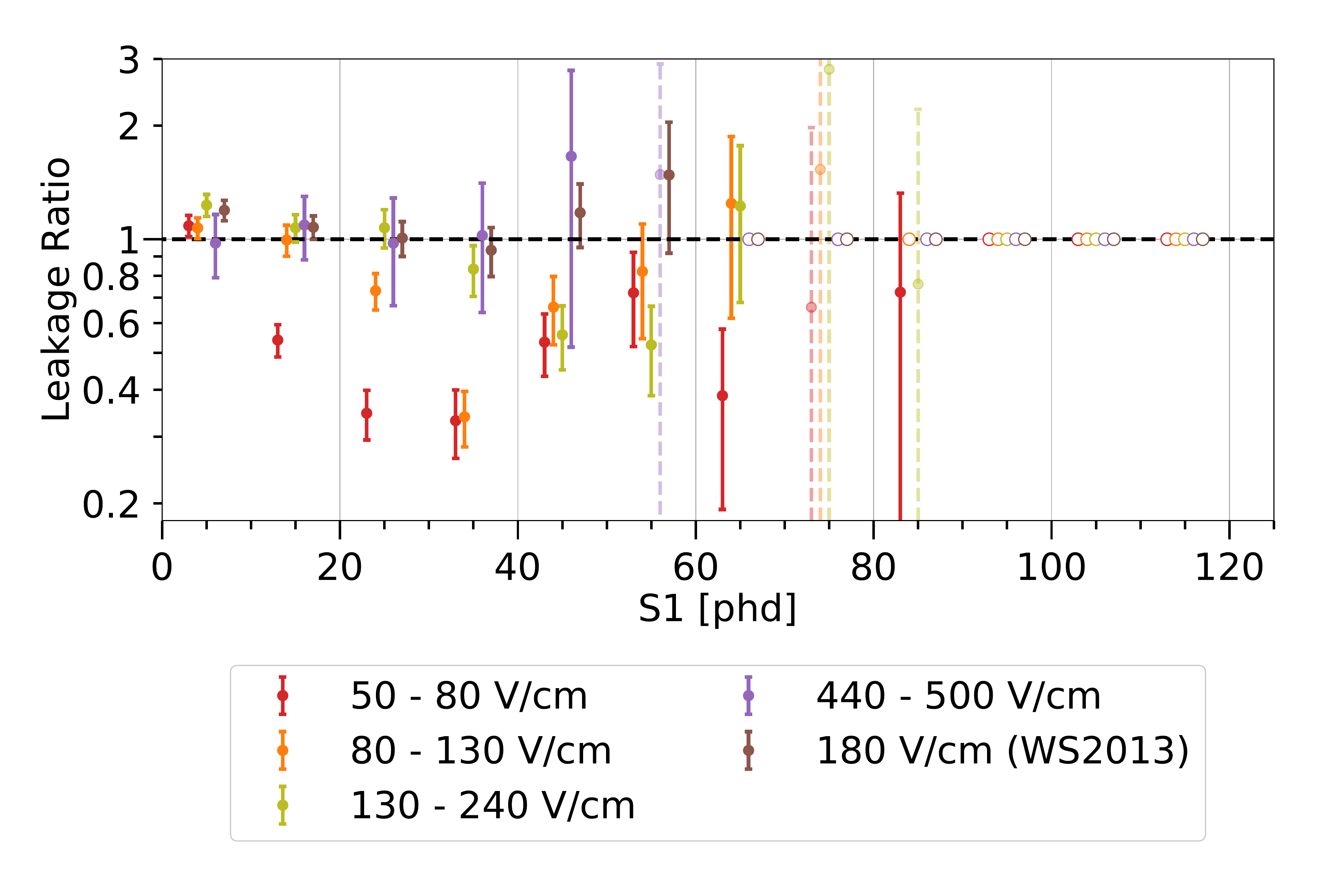}}
\caption{The ratio of two-factor leakage to charge-to-light leakage, for various S1 and drift field bins. Error bars are statistical; see text for details. Open circles represent bins for which charge-to-light discrimination alone gives zero electronic recoils falling below the NR band; as a result, it is impossible to calculate the improvement from two-factor discrimination. Leakage ratios with large error bars are made transparent and plotted as dashed lines to draw the eye toward more precise measurements. The plotted S1 values are slightly shifted relative to their true value (by up to 2~phd) for ease of visualization. The true S1 coordinates are 5~phd, 15~phd, 25~phd, etc. See Fig.~\ref{fig:PSD_Leakage_Ratios_Full} for the leakage ratios across all the field bins in the dataset.}
\label{fig:PSD_Leakage_Ratios}
\end{figure}

The results are shown in Fig.~\ref{fig:PSD_Leakage_Ratios}, where we plot the ratio of the two-factor leakage to the charge-to-light leakage. A marked improvement in discrimination is observed below 50~phd for the lowest electric fields (50--80 and 80--130~V/cm). The 130--240~V/cm field bin is ambiguous: the \mbox{WS2014--16} data show improvement for energies between 30--60~phd, but the WS2013 data at 180~V/cm show no improvement over charge-to-light discrimination. For higher electric fields, there does not seem to be a significant reduction in leakage when using the two-factor discriminant. The most likely explanation for this is that higher electric fields are associated with less recombination. Thus, fewer scintillation photons leave the recoil site, and the S1 pulse shape is dominated by the longer triplet decay time for both nuclear and electronic recoils \cite{Mock2014_NEST_Pulse_Shape}. We also do not observe improvement at higher energies, but this could be due to low statistics; there are plenty of $^{14}$C events in the dataset, but the charge-to-light leakage is so robust that virtually none of them falls below the NR band. Although the leakage values appear to be different than the ones reported in \cite{Akerib2018_LUX_PSD}, this is due to the varying methodology and drift field range. We have confirmed that if we modify our procedure to be identical to the one detailed there, our results are consistent.

\begin{figure}
{\includegraphics[width=3.25in]{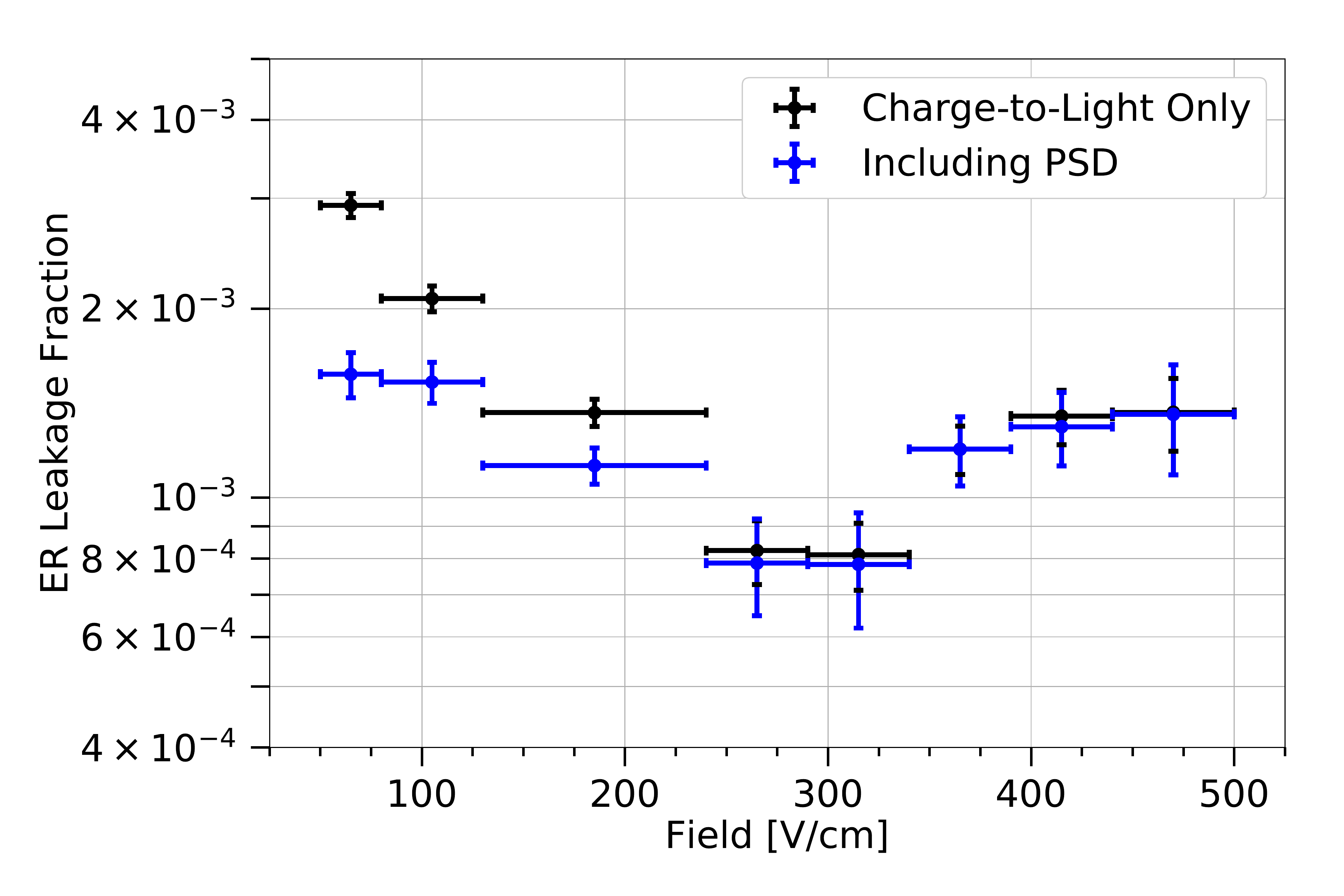}}
\caption{The integrated electronic recoil leakage for a flat recoil energy spectrum from 1--80 S1 photons detected (equivalent to 2--65~keVnr), while varying drift field in \mbox{WS2014--16} data. The leakage is calculated using only the charge-to-light ratio, i.e.~log\textsubscript{10}(S2c/S1), and using both charge-to-light and pulse-shape discrimination in tandem. Both leakage values are based on the ``counting'' method described in Fig.~\ref{fig:Integrated_Leakage_Field}, where we count the number of electronic recoils leaking into the nuclear recoil 50\% acceptance region.}
\label{fig:Integrated_Leakage_Field_PSD}
\end{figure}

We also consider the two-factor leakage across the entire 1--80~phd energy range. Figure~\ref{fig:Integrated_Leakage_Field_PSD} shows these results, as well as a comparison to the charge-to-light only leakage. We see that although there is improvement in discrimination for low fields, the optimal drift field bins are still 240--290~V/cm and 290--340~V/cm. We also show the two-factor leakage in S1 bins in Fig.~\ref{fig:Leakage_Field_Full_PSD}, although we emphasize that this is an estimate. The charge-to-light leakage in S1 bins is calculated with a skew-Gaussian extrapolation, whereas the leakage ratio is calculated by counting electronic recoils in the nuclear recoil acceptance region; thus, it is not exactly consistent to combine the two.

\begin{figure}
{\includegraphics[width=3.25in]{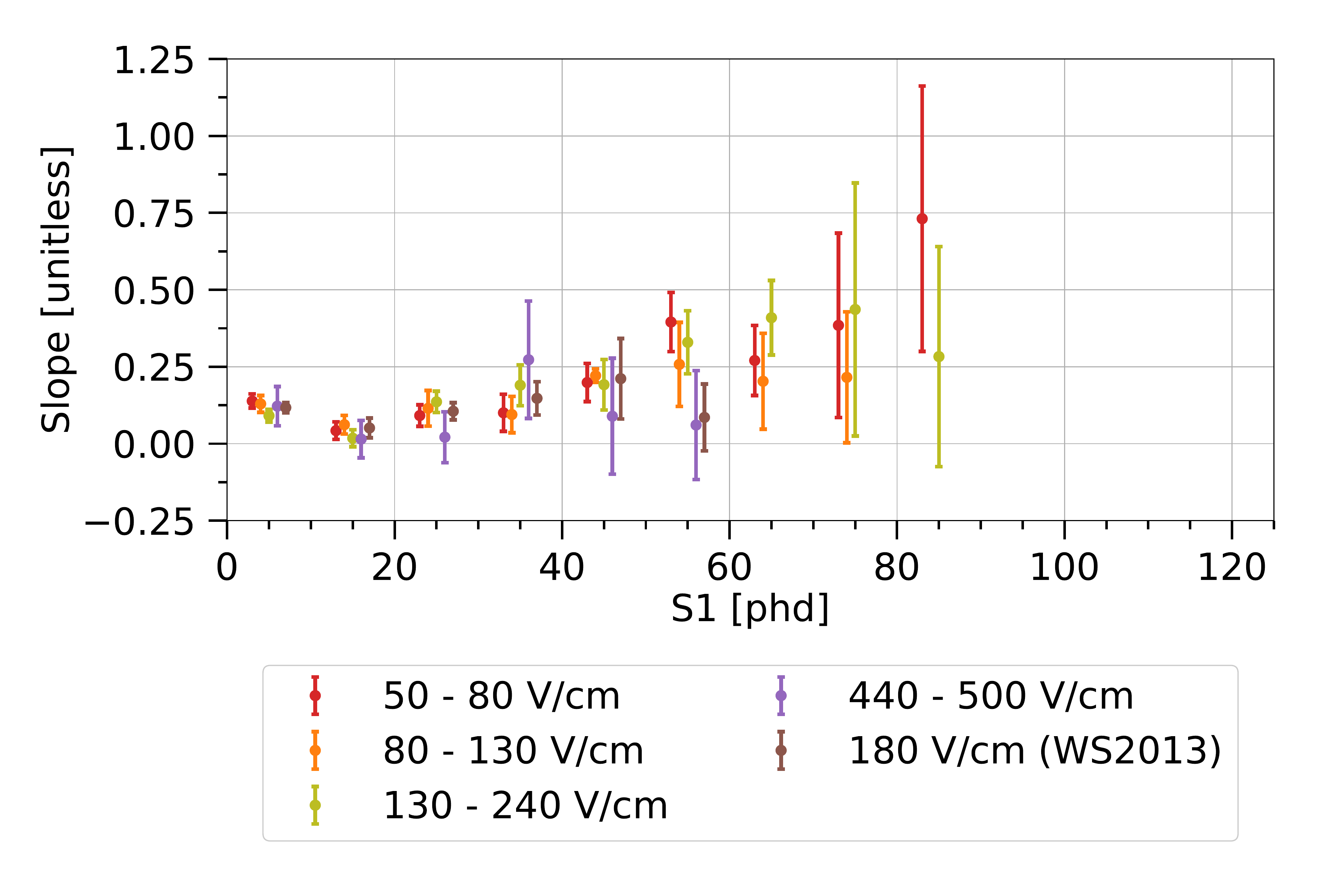}}
\caption{The slope of the two-factor discrimination line in log\textsubscript{10}(S2c/S1) vs.~prompt fraction space, for each S1 and field bin. Missing points represent bins for which charge-to-light discrimination alone gives zero electronic recoils falling below the NR band. The plotted S1 values are slightly shifted relative to their true value (by up to 2~phd) for ease of visualization. The true S1 coordinates are 5~phd, 15~phd, 25~phd, etc. See Fig.~\ref{fig:PSD_Slopes_vs_Field_Full} for the slopes across all the field bins in the dataset.}
\label{fig:PSD_Slopes_vs_Field}
\end{figure}

Figure~\ref{fig:PSD_Slopes_vs_Field} shows how the slope of the discriminating line varies with electric field and S1. The most striking effect is that the slope is almost always positive, meaning that the ER population is tilted toward higher log\textsubscript{10}(S2c/S1) at higher prompt fraction. In addition, there appears to be a weak increase in the slope with energy and no dependence on field. Note that for ease of visualization, we only show five field bins in Figs.~\ref{fig:PSD_Leakage_Ratios}~and~\ref{fig:PSD_Slopes_vs_Field}; the full set of field bins is shown in Figs.~\ref{fig:PSD_Leakage_Ratios_Full}~and~\ref{fig:PSD_Slopes_vs_Field_Full} in Appendix~\ref{app:Additional_Figures}.

\section{Modeling skewness}
\label{sec:Skewness}

\subsection{Noble Element Scintillation Technique}
\label{sec:NEST}

Skewness of the ER band has been observed previously \cite{Lebedenko2009_ZEPLIN_III_First_Science_Run, Aprile2019_XENON1T_Signal_Background_Models}, but no physical motivation for it has emerged.\footnote{Reference~\cite{Aprile2019_XENON1T_Signal_Background_Models} does not directly report skewness. However, they observe that their signal-like mismodeling parameter is fit to a negative value by data. This means that within S1c bins, the S2c distribution is shifted to higher values, an identical effect qualitatively to our observation of positive ER band skewness.} Here, we present one potential explanation by utilizing the Noble Element Scintillation Technique, or NEST \cite{Szydagis2011_NEST_Original, Lenardo2015_NEST_Update, Szydagis2019_NEST_v2.0.1}.

The current stable version of NEST is tagged as NESTv2.0.1. Full details can be found in \cite{Szydagis2019_NEST_v2.0.1}, but for the sake of this paper, we summarize the main principles of how NEST simulates a two-phase liquid/gas xenon time projection chamber. First, the detector is modeled, including parameters such as its size, drift field, $g_1$ and $g_2$, electron lifetime, and information about its PMTs. Then, an energy deposition is simulated with a location in the detector, the species of the incident particle, and the amount of energy deposited. NEST uses empirical fits to world data to determine the average charge and light yield for the interaction. It then simulates the number of excitons and ions produced by the energy deposit, as well as the number of electrons and photons leaving the recoil site. This step uses a recombination model that extends the naive binomial variance with a term that is quadratic in $N_{\text{ions}}$, as multiple analyses \cite{Akerib2016_LUX_Tritium_Calibration_Run3, Akerib2017_LUX_Signal_Yields, Akerib2019_LUX_Beta_Calibrations_Combined, Akerib2020_LUX_ER_Modeling} have concluded that it is necessary to simulate the full magnitude of recombination fluctuations. Finally, the detector response is simulated, and the user can obtain an S1 and S2 signal, as well as auxiliary quantities such as reconstructed position, drift field, and position corrections on the S1 and S2 signals. 

A LUX-specific NEST model, which we will refer to as LUX-NESTv2, has been described in \cite{Akerib2020_LUX_ER_Modeling}. It has had great success in reproducing the median and width of the ER and NR bands in \mbox{WS2014--16} data. The only deficiency has been that it fails to correctly reproduce the skewness of the ER and NR bands. Here, we present a model of skewness that can be inserted into NEST and correctly reproduce the data.

\subsection{ER Skewness}
\label{sec:ER_Skewness}

The skewness of the ER band is critical to discrimination and thus to sensitivity in general, so it is equally critical that LUX-NESTv2 models it correctly. In the present version of LUX-NESTv2, if a user simulates the LUX \mbox{WS2014--16} calibrations of $^3$H and $^{14}$C, they will arrive at an ER band with (small) negative skewness in the WIMP search region. However, the data clearly show that the ER band has positive skewness in this energy range.

In order to rectify this inconsistency, our solution is to add skewness into LUX-NESTv2 at the level of recombination fluctuations. In LUX-NESTv2, after calculating the quanta produced $N_{\text{ions}}$ and $N_{\text{excitons}}$, the code calculates the mean recombination probability $r$ and its variance $\sigma_r^2$; all of these quantities are deterministic and only based on the particle type, energy, and electric field. It then simulates the number of electrons and photons leaving the recoil site using Eq.~\ref{eq:NEST_Default_Ne} and Eq.~\ref{eq:NEST_Default_Nph}, respectively.
\begin{equation}
N_{\text{electrons}} = G\,[\,(1 - r)\,N_{\text{ions}}, \: \sigma_r^2\,] \,,
\label{eq:NEST_Default_Ne}
\end{equation}
where $G[\mu, \sigma^2]$ is a randomly generated number from a Gaussian distribution with mean $\mu$ and variance $\sigma^2$.
\begin{equation}
N_{\text{photons}} = N_{\text{excitons}} + N_{\text{ions}} - N_{\text{electrons}}\,.
\label{eq:NEST_Default_Nph}
\end{equation}
\noindent However, we update this step such that the number of electrons is drawn from a skew-Gaussian distribution, shown in Eq.~\ref{eq:NEST_Skew_Ne}. This scheme preserves the mean and variance of Eq.~\ref{eq:NEST_Default_Ne}. The number of photons leaving the recoil site is still given by Eq.~\ref{eq:NEST_Default_Nph}. For clarity, we emphasize that there are two skewness parameters that will be frequently referenced: $\alpha_R$ is the skewness parameter in the recombination fluctuations model in Eq.~\ref{eq:NEST_Skew_Ne}, while $\alpha_B$ is the skewness parameter of the ER or NR band in log\textsubscript{10}(S2c/S1c) space, as described in Section~\ref{sec:ER_Band}.
\begin{equation}
N_{\text{electrons}} = F\,[\,(1 - r)\,N_{\text{ions}} - \xi_c \, , \: \frac{1}{\omega_c} \sqrt{\sigma_r^2}, \: \alpha_R\,] \,,
\label{eq:NEST_Skew_Ne}
\end{equation}
where \(F[\,\xi, \omega, \alpha]\) is a randomly generated number from a skew-Gaussian distribution given by the PDF in Eq.~\ref{eq:skewgaus},
\begin{equation}
\omega_c = \sqrt{1 - \frac{2}{\pi} \frac{\alpha_R^2}{1 + \alpha_R^2}} \,,
\end{equation}
and
\begin{equation}
\xi_c = \sqrt{\, \sigma_r^2 \: \frac{1 - \omega_c^2}{\omega_c^2}} \,.
\end{equation}

If $\alpha_R$ is sufficiently positive, the results of a LUX-NESTv2 simulation will give $\alpha_B > 0$. However, the skewness of the ER band can only be reproduced if $\alpha_R$ varies with energy and field. The model in Eq.~\ref{eq:Skewness_Model}, where $E$ is the total energy deposited by the electronic recoil and $F$ is the drift field at the recoil site, correctly reproduces data with a certain set of parameter values. This model is empirical. We develop it by determining the $\alpha_R$ that reproduces the correct $\alpha_B$ in bins of drift field and S1c. We observed that the $\alpha_R$ required to match the measured $\alpha_B$ behaves differently in the low-energy and high-energy regimes, i.e.~above and below $E_2$. As a result, we construct a separate model for each energy regime, capturing the energy- and field-dependence of $\alpha_R$ in that regime. The final model is a weighted sum of the two models, in which the weight is an energy-dependent sigmoid function that asymptotically goes to zero and one in the appropriate limits. The transition between the models is field-independent and found to be about 25~keV, which is comparable to the energy at which LUX-NESTv2 transitions from an electronic recoil yields model based on the Doke-Birks model to one based on the Thomas-Imel Box model \cite{Akerib2020_LUX_ER_Modeling}.
\begin{multline}
\alpha_R = \frac{1}{1 + e\,^{(E - E2) / E3}} \left[ \alpha_0 + c_0 \, e^{-F / F_0} \, (1 - e^{-E / E_0}) \right] + \\
\frac{1}{1 + e\,^{-(E - E_2) / E_3}} \left[ c_1 \, e^{-E / E_1} \, e^{-\sqrt{F / F_1}} \right] \,.
\label{eq:Skewness_Model}
\end{multline}

The nine parameters in Eq.~\ref{eq:Skewness_Model} are not obtained by a rigorous optimization, due to the immense computational power that would be required for a nine-dimensional fit. Instead, we proceed as follows. For each parameter $X$, we find a value that approximately matches the data. Using this value, we simulate the $^{14}$C and $^3$H \mbox{WS2014--16} calibrations, and we calculate the ER bands for six field bins equally spaced between 50 and 500~V/cm. In doing so, we neglect the energy weighting and $g_1$ adjustments described in Section~\ref{sec:ER_Band}. Next, we compute the degree to which the simulated ER band skewness is consistent with data by using Eq.~\ref{eq:Skewness_Chi2}, in which $j$ and $k$ iterate over field and S1c bins, respectively, and $\delta$ represents the uncertainty on $\alpha_B$ from the skew-Gaussian fit. By adjusting $X$ slightly and repeating this procedure several times, we obtain a set of points ($X_p$, $\chi^2_p$). Finally, we fit a quadratic function to these points. Defining ($\bar{X}$, $\bar{\chi^2}$) as the vertex of this parabola, we derive our desired quantities: the estimated value of $X$ is $\bar{X}$, and the uncertainty on $X$ is the amount $\delta_X$ such that $X = \bar{X} \pm \delta_X$ implies $\chi^2 = \bar{\chi^2} \pm 1$.
\begin{equation}
\chi^2 = \sum_{i \in \{^{14}\text{C}, ^3\text{H}\}} \sum_{j} \sum_{k} \left[ \frac{(\alpha_{\text{B, Data}} - \alpha_{\text{B, MC}})^2}{\delta_\text{Data}^2 + \delta_\text{MC}^2} \right]_{i,j,k} \,.
\label{eq:Skewness_Chi2}
\end{equation}
The parameter values determined by this procedure are listed in Table~\ref{table:Skewness_Model_Parameters}.

\renewcommand{\arraystretch}{1.2}
\begin{table}[b]
\centering
\caption{The optimal values for the parameters of the electronic recoil skewness model (i.e.~Eq.~\ref{eq:Skewness_Model}), based on LUX \mbox{WS2014--16} $^3$H and $^{14}$C calibration data.\strut}
 \begin{tabular}{|| c || c || c||} 
 \hline
 $\,$ Parameter $\,$ & $\,$ Value $\pm$ Uncertainty $\,$ & $\,$ Units $\,$ \\
 \hline\hline
 $\alpha_0$ & 1.39 $\pm$ 0.03 & ... \\ 
 $c_0$ & 4.0 $\pm$ 0.2 & ... \\ 
 $c_1$ & 22.1 $\pm$ 0.5 & ... \\
 $E_0$ & 7.7 $\pm$ 0.4 & keV \\
 $E_1$ & 54 $\pm$ 2 & keV \\
 $E_2$ & 26.7 $\pm$ 0.5 & keV \\
 $E_3$ & 6.4 $\pm$ 0.9 & keV \\
 $F_0$ & 225 $\pm$ 12 & V/cm \\
 $F_1$ & 71 $\pm$ 4 & V/cm \\
 \hline
\end{tabular}
\label{table:Skewness_Model_Parameters}
\end{table}

\begin{figure}
{\includegraphics[width=3.25in]{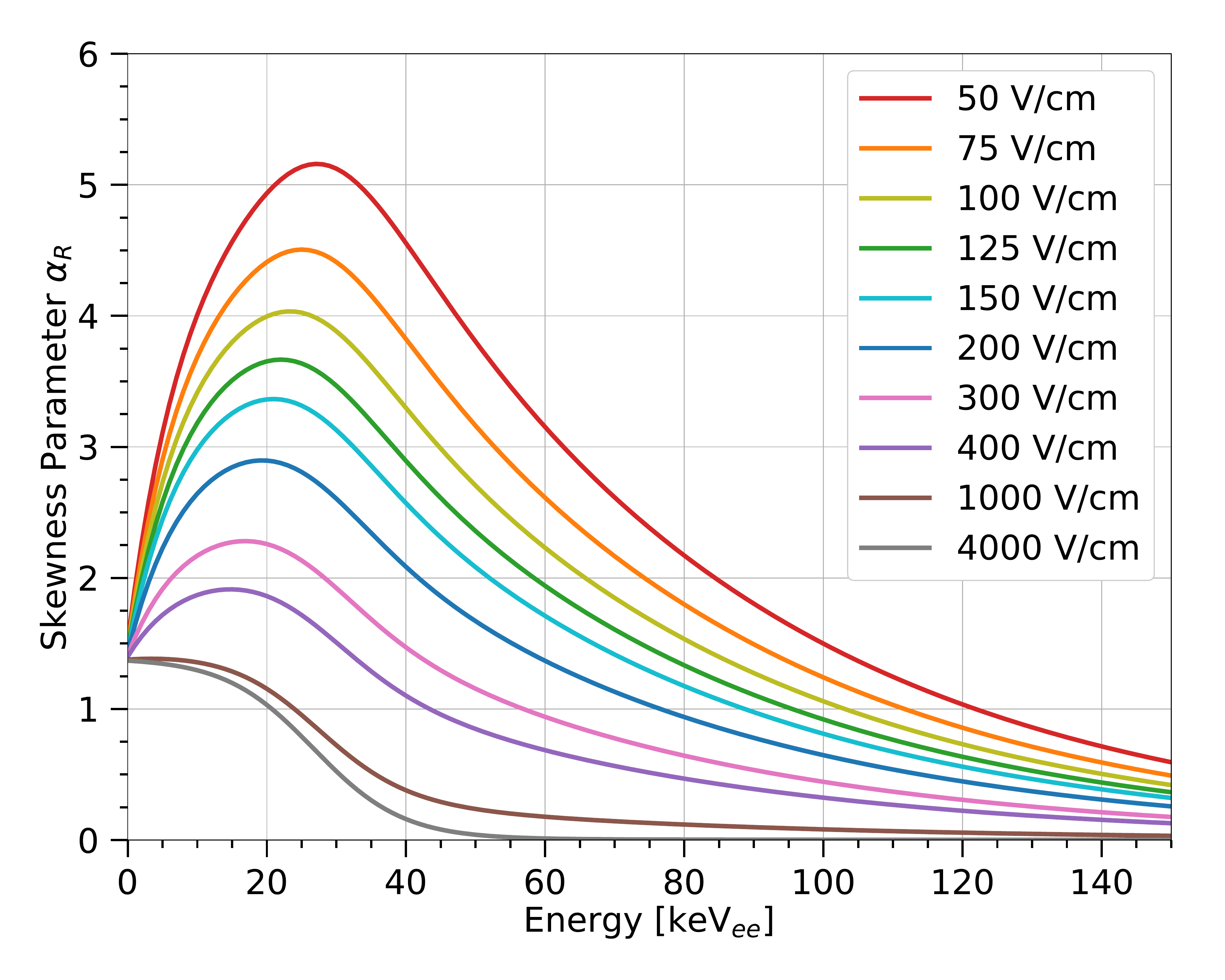}}
\caption{The skewness model for recombination fluctuations in Eq.~\ref{eq:Skewness_Model}.}
\label{fig:Skewness_Model}
\end{figure}

\begin{figure}
{\includegraphics[width=3.25in]{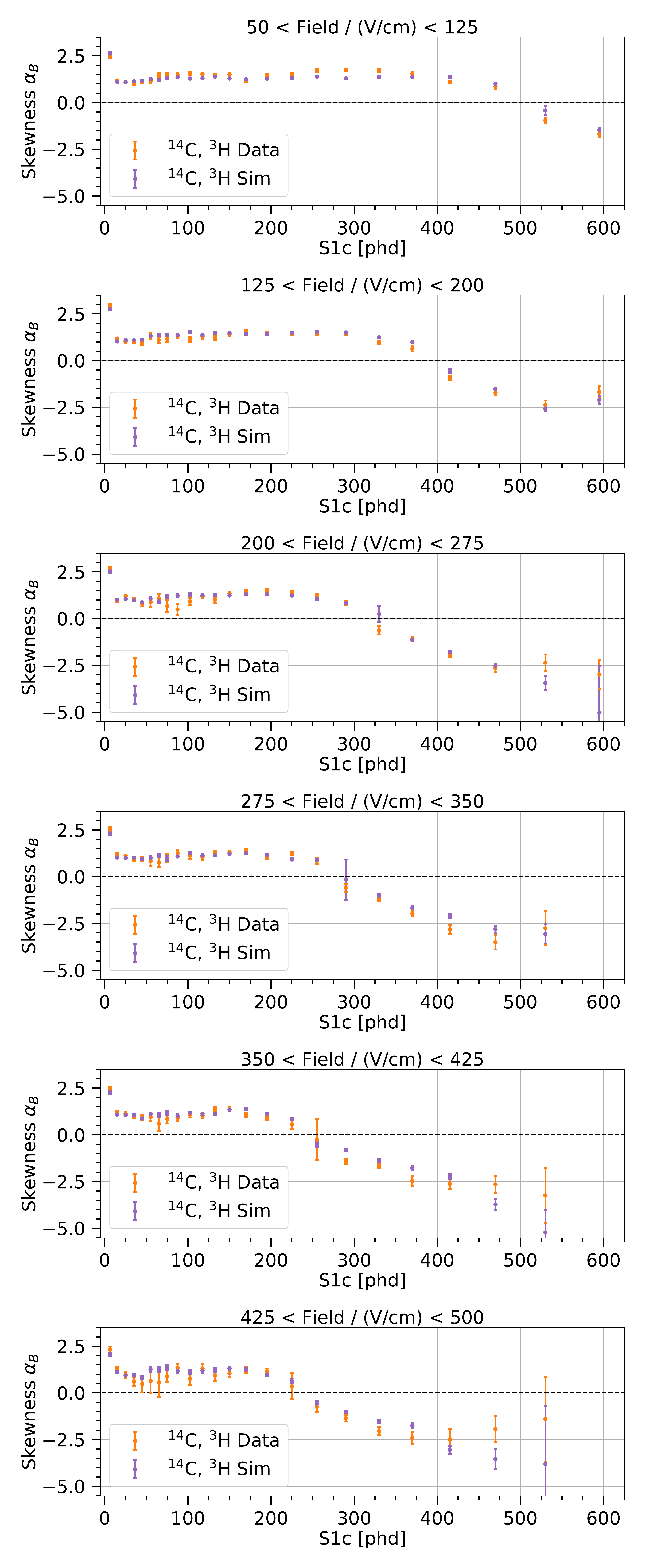}}
\caption{A comparison of the skewness of the ER band in \mbox{WS2014--16} data vs.~simulation from LUX-NESTv2, based on our model in Eq.~\ref{eq:Skewness_Model}. Points below 50~phd are from $^3$H data and simulation, and points above 50~phd are from $^{14}$C data and simulation.}
\label{fig:ER_Skewness_Comparison}
\end{figure}

Figure~\ref{fig:Skewness_Model} shows a plot of Eq.~\ref{eq:Skewness_Model} for a variety of energies and fields, and Fig.~\ref{fig:ER_Skewness_Comparison} shows a comparison of $\alpha_B$ between data and simulation. One observes that the two match well, and that $\alpha_B$ dips below zero at high enough energy. Here, the uncertainty on the skewness is obtained from the fit.

We also observe that our skewness model is successful at matching data from other experiments. See Fig.~\ref{fig:ER_ZEPLIN_III_Comparison} in Appendix~\ref{app:Additional_Figures} for a comparison to \mbox{ZEPLIN-III} data, which reported an average leakage of ${1.3\times10^{-4}}$ at a 3.8~kV/cm drift field \cite{Lebedenko2009_ZEPLIN_III_First_Science_Run, Horn2011_ZEPLIN_III_NR_Yields, Araujo2020_ZEPLIN_III_Revision}. Furthermore, the authors of \cite{Szydagis2020_XENON1T_Rebuttal} used our ER skewness model to accurately simulate $^{37}$Ar calibration data in XENON1T.

\subsection{NR Skewness}
\label{sec:NR_Skewness}

The NR band exhibits skewness, but it is substantially more difficult to model. There are a few reasons for the difficulty: first, skewness is a third-order effect (as mentioned previously, it is associated with the third standardized moment of the distribution), so correctly measuring it requires a substantial amount of data. This is possible for electronic recoils because in \mbox{WS2014--16}, there are over 1.5 million events. On the other hand, there are only about 80,000 nuclear recoils in the data set, so this dataset is prone to large uncertainties and statistical fluctuations. Second, there is a small number of multiple scatters in the nuclear recoil dataset, because occasionally multiple S2 pulses are so close together that they are classified as a single S2 pulse. We cut these out without significantly reducing the single-scatter acceptance, but a small number do persist, and they have a disproportionately high S2 area. This means that although they have a negligible effect on the NR band median and width, they have a considerable effect on the skewness. Including these multiple scatters, which are prevalent at high energy and high electric field, causes the skew-Gaussian fit to be fit at $\alpha_B$ of 3.0 or above.

\begin{figure}
{\includegraphics[width=3.25in]{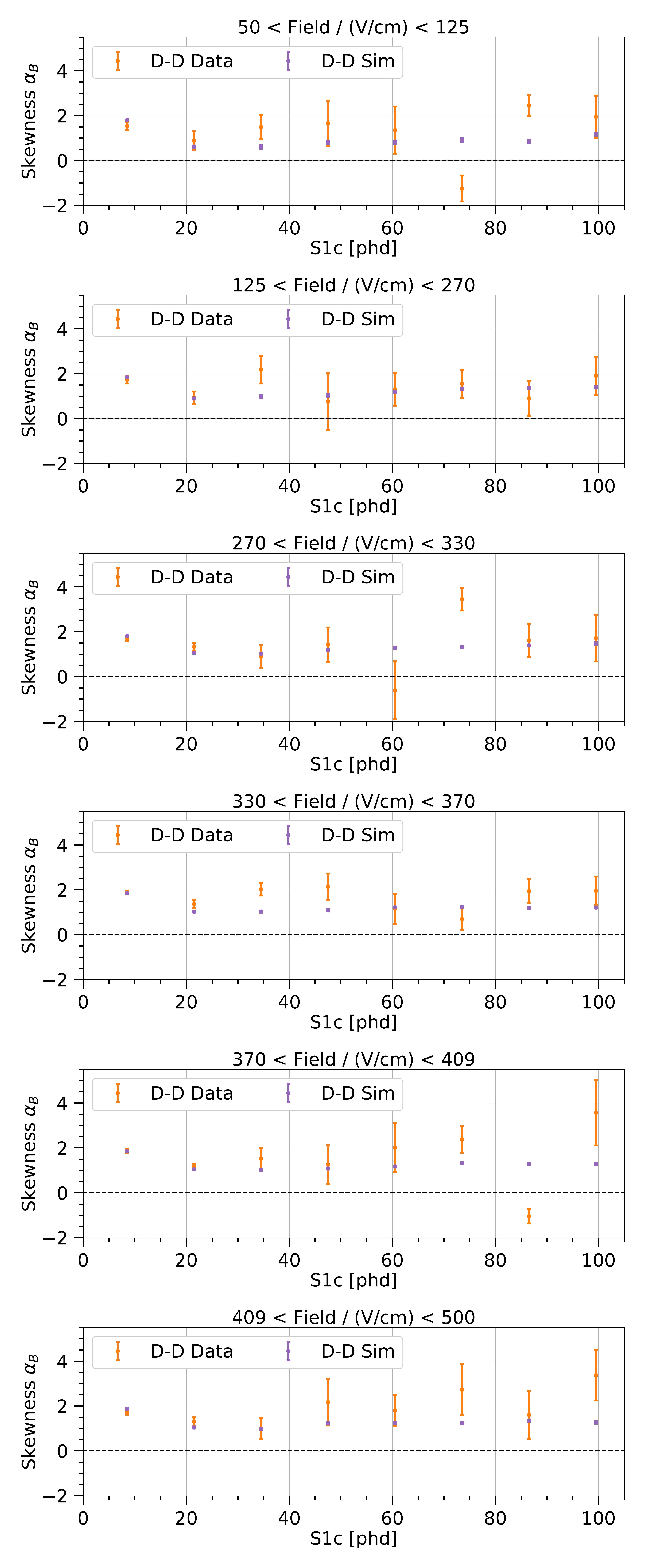}}
\caption{A comparison of the skewness of the NR band in \mbox{WS2014--16} data vs.~simulation from LUX-NESTv2, using $\alpha_R = 20$.}
\label{fig:NR_Skewness_Comparison}
\end{figure}

To account for this, we remove events at high S2 before histogramming log\textsubscript{10}(S2c/S1) and doing the skew-Gaussian fit, resulting in the data points of Fig.~\ref{fig:NR_Skewness_Comparison}. The NR band skewness does not affect leakage if it is defined through a cut-and-count procedure, i.e.~the fraction of electronic recoils falling below the NR band median. However, most experiments use a profile likelihood ratio or a similar hypothesis test, in which case a positive NR skewness would worsen an experiment's sensitivity.

The skewness in NR data is still relatively high, even with this change. We simulate recombination fluctuations with Eq.~\ref{eq:NEST_Skew_Ne}, but we require $\alpha_R \rightarrow \, \infty$. To clarify, the skew-Gaussian PDF (Eq.~\ref{eq:skewgaus}) is such that as $\alpha$ increases, the PDF tends to ``saturate.'' This means that for $\alpha \gtrsim 10$, the PDF does not substantially change; it effectively becomes a unit step function multiplied by a Gaussian. We use $\alpha_R~=~20$ in LUX-NESTv2 to simulate nuclear recoils, and the results are shown in Fig.~\ref{fig:NR_Skewness_Comparison}. The match is moderate; we observe no substantial field or energy dependence.

\section{Fluctuations of the ER Band}
\label{sec:Fluctuations}

The width of the ER band is crucial to understanding particle discrimination; as the width increases, more electronic recoil events leak below the NR band, and detector sensitivity to dark matter deteriorates. It is therefore an integral part of our analysis to examine the effects of different types of fluctuations on the band width, and especially to see their dependence on drift field and energy.

\begin{figure}
{\includegraphics[width=3.25in]{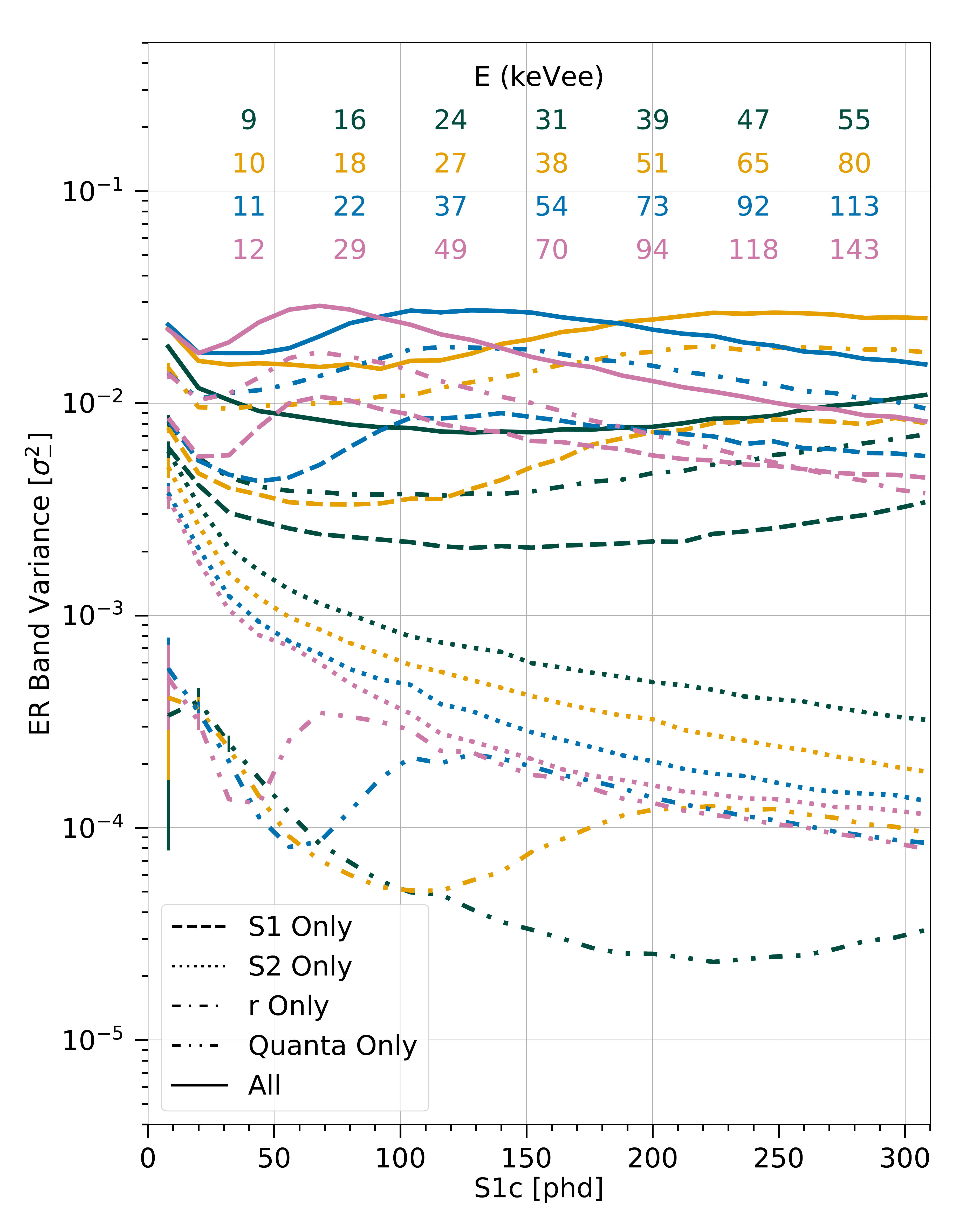}}
\caption{The ER band variance $\sigma_{\text{-}}^2$ for different types of fluctuations and fields, based on simulation from LUX-NESTv2. Color represents electric field: dark green, gold, blue, and magenta represent 180, 500, 1000, and 2000~V/cm, respectively. Line style represents the types of fluctuations that are turned on in the simulation: dot-dot-dashed lines are fluctuations in the number of quanta, dotted lines are S2 fluctuations, dashed lines are S1 fluctuations, dot-dashed lines are recombination fluctuations, and solid lines are all fluctuations turned on simultaneously (i.e.~the default status for LUX-NESTv2). All points have associated error bars, but most are too small to be visible, except for points in the lowest S1c bin. If the fluctuations were uncorrelated, the solid lines would represent the sum of all the other lines for each field, but there are small correlations, so this is not quite true. For a given field, the difference between the solid line and the sum of the other lines is at most 12\% except in the lowest S1c bin, where the total variance can be as much as double the sum of the individual component variances. At the top of the figure, we show the electronic equivalent energy for S1c in multiples of 40~phd, for each electric field.}
\label{fig:Fluctuation_Decomposition}
\end{figure}

LUX-NESTv2 calculates an S1 and S2 signal for each energy deposit, but there are random fluctuations about some mean for these values. We split all these fluctuations into four categories: 1) S1-based fluctuations, including photon detection efficiency, the double-photoelectron effect \cite{Faham2015_Double_PE_Emission, Akerib2020_LUX_Single_Phe}, pulse area smearing, PMT coincidence, and position dependence; 2) S2-based fluctuations, including electron extraction efficiency, photon detection efficiency in gas, the double-photoelectron effect, pulse area smearing, and position dependence; 3) recombination fluctuations; and 4) fluctuations in the number of quanta (i.e.~excitons and ions) produced for a given energy deposit. For each category, we turn off all other fluctuations in LUX-NESTv2, and we simulate 10 million electronic recoils using a flat energy spectrum, LUX detector-specific parameters, a uniform value of ${g_1=0.10}$, and a uniform drift field. We then calculate the ER band as described in Section~\ref{sec:ER_Band}, including the skewness model described in Section~\ref{sec:ER_Skewness}. We repeat this procedure for electric fields of 180, 500, 1000, and 2000~V/cm. Then, we look specifically at $\sigma_{\text{-}}^2$, the band variance due only to the downward fluctuations. The variance is examined rather than the width because if the fluctuations are independent, adding the variances will give the total variance. The results are shown in Fig.~\ref{fig:Fluctuation_Decomposition}.

We observe that the fluctuations in the number of quanta are an insignificant portion of the full ER band variance (a few percent at most), but they do grow with field. The S2-based fluctuations contribute to about \mbox{5--10\%} of the full band variance; they are suppressed by both energy and field. The field-dependent suppression of S2-based fluctuations is explained by the fact that a higher electric field is associated with less recombination, so the S2 signal is larger for a given S1 signal. Similarly, an increased energy leads an increased charge yield and a suppression of S2-based fluctuations. The S1-based fluctuations are significant at all energies and fields, accounting for \mbox{20--30\%} of the total variance. Their field dependence is weak, but they do get stronger with field, for the same reason that S2-based fluctuations are suppressed by an increased field. Finally, the recombination fluctuations are clearly the strongest contributor to band width, consistent with the findings of \cite{Dahl2009_Thesis}. Their field and energy dependence is not easy to summarize quickly, though. At low energies, the recombination fluctuations unambiguously grow with field in this field range. At higher energies, recombination fluctuations begin to shrink with energy in a way that is field-dependent; as a result, the ordering of the fields is not monotonic. For example, looking at just the 2000~V/cm points, recombination fluctuations begin to decrease above $\sim$70~phd and continue their downward trend at higher energies. The 2000~V/cm recombination fluctuations are larger than the recombination fluctuations for any other drift field below 70~phd, but they become the smallest at the highest values of S1. One particularly interesting feature is that at very high energies and fields---specifically, the 2000~V/cm simulation above 250~phd, or 110~keVee---the recombination fluctuations become smaller than the S1 fluctuations, which are dominantly from $g_1$ binomial statistics.

\section*{Conclusion}
\label{sec:Conclusion}

We have explored electronic vs.~nuclear recoil discrimination and shown convincing evidence of improvement at high energies. This means that detectors can enhance their sensitivity to dark matter interactions by increasing their $g_1$ or examining high-energy signals, such as heavier WIMPs or effective field theory interactions. Furthermore, we find that pulse shape discrimination enhances charge-to-light discrimination, but interestingly only for lower fields (below 200~V/cm or so). Combining both types of discrimination, we find that our optimal field range is 240--290~V/cm, which is consistent with the projected capabilities of the upcoming LZ experiment. We also emphasize the importance of understanding recombination fluctuations, both for their effect on the ER band skewness and their importance in the size of the ER band width. Future work will include an understanding of how these detector parameters affect sensitivity to various dark matter models.

\section*{Acknowledgement}

This work was partially supported by the U.S. Department of Energy (DOE) under Award No. DE-AC02-05CH11231, DE-AC05-06OR23100, DE-AC52-07NA27344, DE-FG01-91ER40618, DE-FG02-08ER41549, DE-FG02-11ER41738, DE-FG02-91ER40674, DE-FG02-91ER40688, DE-FG02-95ER40917, DE-NA0000979, DE-SC0006605, DE-SC0010010, DE-SC0015535, and DE-SC0019066; the U.S. National Science Foundation under Grants No. PHY-0750671, PHY-0801536, PHY-1003660, PHY-1004661, PHY-1102470, PHY-1312561, PHY-1347449, PHY-1505868, and PHY-1636738; the Research Corporation Grant No. RA0350; the Center for Ultra-low Background Experiments in the Dakotas (CUBED); and the South Dakota School of Mines and Technology (SDSMT).

Laborat\'{o}rio de Instrumenta\c{c}\~{a}o e F\'{i}sica Experimental de Part\'{i}culas (LIP)-Coimbra acknowledges funding from Funda\c{c}\~{a}o para a Ci\^{e}ncia e a Tecnologia (FCT) through the Project-Grant PTDC/FIS-NUC/1525/2014. Imperial College and Brown University thank the UK Royal Society for travel funds under the International Exchange Scheme (IE120804). The UK groups acknowledge institutional support from Imperial College London, University College London and Edinburgh University, and from the Science \& Technology Facilities Council for PhD studentships R504737 (EL), M126369B (NM), P006795 (AN), T93036D (RT) and N50449X (UU). This work was partially enabled by the University College London (UCL) Cosmoparticle Initiative. The University of Edinburgh is a charitable body, registered in Scotland, with Registration No. SC005336.

This research was conducted using computational resources and services at the Center for Computation and Visualization, Brown University, and also the Yale Science Research Software Core.

We gratefully acknowledge the logistical and technical support and the access to laboratory infrastructure provided to us by SURF and its personnel at Lead, South Dakota. SURF was developed by the South Dakota Science and Technology Authority, with an important philanthropic donation from T. Denny Sanford. SURF is a federally sponsored research facility under Award Number DE-SC0020216.

\bibliography{Bibliography} 

\appendix

\section{Additional Figures}
\label{app:Additional_Figures}

\begin{figure*}
{\includegraphics[width=6.5in]{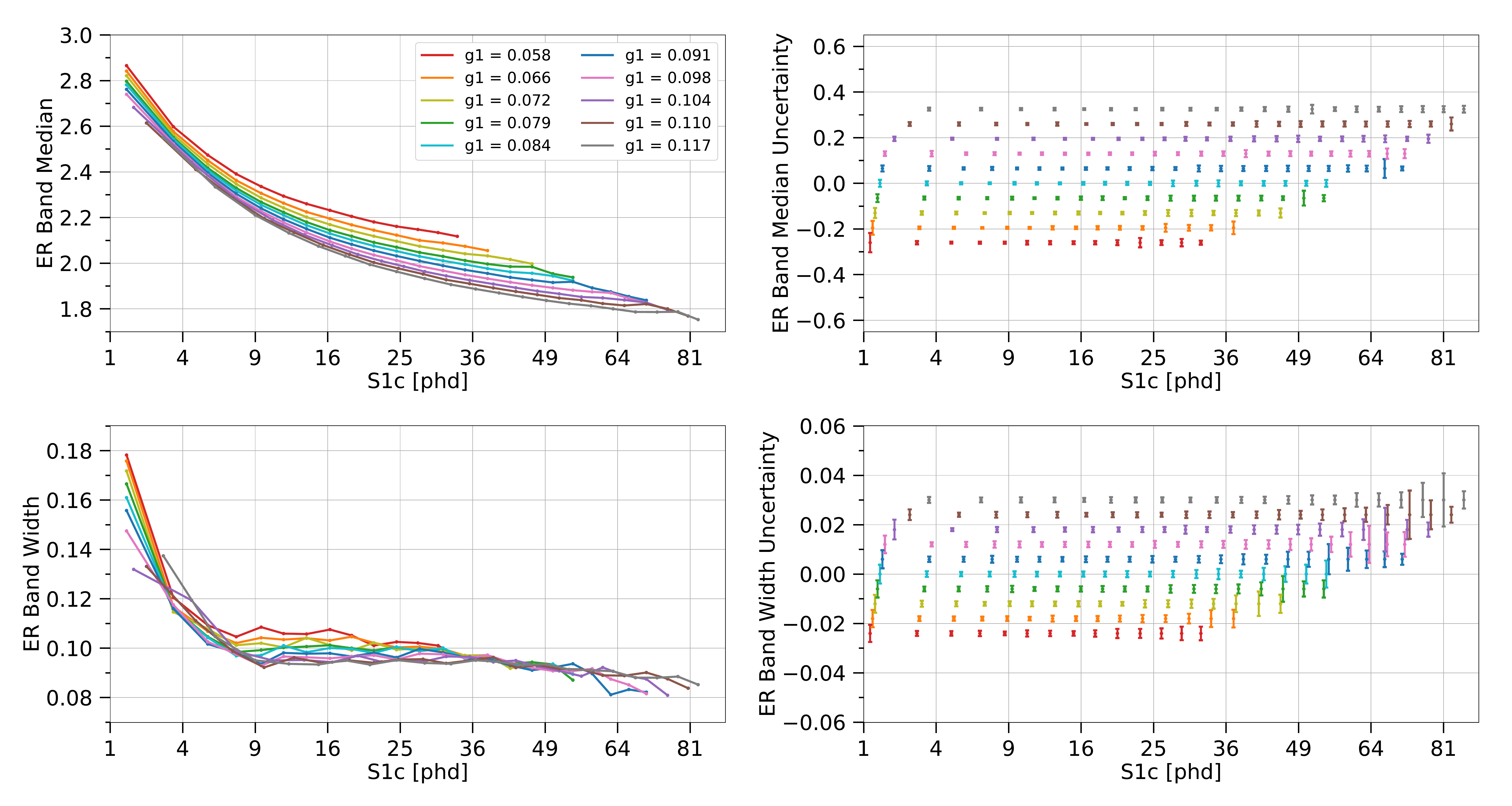}}
\caption{The median and width $\sigma_{\text{-}}$ of the ER band for several values of $g_1$, using WS2013 data. The left plots show measurements, and the right plots display error bars corresponding to these measurements. The S1c axis is proportional to $\sqrt{\text{S1c}}$. In each row, the $y$-axes have the same range; the size of the error bars on the right plot can be directly translated to the points on the left plot. For ease of visualization on the right plots, the S1c values are slightly shifted relative to their true value, and the error bars are centered at a different $y$-value for each $g_1$. Note the S1c range varies for each $g_1$ because as $g_1$ decreases, the $^3$H end point in S1c space decreases.}
\label{fig:ER_Medians_Widths_g1_Full}
\end{figure*}

\begin{figure*}
{\includegraphics[width=6.5in]{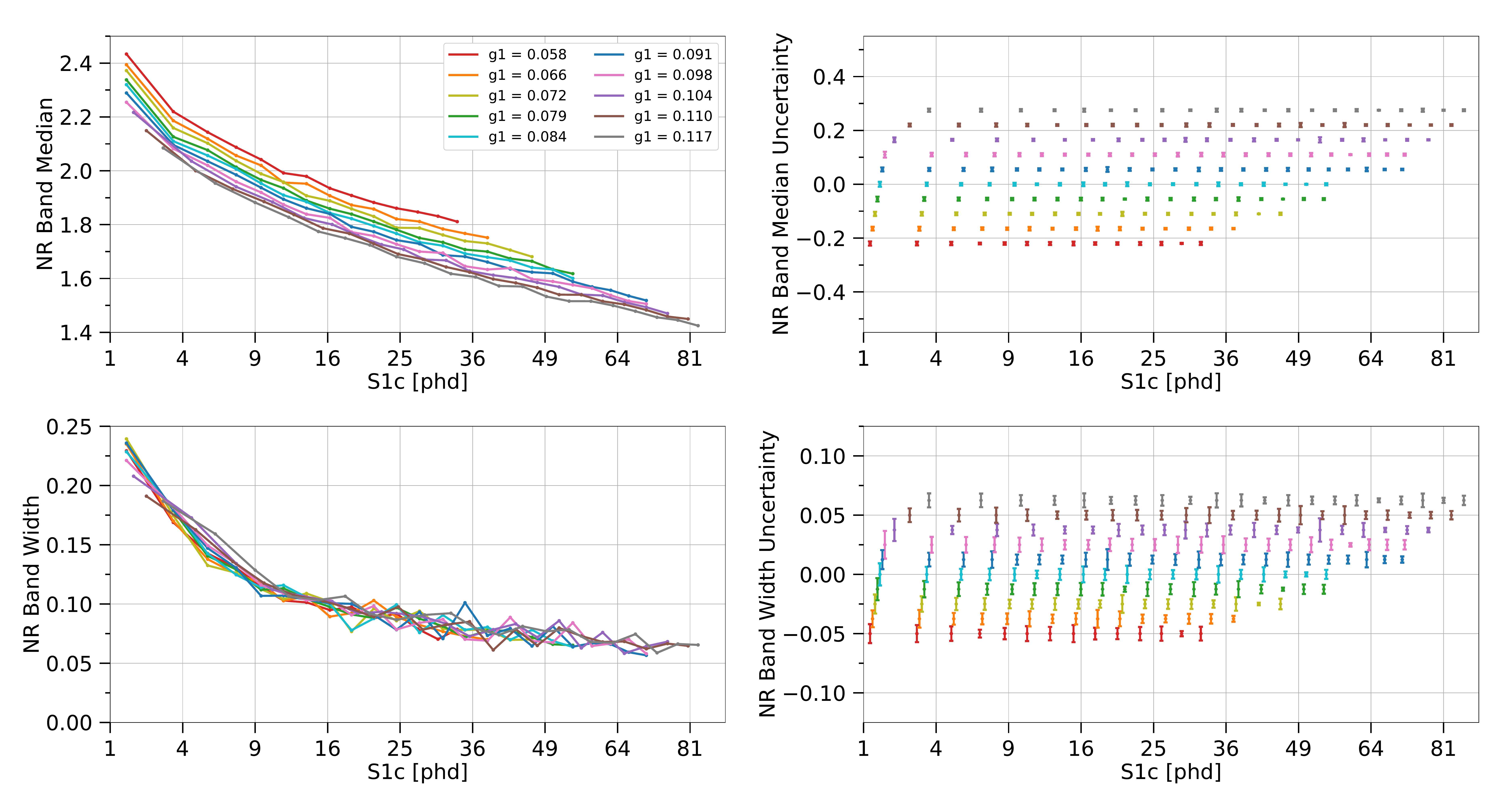}}
\caption{The median and width of the NR band for several values of $g_1$, using WS2013 data. The left plots show measurements, and the right plots display error bars corresponding to these measurements. The S1c axis is proportional to $\sqrt{\text{S1c}}$. In each row, the $y$-axes have the same range; the size of the error bars on the right plot can be directly translated to the points on the left plot. For ease of visualization on the right plots, the S1c values are slightly shifted relative to their true value, and the error bars are centered at a different $y$-value for each $g_1$. Note the S1c range varies for each $g_1$ because as $g_1$ decreases, the D-D end point in S1c space decreases.}
\label{fig:NR_Medians_Widths_g1_Full}
\end{figure*}

\begin{figure*}
{\includegraphics[width=6.5in]{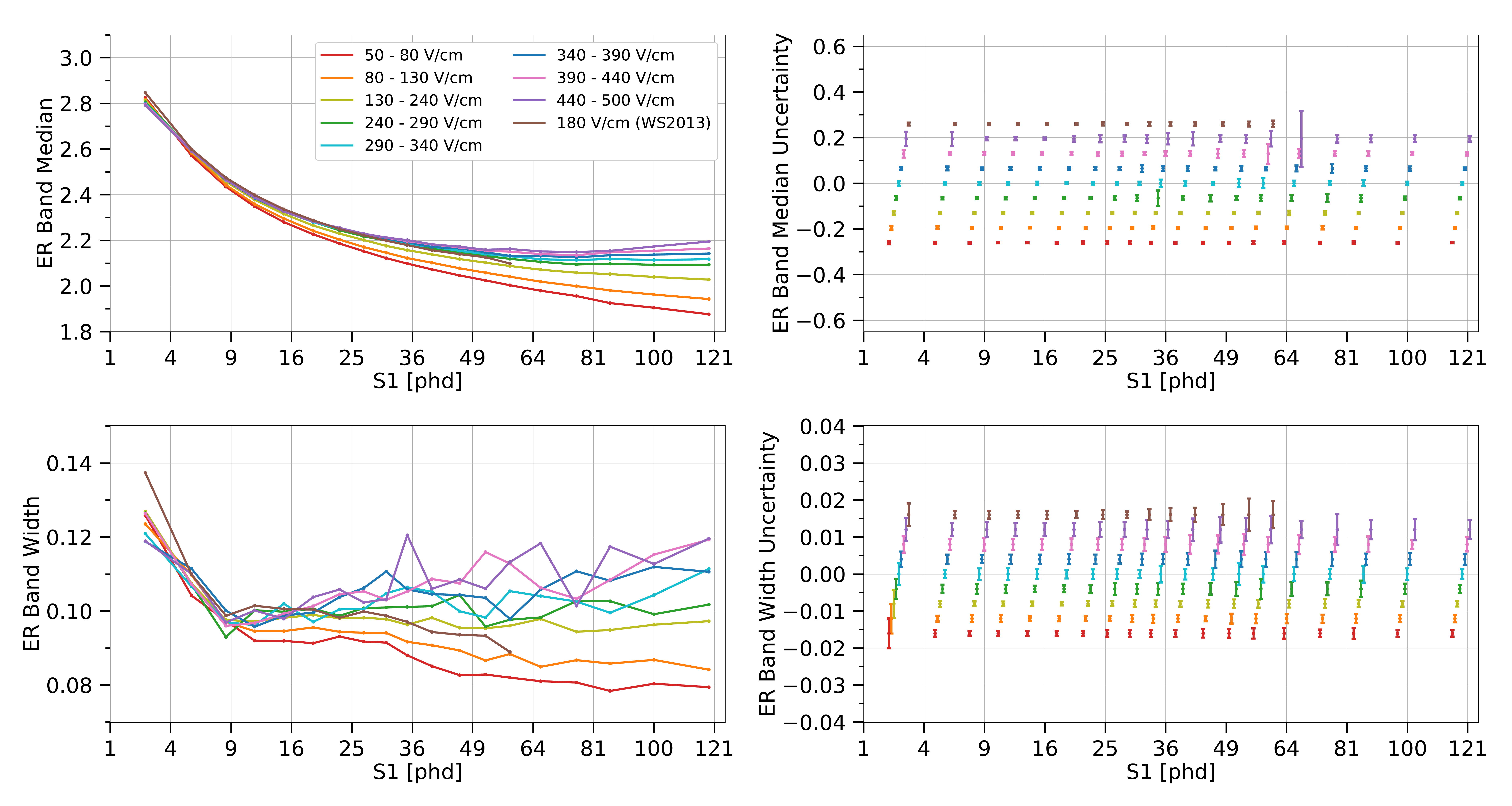}}
\caption{The median and width $\sigma_{\text{-}}$ of the ER band for several drift fields. The left plots show measurements, and the right plots display error bars corresponding to these measurements. The S1 axis is proportional to $\sqrt{\text{S1}}$. The ER band for WS2013 is adjusted so $g_2$ is consistent for the WS2013 and \mbox{WS2014--16} results. In each row, the $y$-axes have the same range; the size of the error bars on the right plot can be directly translated to the points on the left plot. For ease of visualization on the right plots, the S1 values are slightly shifted relative to their true value, and the error bars are centered at a different $y$-value for each field bin.}
\label{fig:ER_Medians_Widths_Field_Full}
\end{figure*}

\begin{figure*}
{\includegraphics[width=6.5in]{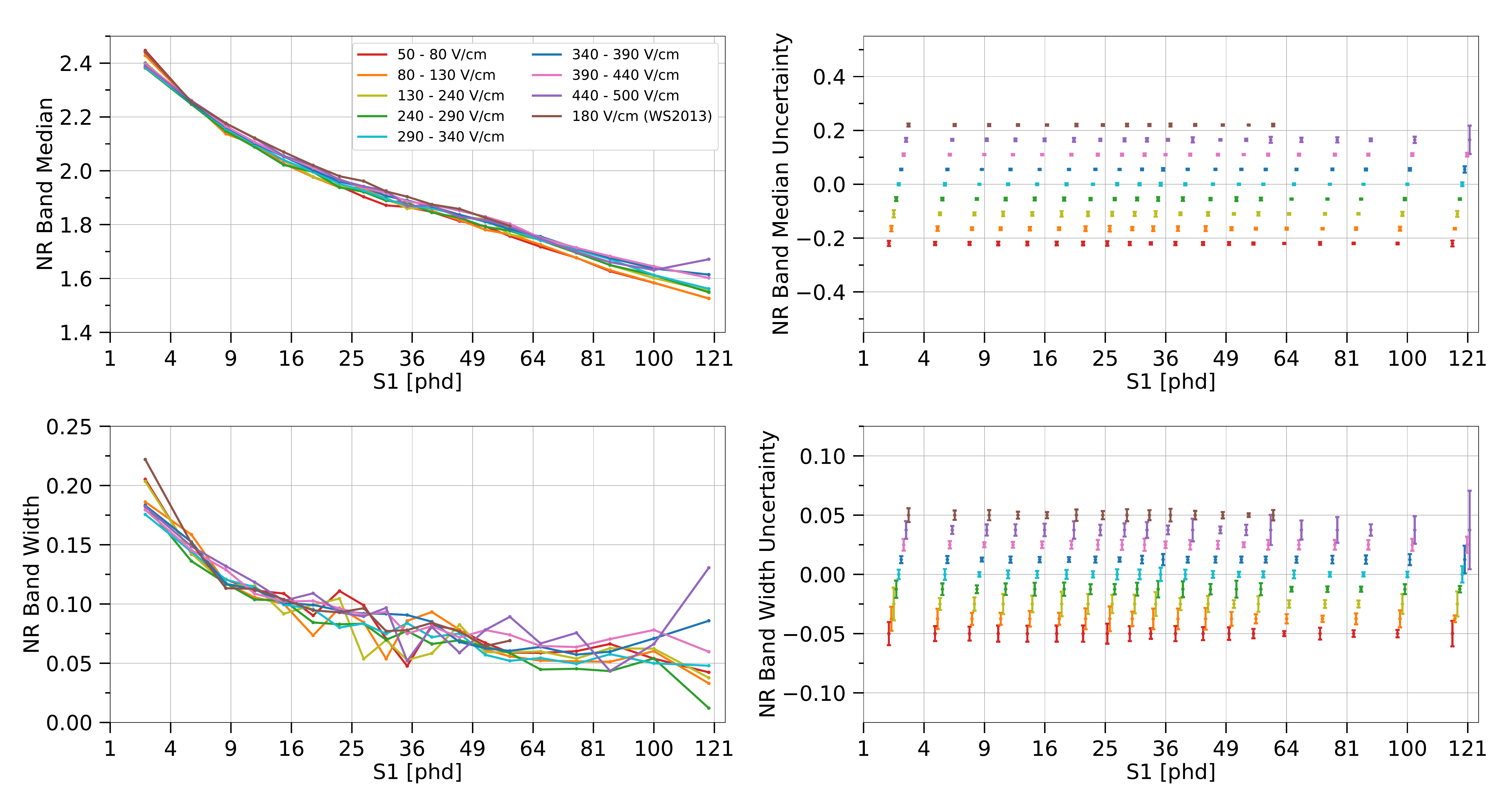}}
\caption{The median and width $\sigma_{\text{-}}$ of the NR band for several drift fields. The left plots show measurements, and the right plots display error bars corresponding to these measurements. The S1 axis is proportional to $\sqrt{\text{S1}}$. The NR band for WS2013 is adjusted so $g_2$ is consistent for the WS2013 and \mbox{WS2014--16} results. In each row, the $y$-axes have the same range; the size of the error bars on the right plot can be directly translated to the points on the left plot. For ease of visualization on the right plots, the S1 values are slightly shifted relative to their true value, and the error bars are centered at a different $y$-value for each field bin.}
\label{fig:NR_Medians_Widths_Field_Full}
\end{figure*}

\begin{figure}
{\includegraphics[width=3.25in]{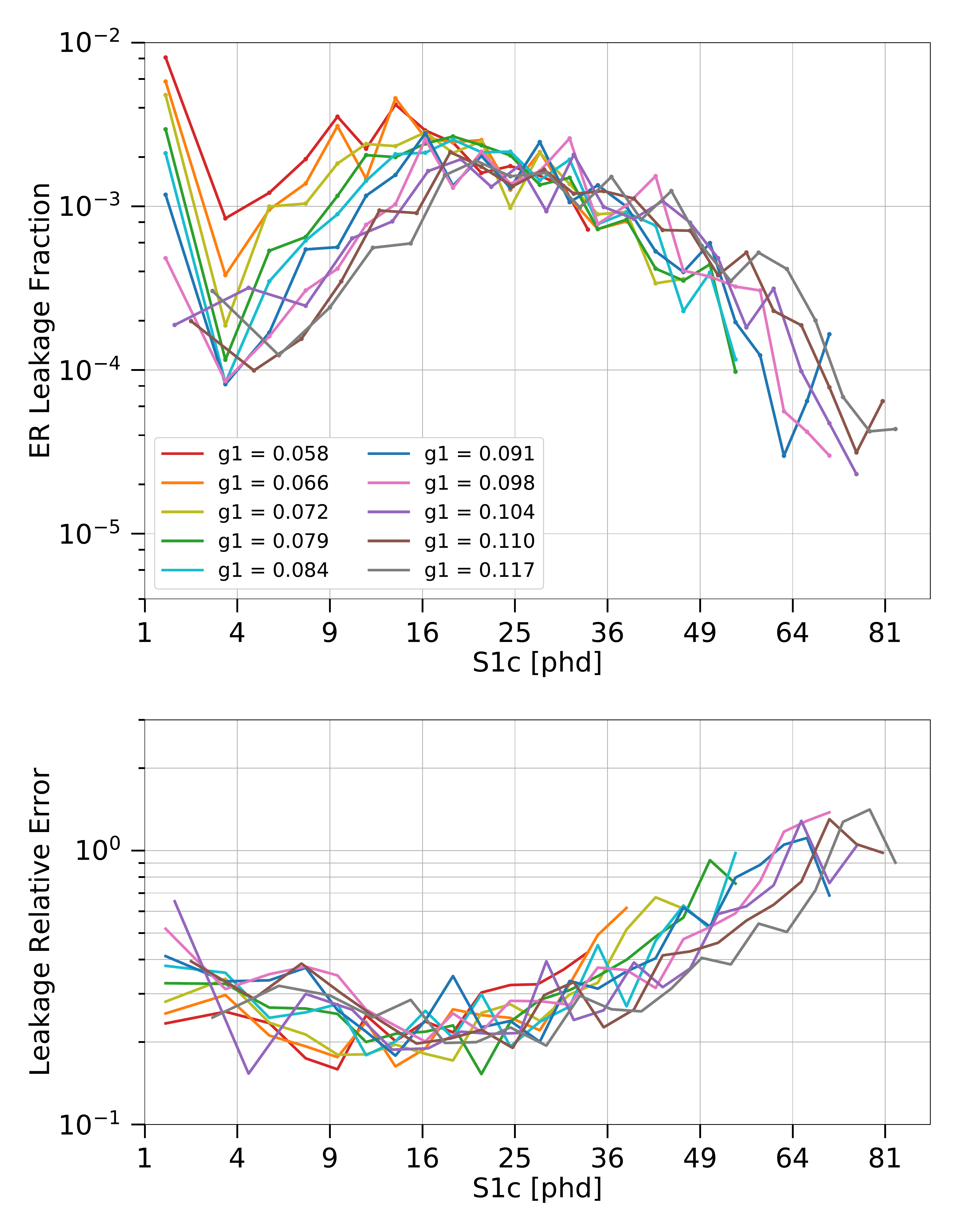}}
\caption{\textit{(Top)} The electronic recoil leakage fraction for a flat energy spectrum in S1c bins, for various values of $g_1$, calculated from a skew-Gaussian extrapolation of the ER band below the NR band median. The S1c axis is proportional to $\sqrt{\text{S1c}}$. The leakage fraction calculated in this way is consistent with the real counted leakage, except in the lowest S1c bin; see Fig.~\ref{fig:Leakage_Lowest_g1} for a comparison between the two leakage calculations in this S1c bin. \textit{(Bottom)} The relative error on these leakage fraction values, defined as: $\text{leakage\_fraction\_error}~/~\text{leakage\_fraction}$. Note that the leakage relative error can be greater than 1, indicating that the leakage fraction is consistent with 0.}
\label{fig:Leakage_g1_Full}
\end{figure}

\begin{figure}
{\includegraphics[width=3.25in]{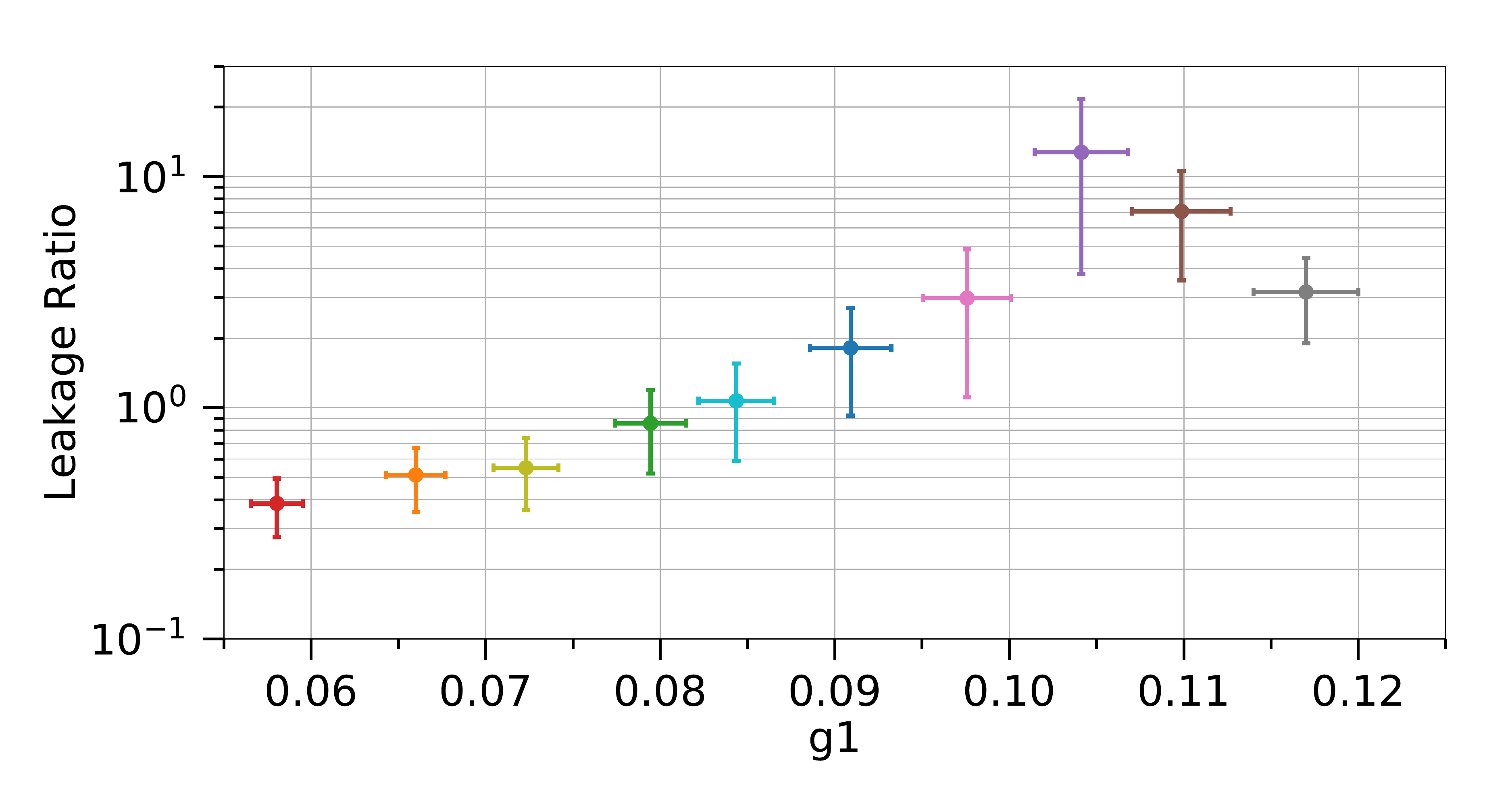}}
\caption{The ratio of the real (counted) electronic recoil leakage fraction to the estimated leakage fraction from a skew-Gaussian extrapolation of log\textsubscript{10}(S2c/S1c) in the lowest S1c bin. This ratio is shown vs.~$g_1$, using WS2013 data. The color of each data point is degenerate with the $g_1$ value; it is included for consistency with other figures.}
\label{fig:Leakage_Lowest_g1}
\end{figure}

\begin{figure}
{\includegraphics[width=3.25in]{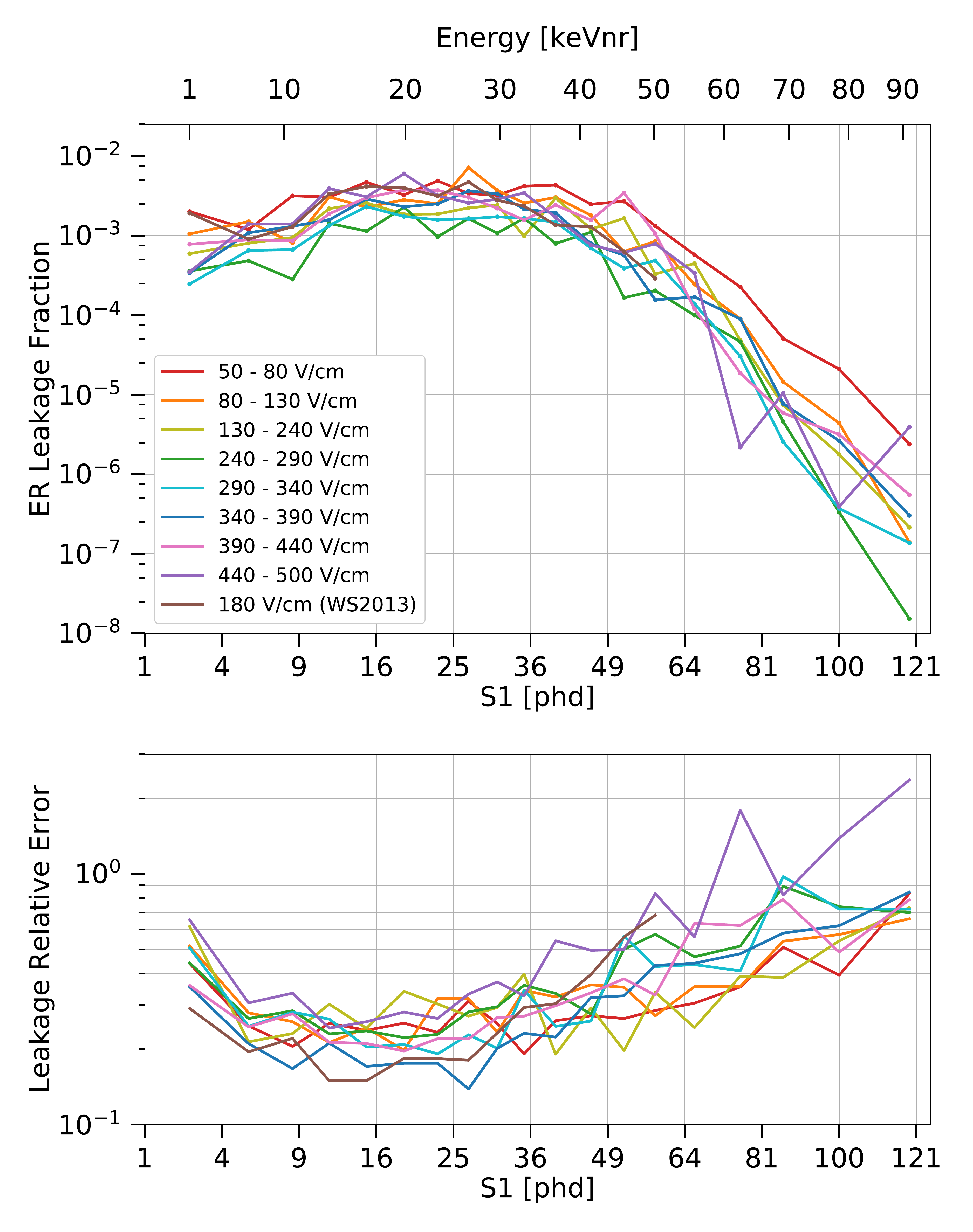}}
\caption{\textit{(Top)} The electronic recoil leakage fraction for a flat energy spectrum in S1 bins, for various values of drift field, calculated from a skew-Gaussian extrapolation of the ER band below the NR band median. The S1 axis is proportional to $\sqrt{\text{S1}}$. The equivalent nuclear recoil energy for an S1 is calculated by using the S1 and S2c at the median of the NR band; this varies by field, but not significantly, so we report the energy averaged over the eight field bins. The leakage fraction calculated in this way is consistent with the real counted leakage, except in the lowest S1 bin; see Fig.~\ref{fig:Leakage_Lowest_Field} for a comparison between the two leakage calculations in this S1 bin. \textit{(Bottom)} The relative error on these leakage fraction values, defined as: $\text{leakage\_fraction\_error}~/~\text{leakage\_fraction}$. Note that the leakage relative error can be greater than 1, indicating that the leakage fraction is consistent with 0.}
\label{fig:Leakage_Field_Full}
\end{figure}

\begin{figure}
{\includegraphics[width=3.25in]{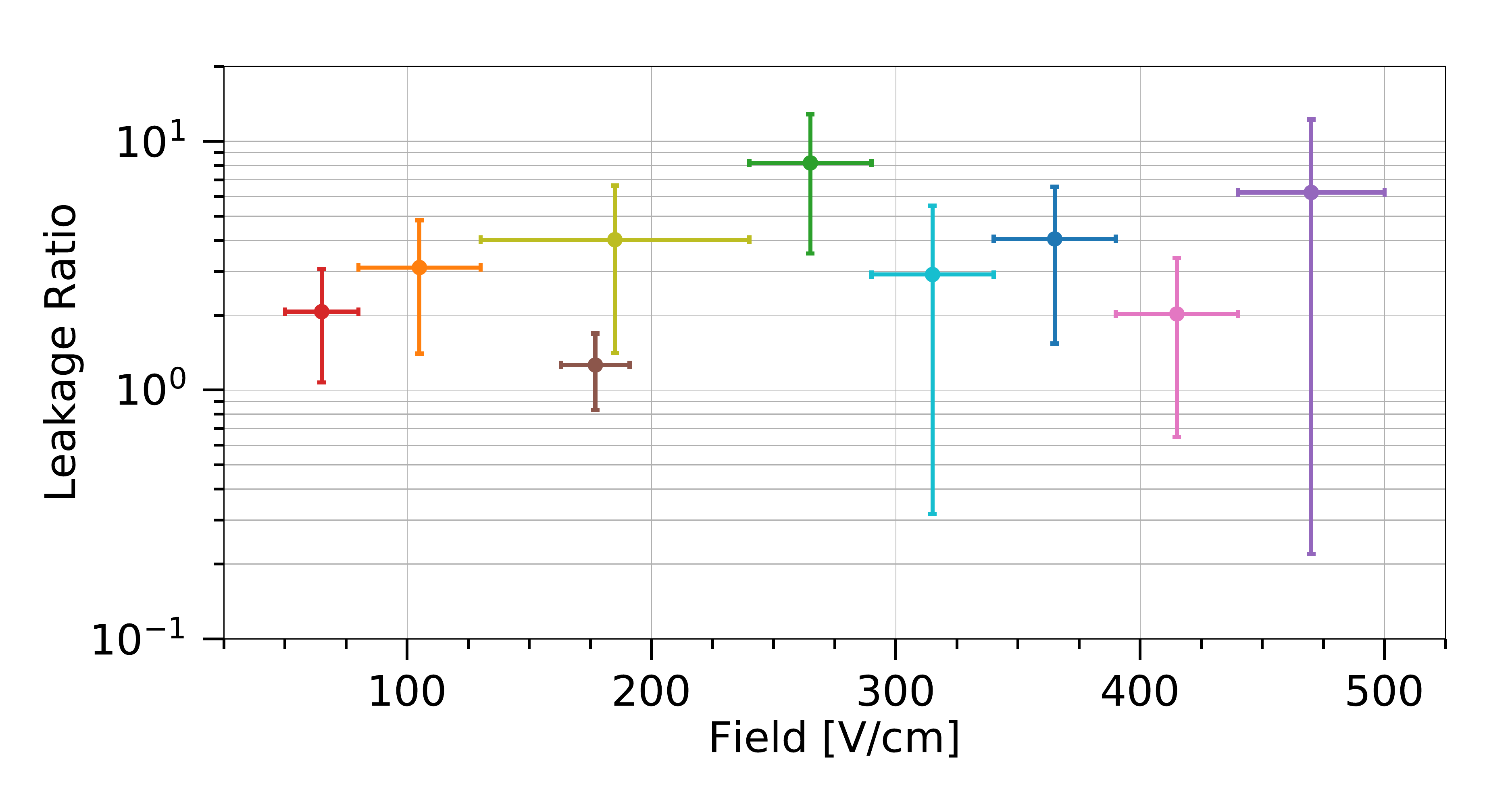}}
\caption{The ratio of the real (counted) electronic recoil leakage fraction to the leakage fraction calculated from a skew-Gaussian extrapolation of log\textsubscript{10}(S2c/S1) in the lowest S1 bin. This ratio is shown vs.~drift field. The color of each data point is degenerate with the field value; it is included for consistency with other figures.}
\label{fig:Leakage_Lowest_Field}
\end{figure}

\begin{figure}
{\includegraphics[width=3.25in]{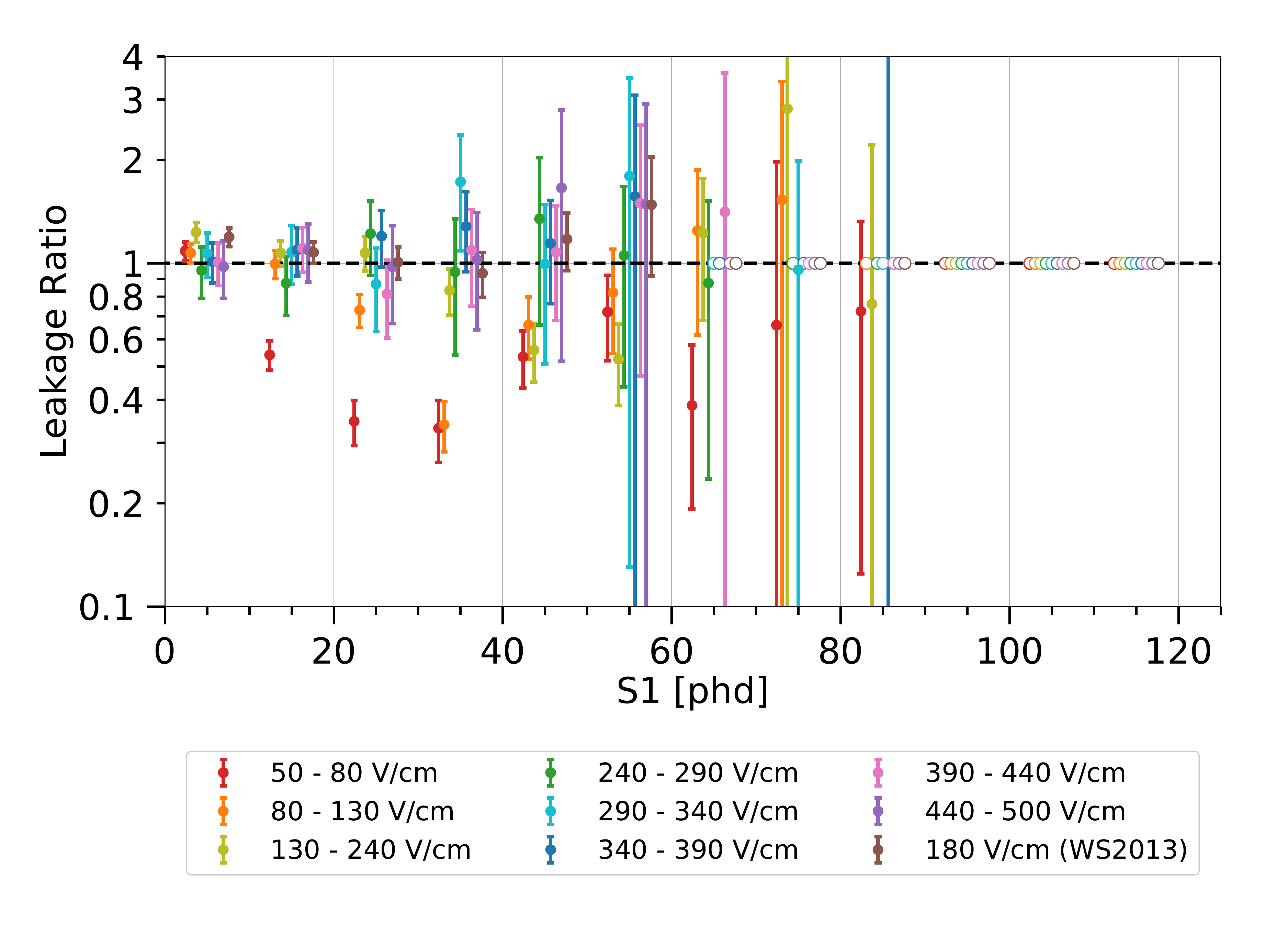}}
\caption{The ratio of two-factor leakage to charge-to-light leakage, for various S1 and drift field bins. Error bars are statistical; see text for details. Open circles represent bins for which charge-to-light discrimination alone gives zero electronic recoils falling below the NR band; as a result, it is impossible to calculate the improvement from two-factor discrimination. The plotted S1 values are slightly shifted relative to their true value (by up to 2.6~phd) for ease of visualization. The true S1 coordinates are 5~phd, 15~phd, 25~phd, etc.}
\label{fig:PSD_Leakage_Ratios_Full}
\end{figure}

\begin{figure}
{\includegraphics[width=3.25in]{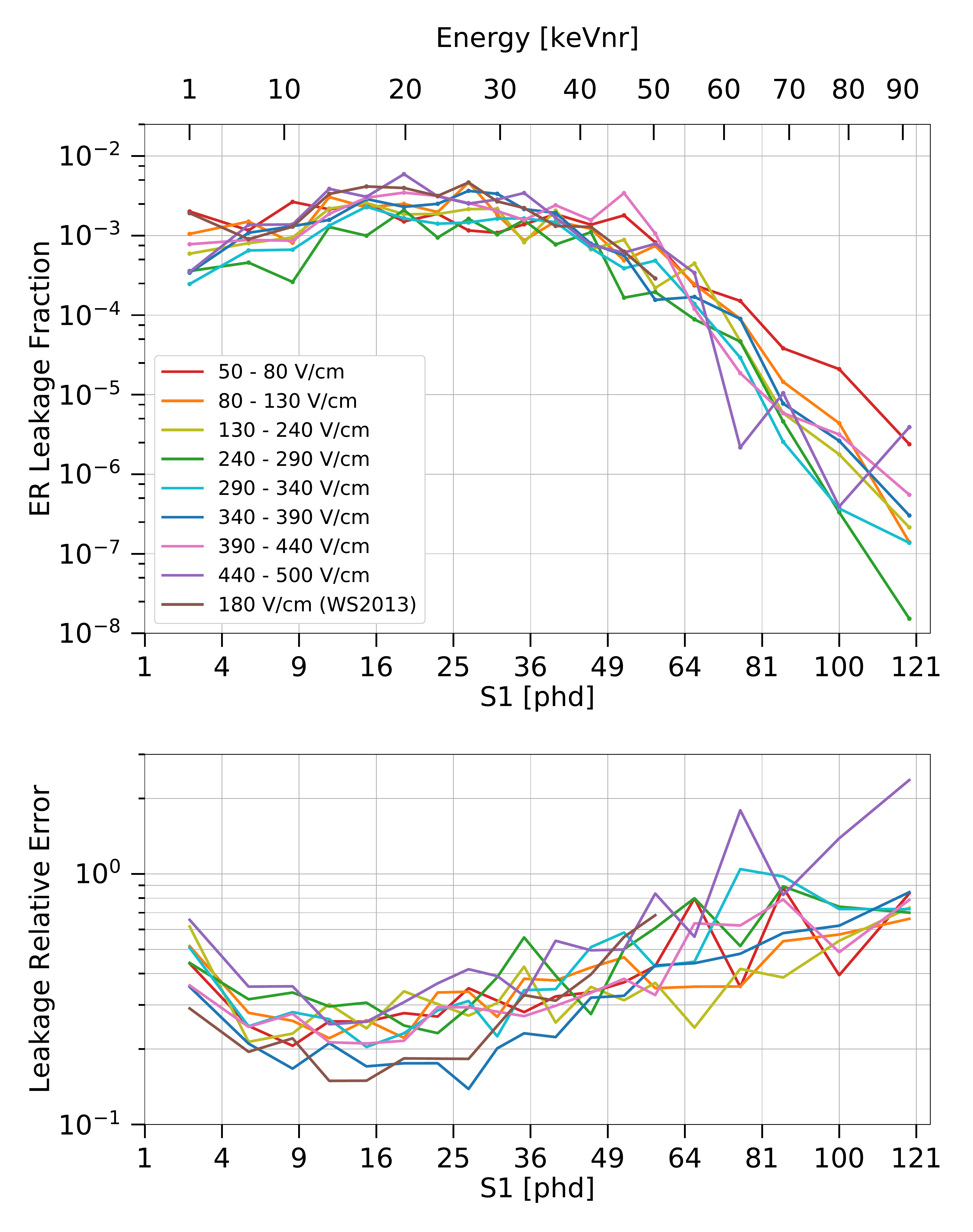}}
\caption{\textit{(Top)} An estimate of the electronic recoil two-factor leakage fraction for a flat energy spectrum in S1 bins, for various values of drift field. The leakage fraction is estimated by multiplying the charge-to-light leakage by the ratio of two-factor to charge-to-light leakage (i.e.~the results in Figs.~\ref{fig:Leakage_Field_Full} and \ref{fig:PSD_Leakage_Ratios_Full}, respectively). This is an estimate for two reasons: first, the two calculations use different S1 bins; second, the charge-to-light leakage is a skew-Gaussian extrapolation, while the ratio is based on counting individual events, so they are not perfectly consistent. The S1 axis is proportional to $\sqrt{\text{S1}}$. The equivalent nuclear recoil energy for an S1 is calculated by using the S1 and S2c at the median of the NR band; this varies by field, but not significantly, so we report the energy averaged over the eight field bins. \textit{(Bottom)} The relative error on these leakage fraction values, defined as: $\text{leakage\_fraction\_error}~/~\text{leakage\_fraction}$. Note that the leakage relative error can be greater than 1, indicating that the leakage fraction is consistent with 0.}
\label{fig:Leakage_Field_Full_PSD}
\end{figure}

\begin{figure}
{\includegraphics[width=3.25in]{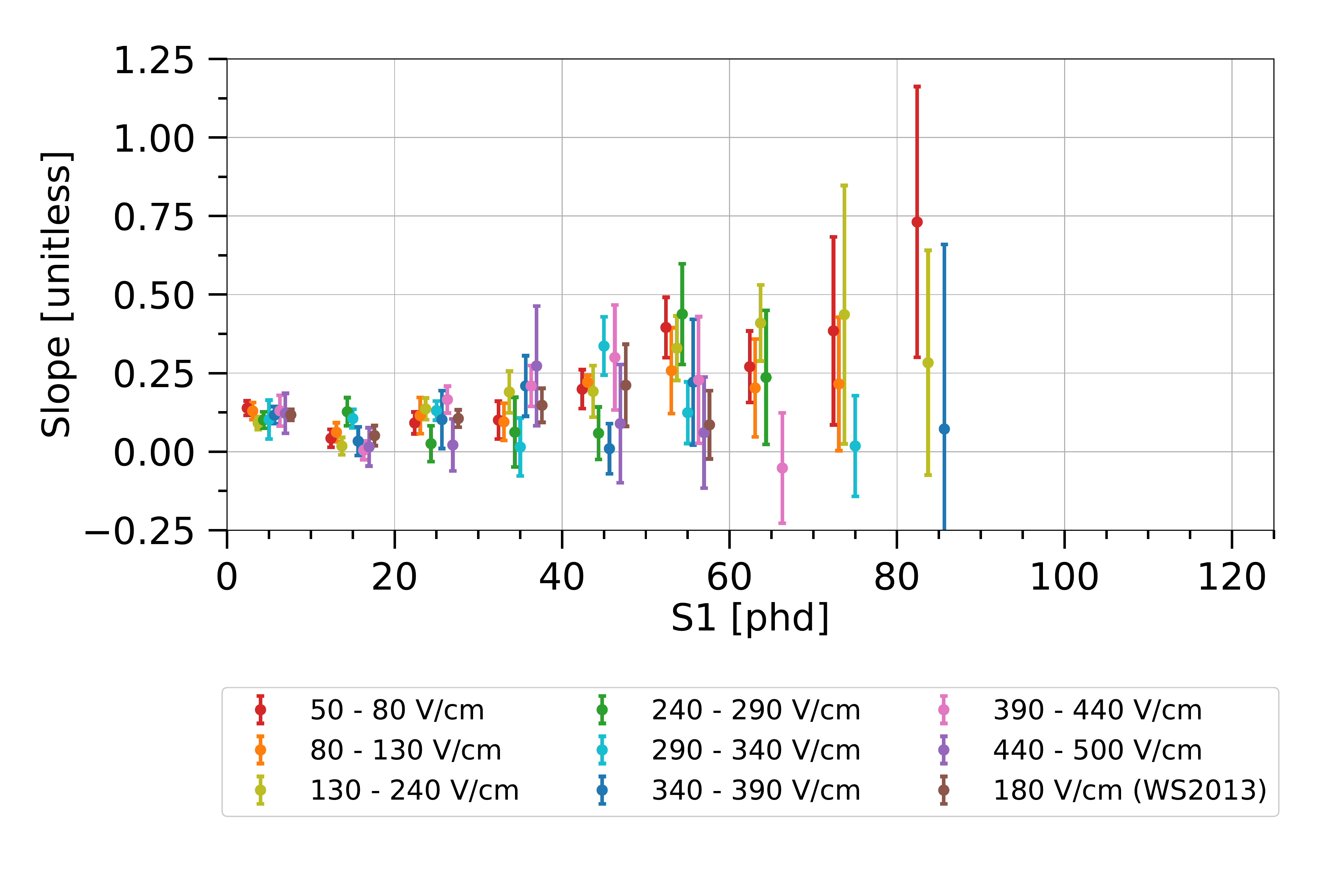}}
\caption{The slope of the two-factor discrimination line in log\textsubscript{10}(S2c/S1) vs.~prompt fraction space, for each S1 and field bin. Missing points represent bins for which charge-to-light discrimination alone gives zero electronic recoils falling below the NR band. The plotted S1 values are slightly shifted relative to their true value (by up to 2.6~phd) for ease of visualization. The true S1 coordinates are 5~phd, 15~phd, 25~phd, etc.}
\label{fig:PSD_Slopes_vs_Field_Full}
\end{figure}

\begin{figure}
{\includegraphics[width=3.25in]{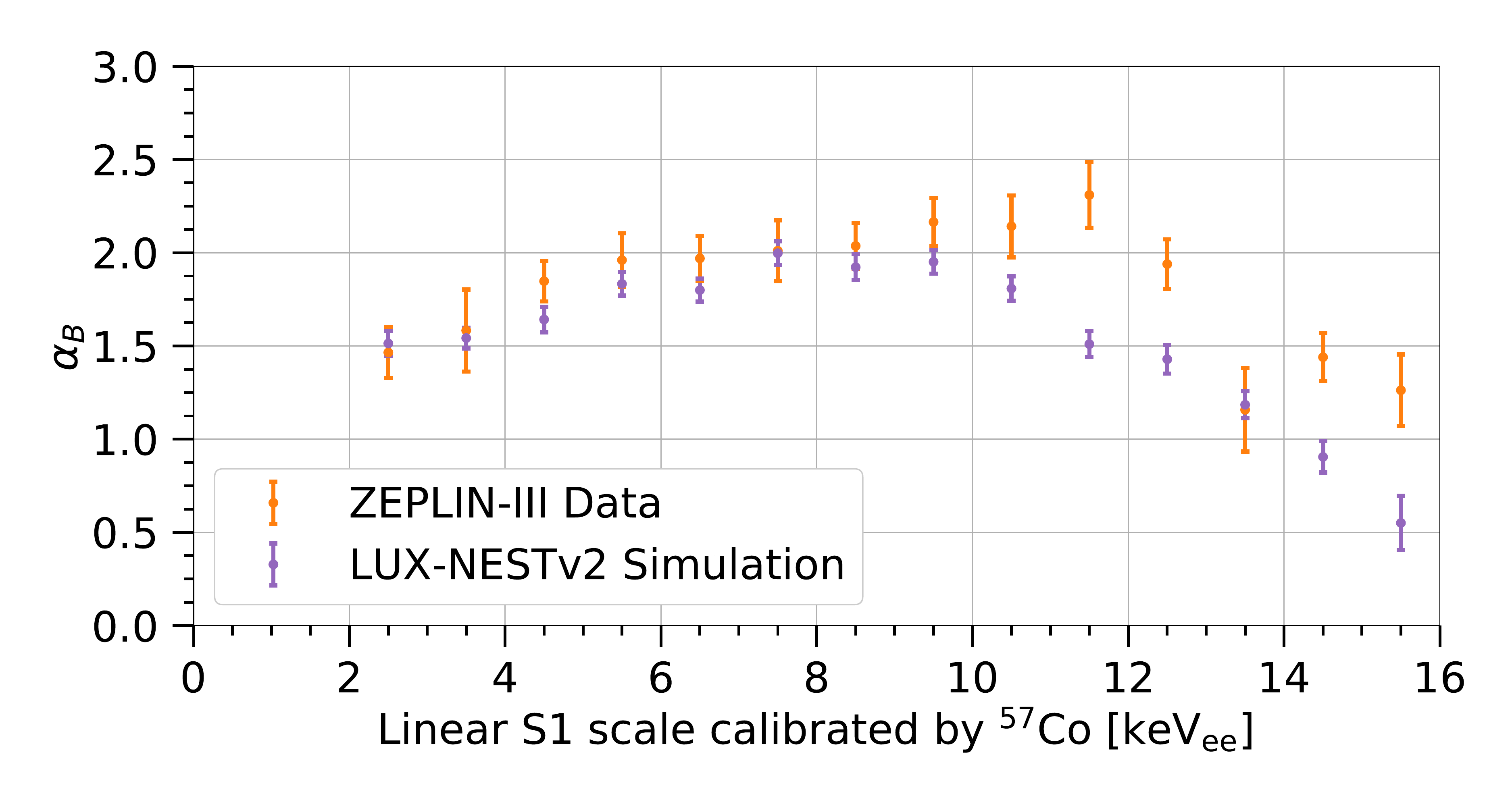}}
\caption{A comparison of the skewness of the ER band in \mbox{ZEPLIN-III} data \cite{Lebedenko2009_ZEPLIN_III_First_Science_Run, Araujo2020_ZEPLIN_III_Revision} vs.~our simulation of the \mbox{ZEPLIN-III} data using LUX-NESTv2 and our model in Eq.~\ref{eq:Skewness_Model}. We use a constant light yield of 1.881~phd/keVee to convert the simulated S1 signal to the energy scale used by \mbox{ZEPLIN-III}. The data points match our simulation within 1~standard deviation below 10~keVee. We do not reproduce the highest-energy data with complete accuracy, but we do observe the same qualitative decrease in skewness between 10 and 16~keVee.}
\label{fig:ER_ZEPLIN_III_Comparison}
\end{figure}


\end{document}